\documentclass[letter,preprint,aps,superscriptaddress,nofootinbib,tightenlines,preprintnumbers,cites]{revtex4}
\usepackage{graphicx}
\usepackage{enumerate}
\usepackage{bm}
\usepackage{slashed}
\usepackage{multirow}
\usepackage{adjustbox}
\usepackage{mathtools}
\usepackage{verbatim}
\usepackage{color}
\usepackage{amsmath,amssymb}
\usepackage{epsfig}
\usepackage{hhline}
\usepackage{placeins}

\newcommand\psidag{\psi^{\dagger}}
\newcommand\tautoinfty{\underset{\tau \to\infty}{\longrightarrow}}
\newcommand\Eq[1]{Eq.~(\ref{eq:#1})}
\newcommand\Eqs[2]{Eqs.~(\ref{eq:#1},\ref{eq:#2})}
\newcommand\Fig[1]{Fig.~\ref{fig:#1}}

\newcommand\Sec[1]{Sec.~\ref{sec:#1}}

\newcommand\Tab[1]{Table~\ref{tab:#1}}
\newcommand{\be}{\begin{equation}}
\newcommand{\ee}{\end{equation}}
\newcommand\beq{\begin{eqnarray}}
\newcommand\eeq{\end{eqnarray}}

\newcommand{\bfx}{{\mathbf x}}

\newcommand\Ncfg{N_{\mbox{\tiny cfg}}}

\newcommand{\calC}{{\mathcal{ C}}}
\newcommand{\calD}{{\mathcal{ D}}}
\newcommand{\calE}{{\mathcal{ E}}}

\newcommand{\calI}{{\mathcal{ I}}}

\newcommand{\calO}{{\mathcal{ O}}}
\newcommand{\calM}{{\mathcal{ M}}}
\newcommand{\calN}{{\mathcal{ N}}}

\newcommand{\calR}{{\mathcal{ R}}}

\newcommand{\calT}{{\mathcal{ T}}}

\newcommand{\calV}{{\mathcal{ V}}}

\newcommand{\calL}{{\mathcal{ L}}}

\newcommand{\mybar}[1]%
        {\kern 0.6pt\overline{\kern -0.6pt#1\kern -0.6pt}\kern 0.6pt}
\begin{document}
\unitlength = 1mm

\title{Lattice methods and effective field theory}
\author{Amy N. Nicholson}
 \email{anicholson@berkeley.edu} 
\affiliation{Department of Physics,
University of California, Berkeley, Berkeley CA 94720, USA}

 \date{\today}

\begin{abstract}
Lattice field theory is a non-perturbative tool for studying properties of strongly interacting field theories, which is particularly amenable to numerical calculations and has quantifiable systematic errors. In these lectures we apply these techniques to nuclear Effective Field Theory (EFT), a non-relativistic theory for nuclei involving the nucleons as the basic degrees of freedom. The lattice formulation of \cite{EKLN1,EKLN4} for so-called pionless EFT is discussed in detail, with portions of code included to aid the reader in code development. Systematic and statistical uncertainties of these methods are discussed at length, and extensions beyond pionless EFT are introduced in the final Section.
\end{abstract}

\maketitle

\tableofcontents
 
\section{\label{sec:intro}Introduction}
Quantitative understanding of nuclear physics at low energies from first principles remains one of the most challenging programs in contemporary theoretical physics research. While physicists have for decades used models combined with powerful numerical techniques to successfully reproduce known nuclear structure data and make new predictions, currently the only tools available for tackling this problem that have direct connections to the underlying theory, Quantum Chromodynamics (QCD), as well as quantifiable systematic errors, are Lattice QCD and Effective Field Theory (EFT). In principle, when combined these techniques may be used to not only quantify any bias introduced when altering QCD in order to make it computationally tractable, but also to better understand the connection between QCD and nuclear physics.

The lattice is a tool for discretizing a field theory in order to reduce the path integral, having an infinite number of degrees of freedom, to a finite-dimensional ordinary integral. After rendering the dimension finite (though extremely large), the integral may then be estimated on a computer using Monte Carlo methods. Errors introduced through discretization and truncation of the region of spacetime sampled are controlled through the spatial and temporal lattice spacings, $b_s,b_{\tau}$, and the number of spatial and temporal points, $L,N_{\tau}$. Thus, these errors may be quantified through the lattice spacing dependence of the observables, and often may be removed through extrapolation to the continuum and infinite volume limits.

LQCD is a powerful and advanced tool for directly calculating low-energy properties of QCD. However, severe computational issues exist when calculating properties of systems with nucleons. Unfortunately, these problems grow rapidly with the number of nucleons in the system. 

The first issue is the large number of degrees of freedom involved when using quark fields to create nucleons. In order to calculate a correlation function for a single nucleon in LQCD using quarks (each of which has twelve internal degrees of freedom given by spin and color), one has to perform all possible Wick contractions of the fields in order to build in fermion antisymmetrization. For example, to create a proton using three valence quark operators requires the calculation of two different terms corresponding to interchanging the two up quark sources. The number of contractions involved for a nuclear correlation function grows with atomic number $Z$ and mass number $A$ as $(A+Z)!(2A-Z)!$. For He$_4$ this corresponds to $\sim 5 \times 10^5$ terms\footnote{This is a very na\"ive estimate; far more sophisticated algorithms exist with power-law scaling.}!

The second major problem occurs when performing a stochastic estimate of the path integral. A single quark propagator calculated on a given gauge field configuration may be a part of either a light meson or a heavy nucleon. However, the difference cannot be determined until correlations with the other quark fields present are built in by summing over a sufficiently large number of these field configurations\footnote{This interpretation of the signal-to-noise problem has been provided by David B. Kaplan.}. This leads to large fluctuations from configuration to configuration, and a stochastic signal-to-noise ratio, $\calR$, which degrades exponentially with the number of nucleons in the system,
\beq
\calR \sim e^{-A(M-3/2 m_{\pi})\tau} \ ,
\eeq
where $M$ is the nucleon mass and $m_{\pi}$ is the pion mass \cite{Lepage:1989hd}. This is currently the major limiting factor for the size of nuclear which can be probed using LQCD. The best calculations we have from LQCD using multiple nucleons to date are in the two-nucleon sector \cite{Berkowitz:2015eaa,Kurth:2015cvl,Nicholson:2015pys,Orginos:2015aya,Detmold:2015daa,Chang:2015qxa,Beane:2015yha,Beane:2014sda,Beane:2014ora,Beane:2013br,Beane:2012vq,Beane:2011iw,Beane:2009py,Yamazaki:2015vjn,Yamazaki:2015asa,Yamazaki:2013rna,Yamazaki:2012fn,Yamazaki:2012hi,Doi:2015uvd,Doi:2015oha,Ishii:2006ec,Murano:2013xxa,Aoki:2014mia,Murano:2013gta,HALQCD:2012aa,Inoue:2010hs}, while fewer calculations have been performed for three and four nucleon systems \cite{Beane:2009gs,Beane:2012vq,Beane:2014ora,Beane:2014sda,Chang:2015qxa,Yamazaki:2015vjn,Yamazaki:2015asa,Yamazaki:2013rna,Yamazaki:2012fn,Yamazaki:2012hi,Doi:2011gq}; however, even for two nucleon systems unphysically large pion masses must be used in order to reduce the noise problem. We will discuss signal-to-noise problems in more detail in \Sec{SNR}. 

Starting from an EFT using nucleons as the fundamental degrees of freedom greatly reduces the consequences from both of these issues. EFTs also enjoy the same benefit as the lattice over traditional model techniques of having quantifiable systematic errors, this time controlled by the cutoff of the EFT compared to the energy regime studied. For chiral EFTs this scale is generally $\Lambda_{\chi} \sim m_{\rho} \sim 700$ MeV. Systematic errors can be reduced by going to higher orders in an expansion of $p/\Lambda_{\chi}$, where $p$ is the momentum scale probed, with the remaining error given by the size of the first order which is not included. In a potential model there is no controlled expansion, and it is generally unknown how much the results will be affected by leaving out any given operator. In addition, field theories provide a rigorous mathematical framework for calculating physical processes, and can be directly translated into a lattice scheme.

In these lecture notes we will explore the use of lattice methods for calculating properties of many-body systems starting from nuclear EFT, rather than QCD. Our discussion will begin with understanding a very basic nuclear EFT, pionless EFT, at leading order. We will then proceed to discretize this theory and set up a framework for performing Monte Carlo calculations of our lattice theory. We will then discuss how to calculate observables using the lattice theory, and how to understand their associated statistical uncertainties. Next we will discuss quantifying and reducing systematic errors. Then we will begin to add terms to our theory going beyond leading order pionless EFT. Finally, we will discuss remaining issues and highlight some successes of the application of these methods by several different groups.

\section{Basics of Effective Field Theory and Lattice Effective Field Theory}
\subsection{\label{sec:EFT}Pionless Effective Field Theory}
To develop an EFT we will first write down all possible operators involving the relevant degrees of freedom within some energy range (determined by the cutoff) that are consistent with the symmetries of the underlying theory. Each operator will be multiplied by an unknown low-energy constant which may be fixed by comparing an observable with experiment or lattice QCD. In order to reduce this, in principle, infinite number of operators to a finite number we must also establish a power-counting rule for neglecting operators that do not contribute within some desired accuracy. This is a notoriously difficult problem for nuclear physics, and is in general observable and renormalization scheme dependent. Here, we will only briefly touch upon two common power-counting schemes, the so-called Weinberg and KSW expansions \cite{Weinberg:1990rz,Weinberg:1991um,Kaplan:1996xu,Kaplan:1998tg,Kaplan:1998we}. For reviews of these and other power-counting schemes, see \cite{Epelbaum:2008ga,Epelbaum:2010nr,Machleidt:2011zz}.

The simplest possible nuclear EFT involves non-relativistic nucleon fields interacting via delta functions. This is known as a pionless EFT, and is only relevant for energy scales up to a cutoff $\Lambda \sim m_{\pi}$. Below this scale, the finite range of pion exchange cannot be resolved, and all interactions appear to be point-like. In this discussion we will closely follow that of Ref.~\cite{Kaplan:2005es}. For the moment, let's just consider a theory of two-component (spin up/down) fermion fields, $\psi$, with the following Lagrangian,
\beq
\label{eq:leff}
\mathcal{L}_{\mbox{\tiny eff}} = \psi^{\dagger}\left( i \partial_{\tau} + \frac{\nabla^2}{2M}\right) \psi + g_0 \left(\psi^{\dagger}\psi\right)^2 + \frac{g_2}{8}\left[ \left(\psi\psi\right)^{\dagger}\left(\psi\overleftrightarrow{\nabla}^2\psi\right)+ \mbox{\tiny h.c.}\right]+\cdots \ ,
\eeq
where
\beq
\overleftrightarrow{\nabla}^2 \equiv \overleftarrow{\nabla}^2-2\overleftarrow{\nabla} \cdot \overrightarrow{\nabla}+\overrightarrow{\nabla}^2 \ ,
\eeq
$M$ represents the nucleon mass, $g_0, g_2, \ldots$ are unknown, low-energy constants (LECs) which may be fixed by comparing to experimental or LQCD results, and all spin indices are suppressed. Because the effective theory involves dynamical degrees of freedom that are only relevant up to a certain scale, we must define a cutoff, $\Lambda$, above which the theory breaks down. In general, the LECs scale as $\Lambda^{-\mbox{\tiny dim}(\mathcal{O})}$, where dim$(\mathcal{O})$ represents the dimension of the operator associated with the LEC. According to na\"ive power counting, the $g_2$ term in \Eq{leff} should be suppressed relative to the $g_0$ term, because adding a derivative to an operator increases its dimension. One should be careful in practice, however, because na\"ive power counting does not always hold, as we will see several times throughout these lectures. 

\subsubsection{\label{sec:scatamp}Two particle scattering amplitude}
In order to set the coefficients $g_0, g_2, \ldots$, we may look to experimental scattering data. In particular, if we wish to set the $g_0$ coefficient we should consider two-particle $s$-wave scattering because the operator associated with $g_0$ contains no derivatives. $g_2$ and other LECs may be set using $p$- and higher-wave scattering data. Recall that the S-matrix for non-relativistic scattering takes the following form:
\beq
S=1+\frac{iMp}{2\pi}A \ ,
\eeq
where $p$ is the scattering momentum and $A$ is the scattering amplitude. For $s$-wave scattering the amplitude may be written as,
\beq
\label{eq:Apcotd}
A=\frac{4\pi}{M} \frac{1}{p\cot\delta - ip} \ ,
\eeq
where $\delta$ is the $s$-wave scattering phase shift. Given a short-range two-body potential, the scattering phase shift has a well-known expansion for low momenta, called the effective range expansion,
\beq
\label{eq:ere}
p\cot\delta = -\frac{1}{a}+\frac{1}{2}r_0p^2+r_1p^4+\cdots \ ,
\eeq
where $a$ is the scattering length, $r_0$ is the effective range, and $r_1$ and higher order terms are referred to as shape parameters. The effective range and shape parameters describe the short-range details of the potential, and are generally of order of the appropriate power of the cutoff in a naturally tuned scenario. 

\begin{figure}
\begin{center}
\includegraphics[width=0.3\linewidth]{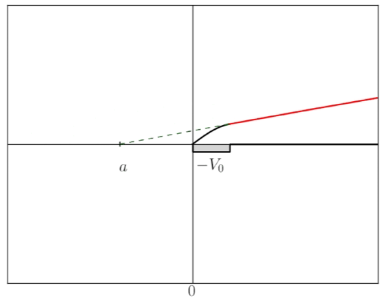}
\includegraphics[width=0.3\linewidth]{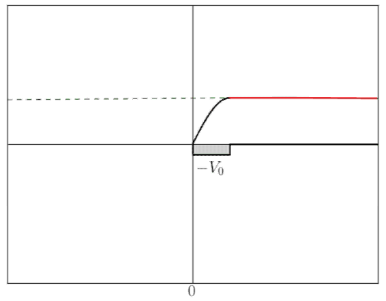}
\includegraphics[width=0.3\linewidth]{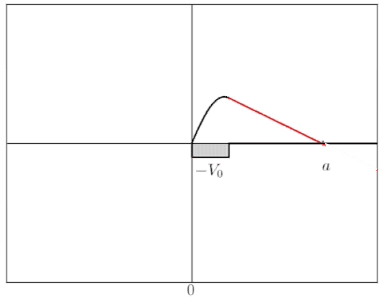}
\caption{\label{fig:a0}Sketches of two-body radial wavefunctions vs. $r$ corresponding to various scattering lengths. From left to right: $a<0$, $a\to\infty$,$a>0$.}
\end{center}
\end{figure}

The scattering length may be used to describe the asymptotic behavior of the radial wavefunction. In particular, consider two-particles interacting via an attractive square-well potential. If the square-well is sufficiently strongly attractive, the wavefunction turns over and goes to zero at some finite characteristic length. This means the system is bound and the size of the bound state is given by the scattering length, $a$. On the other hand, if the wavefunction extends over infinite space, then the system is in a scattering state and the scattering length may be determined as the distance from the origin where the asymptote of the wavefunction intersects the horizontal axis (see \Fig{a0}). This implies that the scattering length in the case of a scattering state is negative. If the potential is tuned to give a system which is arbitrarily close to the crossover point from a bound state to a scattering state, corresponding to infinite scattering length, the state is described as being near unitarity, because the unitarity bound on the scattering cross section is saturated at this point. Note that this implies that the scattering length may be any size and is not necessarily associated with the scale set by the cutoff. However, such a scenario requires fine-tuning of the potential. Such fine-tuning is well-known to occur in nuclear physics, with the deuteron and neutron-neutron $s$-wave scattering being notable examples.

A many-body system composed of two-component fermions with an attractive interaction is known to undergo pairing between the species (higher $N$-body interactions are prohibited by the Pauli exclusion principle), such as in neutron matter, found in the cores of neutron stars, which is composed of spin up and spin down neutrons. At low temperature, these bosonic pairs condense into a coherent state. If the interaction is only weakly attractive, the system will form a BCS state composed of widely separated Cooper pairs, where the average pair size is much larger than the average interparticle spacing. On the other hand, if the interaction is strongly attractive then the pairs form bosonic bound states which condense into a Bose-Einstein condensate. The crossover between these two states corresponds to the unitary regime, and has been studied extensively in ultracold atom experiments, where the interaction between atoms may be tuned using a Feshbach resonance. In this regime, the average pair size is equal to the interparticle spacing (given by the inverse density), which defines the only scale for the system. Thus, all dimensionful observables one wishes to calculate for this system are determined by the appropriate power of the density times some dimensionless constant.  For a review of fermions in the unitary regime, see e.g., \cite{AtomsReview1,AtomsReview2}.

\subsubsection{\label{sec:couplings}Two-body LECs}
\begin{figure}
\begin{center}
\includegraphics[width=\linewidth]{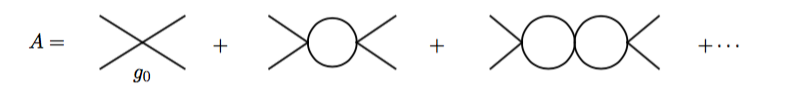}
\caption{\label{fig:bubblesum}Two-body scattering amplitude represented as a sum of bubble diagrams corresponding to a single contact interaction with coupling $g_0$.}
\end{center}
\end{figure}

\begin{figure}
\begin{center}
\includegraphics[width=\linewidth]{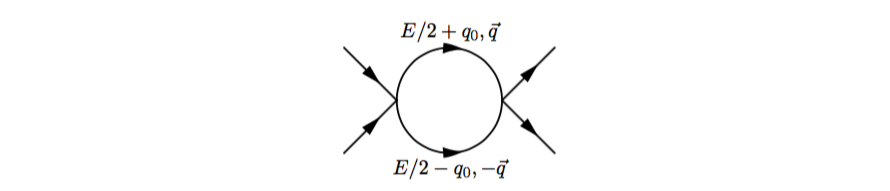}
\caption{\label{fig:loop}Feynman diagram for a single bubble in \Fig{bubblesum}, giving rise to the loop integral \Eq{loop}.}
\end{center}
\end{figure}

Returning to our task of setting the couplings using scattering parameters as input, we might consider comparing \Eq{leff} and \Eq{ere}, to determine the LEC $g_0$ using the scattering length, $g_2$ using the effective range, and so forth. To see how this is done in practice we may compute the scattering amplitude $A$ in the effective theory, and match the coefficients to the effective range expansion. Let's begin using only the first interaction term in the effective theory, corresponding to $g_0$. Diagrammatically, the scattering amplitude may be written as the sum of all possible bubble diagrams (see \Fig{bubblesum}). Because the scattering length may take on any value, as mentioned previously, we cannot assume that the coupling $g_0$ is small, so we should sum all diagrams non-perturbatively. The first diagram in the sum is given by the tree level result, $g_0$. If we assume that the system carries energy $E=p^2/M$, then the second diagram may be labeled as in \Fig{loop}, and gives rise to the loop integral,
\beq
\label{eq:loop}
I_0 = i\int \frac{d^4q}{(2\pi)^4}\frac{1}{\left(E/2+q_0-\frac{q^2}{2M}-i\epsilon\right)\left(E/2-q_0-\frac{q^2}{2M}+i\epsilon\right)} \ .
\eeq 
Performing the integral over $q_0$ and the solid angle gives
\beq
\label{eq:loopEFT}
I_0 &=& \frac{1}{2\pi^2}\int^{\pi\Lambda/2} dq\frac{q^2}{\left(E-\frac{q^2}{M}\right)} \\
&=& \frac{M}{2\pi^2}\left[\frac{\pi\Lambda}{2}-\sqrt{ME}\tanh^{-1}\left(\frac{\Lambda}{\sqrt{ME}}\right)\right] \ ,
\eeq
where I have introduced a hard momentum cutoff, $\Lambda$. Removing the cutoff by taking it to infinity results in
\beq
I_0 \underset{\Lambda\to\infty}{\longrightarrow}\frac{M}{4\pi}\left[\Lambda+ip\right] \ .
\eeq
Because the interaction is separable, the $n$th bubble diagram is given by $n$ products of this loop function. Thus, the scattering amplitude is factorizable, and may be written
\beq
\label{eq:Abubble}
A&=&g_0\left[1+\sum_n\left(g_0I_0\right)^n\right] \\
&=& \frac{g_0}{1-g_0I_0} \ .
\eeq
We may now compare \Eqs{Apcotd}{ere} and \Eq{Abubble} to relate the coupling $g_0$ to the scattering phase shift. This is easiest to do by equating the inverse scattering amplitudes,
\beq
\frac{1}{A} &=& \frac{1}{g_0}-\frac{M}{4\pi}\Lambda -\frac{iMp}{4\pi} = -\frac{M}{4\pi a}-\frac{iMp}{4\pi} \ ,
\eeq
where I have used \Eq{ere} cut off at leading order. We now have the relation
\beq
g_0 = \frac{4\pi}{M} \frac{1}{\Lambda-1/a} 
\eeq
between the coupling and the physical scattering length. 


\begin{figure}
\begin{center}
\includegraphics[width=0.5\linewidth]{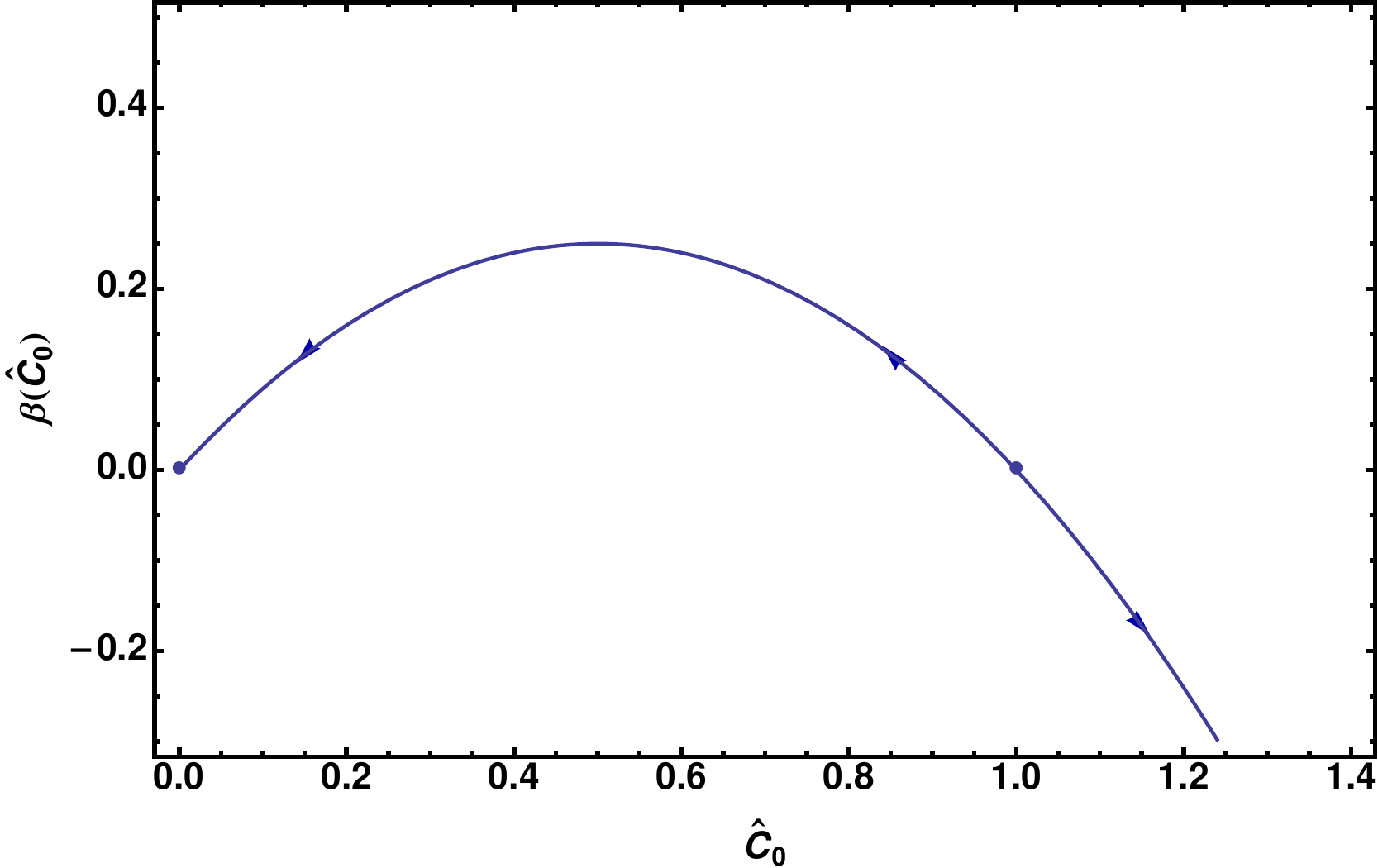}
\end{center}
\caption{\label{fig:beta}Beta function (\Eq{beta}) for the two-body contact interaction. Arrows represent the direction of flow toward the IR.}
\end{figure}

Note that the coupling runs with the scale $\Lambda$; the particular dependence is determined by the regularization and renormalization scheme chosen. In order to understand the running of the coupling we may examine the beta function. To do so we first define a dimensionless coupling,
\beq
\hat{g}_0 \equiv -\frac{M\Lambda}{4\pi}g_0 \ ,
\eeq
then calculate
\beq
\label{eq:beta}
\beta\left(\hat{g}_0\right) \equiv \Lambda \frac{\partial \hat{g}_0}{\partial \Lambda} = -\frac{a\Lambda}{\left(a\Lambda-1\right)^2} = -\hat{g}_0\left(\hat{g}_0-1\right) \ .
\eeq
This function is a simple quadratic that is plotted in \Fig{beta}. The beta function has two zeroes, $\hat{g}_0 = 0,1$, corresponding to fixed points of the theory. At a fixed point, the coupling no longer runs with the scale $\Lambda$, and the theory is said to be scale-invariant (or conformal, given some additional conditions). This means that there is no intrinsic scale associated with the theory. The fixed point at $\hat{g}_0=0$ is a trivial fixed point, and corresponds to a non-interacting, free field theory (zero scattering length). The other, non-trivial fixed point at $\hat{g}_0=1$ corresponds to a strongly interacting theory with infinite scattering length; this is the unitary regime mentioned previously. Here, not only does the scattering length go to infinity, as does the size of the radial wavefunction, but the energy of the bound state (as approached from $\hat{g}_0 > 1$) goes to zero and all relevant scales have vanished. Note that this is an unstable fixed point; the potential must be finely tuned to this point or else the theory flows away from unitarity as $\Lambda \to 0$ (IR limit).

Generally perturbation theory is an expansion around free field theory, corresponding to a weak coupling expansion. This is the approach used as part of the Weinberg power counting scheme for nuclear EFT \cite{Weinberg:1990rz,Weinberg:1991um}. However, in some scattering channels of interest for nuclear theory the scattering length is indeed anomalously large, such as the $^1S_0$ and $^3S_1$ nucleon-nucleon scattering channels, where
\beq
a_{^1S_0} &\sim& -24 \mbox{ fm} \ , \\
a_{^3S_1} &\sim& 5 \mbox{ fm} \ .
\eeq
Such large scattering lengths suggest that an expansion around the strongly coupled fixed-point of unitarity may be a better starting point and lead to better convergence. This approach was taken by Kaplan, Savage, and Wise and led to the KSW power-counting scheme \cite{Kaplan:1998we,Kaplan:1998tg,Kaplan:1996xu}. Unfortunately, nuclear physics consists of many scales of different sizes and a consistent power-counting framework with good convergence for all observables has yet to be developed; in general the convergence of a given scheme depends on the scattering channels involved. 

Because nuclear physics is not weakly coupled in all channels, non-perturbative methods, such as lattice formulations, will be favorable for studying few- and many-body systems, where two-body pairs may interact through any combination of channels simultaneously. Due to the scale-invariant nature of the unitary regime, it provides a far simpler testbed for numerical calculations of strongly-interacting theories, so we will often use it as our starting point for understanding lattice EFT methods. 

\subsection{\label{sec:LEFT}Lattice Effective Field Theory}

Our starting point for building a lattice EFT will be the path integral formulation of quantum field theory in Euclidean spacetime. The use of Euclidean time allows the exponent of the path integral to be real (in certain cases), a property which will be essential to our later use of stochastic methods for its evaluation. Given a general theory for particles $\psi,\psi^{\dagger}$ obeying a Lagrangian density \beq 
\mathcal{L}(\psi^{\dagger},\psi) = \psi^{\dagger}\left( \partial_{\tau}-\mu \right)\psi+ \mathcal{H}\left[\psi^{\dagger}, \psi\right] \ ,
\eeq
where $\tau$ is the Euclidean time, $\mu$ the chemical potential, and $\mathcal{H}$ is the Hamiltonian density, the Euclidean path integral is given by
\beq
Z=\int \mathcal{D} \psi^{\dagger}\mathcal{D} \psi e^{-\int d\tau d^3x\left[\mathcal{L}(\psi^{\dagger},\psi)\right]} \ .
\eeq
If the integral over Euclidean time is compact, then the finite time extent $\beta$ acts as an inverse temperature, and we may draw an analogy with the partition function in statistical mechanics, $Z = tr\left[e^{-H\beta}\right]$. This analogy is often useful when discussing lattice formulations of the path integral. In this work we will generally consider $\mu=0$ and create non-zero particle density by introducing sources and sinks for particles and calculating correlation functions. 

We discretize this theory on a square lattice consisting of $L^3 \times N_{\tau}$ points, where $L$ is the number of points in all spatial directions, and $N_{\tau}$ is the number of temporal points. We will focus on zero temperature physics, corresponding to large $N_{\tau} ~$\footnote{The explicit condition on $N_{\tau}$ required for extracting zero temperature observables will be discussed in \Sec{observables}}. We must also define the physical distance between points, the lattice spacings $b_s, b_{\tau}$, where $b_{\tau} = b_s^2/M$ by dimensional analysis for non-relativistic theories. The fields are now labeled by discrete points, $\psi(\vec{x},\tau) \to \psi_{\vec{n},\tau}$, and continuous integrals are replaced by discrete sums, $\int d^3x \to \sum_{\vec{n},\tau}^{L,N_{\tau}}$.

\subsubsection{Free field theory}
To discretize a free field theory, we must discuss discretization of derivatives. The simplest operator which behaves as a single derivative in the continuum limit is a finite difference operator,
\beq
\partial_{\hat{k}}^{(L)} f_j = \frac{1}{b_s}\left[f_{j+\hat{k}}-f_j\right] \ ,
\eeq
where $\hat{k}$ is a unit vector in the $k$-direction. The discretized second derivative operator must involve two hops, and should be a symmetric operator to behave like the Laplacian. A simple possibility is
\beq
\nabla_L^2 f_j = \sum_{k} \frac{1}{b_s^2}\left[ f_{j+\hat{k}}+f_{j-\hat{k}}-2 f_j \right] \ .
\eeq
We can check the continuum limit by inspecting the corresponding kinetic term in the action,
\beq
S_{\mbox{\tiny KE}} \propto \sum_j \psi_j^{\dagger} \nabla_L^2 \psi_j \ .
\eeq
The fields may be expanded in a plane wave basis,
\beq
\psi_j = \sum_{k=-L/2}^{L/2} \psi_k e^{-\frac{2\pi i}{L} j \cdot k} \ ,
\eeq
for spatial indices, $j$, leading to
\beq
\sum_j \psi_j^{\dagger} \nabla_L^2 \psi_j = \frac{1}{b_s^2}\sum_j \sum_{k'}\sum_k \psi_{k'}^{\dagger} \psi_k \left[ e^{\frac{2\pi i}{L}j \cdot k'} e^{\frac{-2\pi i}{L} j\cdot k} \right] \left[e^{\frac{-2\pi i}{L}k}+e^{\frac{2\pi i}{L} k}-2\right] \ .
\eeq
After performing the sum over $j$ the first piece in brackets gives $\delta_{kk'}$, while the second is proportional to $\sin^2(k\pi/L)$, resulting in,
\beq
\sum_j \psi_j^{\dagger} \nabla_L^2 \psi_j = -\frac{4}{b_s^2}\sum_k \psi_k^{\dagger}\psi_k \sin^2\left(\frac{k\pi}{L}\right) \ .
\eeq 
Finally, expanding the sine function for small $k/L$ gives,
\beq
\label{eq:kinetic}
\sum_j \psi_j^{\dagger} \nabla_L^2 \psi_j = \sum_k \psi_k^{\dagger}\psi_k &&\left[\underbrace{-\left(\frac{2\pi k}{b_s L}\right)^2+\frac{b_s^2}{12}\left(\frac{2\pi k}{b_s L}\right)^4+ \cdots }\right] \ ,\cr
&& \hspace{6mm} -p^2 + \frac{b_s^2}{12} p^4 + \cdots \underset{b_s\to 0}{\longrightarrow} -p^2 
\eeq
where I've used the finite volume momentum $p = \frac{2\pi k}{b_s L}$ to rewrite the expression in square brackets. Thus, we have the correct continuum limit for the kinetic operator. Note that for larger momenta, approaching the continuum limit requires smaller $b_s$. However, this is only one possibility for a kinetic term. We can always add higher dimension operators (terms with powers of $b_s$ in front of them), in order to cancel leading order terms in the expansion \Eq{kinetic}. This is a form of what's called improvement of the action, and will be discussed in more detail in \Sec{systematic}.

Adding a temporal derivative term, 
\beq
\partial_{\tau}^{(L)} \psi_{\vec{x},\tau} = \frac{1}{b_{\tau}}\left[\psi_{\vec{n},\tau}- \psi_{\vec{n},\tau-1}\right] \ ,
\eeq
we can now write down a simple action for a non-relativistic free-field theory,
\beq
S_{\mbox{\tiny free}} = \sum_{\tau,\tau'} \frac{1}{b_{\tau}}\psi_{\tau'}^{\dagger}\left[K_0\right]_{\tau,\tau'}\psi_{\tau} \ ,
\eeq
where I've defined a matrix $K_0$ whose entries are $L^3 \times L^3$ blocks,
\beq
K_0 \equiv \left(\begin{array}{ccccccc}
D & -1 & 0 & 0 & . & . & . \\
0 & D & -1 & 0 & . & . &.  \\
0 & 0 & D & -1 &  .&.  & . \\
. & .&. &. & . & & \\ 
. & .& .& .& & . & \\ 
1 & .&. & .& & & . \\ 
\end{array} \right)
\eeq
where $D \equiv 1-\frac{b_s^2 \nabla_L^2}{2}$ contains the spatial Laplacian, and therefore connects fields on the same time slice (corresponding to diagonal entries of the matrix $K_0$), while the temporal derivative contributes the off-diagonal pieces. Note that the choice of ``1" in the lower left corner corresponds to anti-periodic boundary conditions, appropriate for fermionic fields. For zero temperature calculations the temporal boundary conditions are irrelevant, and it will often be useful to choose different temporal boundary conditions for computational or theoretical ease. 

\subsubsection{Interactions}
Now let's discuss adding interactions to the theory. We'll focus on the first term in a nuclear EFT expansion, the four-fermion interaction:
\beq
\mathcal{L}_{\mbox{\tiny int}} = \sum_n g_0 \psi_{n,\uparrow} \psi_{n,\uparrow} \psi_{n,\downarrow} \psi_{n,\downarrow} \ ,
\eeq
where $(\uparrow,\downarrow)$ now explicitly label the particles' spins (or alternatively, flavors). Because anti-commuting fields cannot easily be accommodated on a computer, they must be integrated out analytically. The only Grassmann integral we know how to perform analytically is a Gaussian, so the action must be bilinear in the fields. One trick for doing this is called a Hubbard-Stratonovich (HS) transformation, in which auxiliary fields are introduced to mediate the interaction. The key is to use the identity,
\beq
e^{b_{\tau}g_0\psi_{\uparrow}^{\dagger}\psi_{\uparrow}\psi_{\downarrow}^{\dagger}\psi_{\downarrow}} = \frac{1}{\sqrt{2\pi}}\int_{-\infty}^{\infty}d\phi^{-\phi^2/2-\phi\sqrt{b_{\tau}g_0}\left(\psi_{\uparrow}^{\dagger}\psi_{\uparrow}+\psi_{\downarrow}^{\dagger}\psi_{\downarrow}\right)} \ ,
\eeq
where I have dropped the spacetime indices for brevity. This identity may be verified by completing the square in the exponent on the right hand side and performing the Gaussian integral over the auxiliary field $\phi$. This form of HS transformation has the auxiliary field acting in what is called the density channel $\left(\psi_{\uparrow}^{\dagger}\psi_{\uparrow}+\psi_{\downarrow}^{\dagger}\psi_{\downarrow}\right)$. It is also possible to choose the so-called BCS channel, $\left(\psi_{\uparrow}^{\dagger}\psi_{\downarrow}^{\dagger}+\psi_{\uparrow}\psi_{\downarrow}\right)$, the usual formulation used in BCS models, however this causes a so-called sign problem when performing Monte Carlo sampling, as will be discussed in detail in \Sec{sign}. Transformations involving non-Gaussian auxiliary fields may also be used, such as
\beq
Z_2 \mbox{ field: } &&\frac{1}{2} \sum_{\phi=\pm 1}e^{-\phi\sqrt{b_{\tau}g_0}\left(\psi_{\uparrow}^{\dagger}\psi_{\uparrow}+\psi_{\downarrow}^{\dagger}\psi_{\downarrow}\right)} \cr
\mbox{compact continuous: } && \frac{1}{2\pi} \int_{-\pi}^{\pi}e^{-\sin \phi\sqrt{b_{\tau}g_0}\left(\psi_{\uparrow}^{\dagger}\psi_{\uparrow}+\psi_{\downarrow}^{\dagger}\psi_{\downarrow} \right)}  \ .
\eeq
These formulations may have different pros and cons in terms of computational and theoretical ease for a given problem, and should be chosen accordingly. For example, the $Z_2$ interaction is conceptually and computationally the simplest interaction, however, it also induces explicit $4-$ and higher-body interactions in systems involving more than two-components which may not be desired. 

\subsubsection{Importance sampling}
The action may now be written with both kinetic and interaction terms,
\beq
\label{eq:actiongen}
S=\frac{1}{b_{\tau}}\sum_{\tau,\tau'}\psi_{\tau'}^{\dagger}\left[K(\phi)\right]_{\tau'\tau}\psi_{\tau} \ ,
\eeq
where the matrix $K$ includes blocks which depend on the auxiliary field $\phi$, and also contains non-trivial spin structure that has been suppressed. The partition function can be written
\beq
Z = \int \mathcal{D}\phi \mathcal{D}\psi^{\dagger}\mathcal{D}\psi \rho[\phi]e^{-S[\phi,\psi^{\dagger}\psi]} \ ,
\eeq
where the integration measure for the $\phi$ field, $\rho[\phi]$, depends on the formulation chosen,
\beq
\rho[\phi] = \left\{ \begin{array}{cc}
\prod_n e^{-\phi_n^2/2} & \mbox{Gaussian}\\
\prod_n \frac{1}{2}\left(\delta_{\phi_{n,1}}+\delta_{\phi_{n,-1}}\right) & Z_2 \\
\prod_n \left(\theta(-\pi+\phi_n)\theta(\pi-\phi_n)\right) & \mbox{compact continuous} 
\end{array}\right. \ .
\eeq

With the action in the bilinear form of \Eq{actiongen}, the $\psi$ fields can be integrated out analytically, resulting in
\beq
\label{eq:prob}
Z_{\phi}=\int \calD \phi P[\phi] \qquad P[\phi] \equiv \rho[\phi]\det K[\phi] \ .
\eeq
Observables take the form
\beq
\langle \calO \rangle = \frac{1}{Z} \int \calD \phi P[\phi]\calO[\phi] \ .
\eeq

Through the use of discretization and a finite volume, the path integral has been converted into a standard integral with finite dimension. However, the dimension is still much too large to imagine calculating it on any conceivable computer, so we must resort to Monte Carlo methods for approximation. The basic idea is to generate a finite set of $\phi$ field configurations of size $\Ncfg$ according to the probability measure $P[\phi]$, calculate the observable on each of these configurations, then take the mean as an approximation of the full integral,
\beq
\langle \calO \rangle \approx \frac{1}{\Ncfg}\sum_n^{\Ncfg}\calO(\phi_n) \ .
\eeq
Assuming the central limit theorem holds, for $\Ncfg$ large enough (a non-trivial condition, as will be discussed in \Sec{overlap}), the distribution of the mean approaches a Gaussian, and the error on the mean falls off with the square root of the sample size. 

There are several algorithms on the market for generating field configurations according to a given probability distribution, and I will only briefly mention a few. Lattice calculations are particularly tricky due to the presence of the determinant in \Eq{prob}, which is a highly non-local object and is very costly to compute. One possible algorithm to deal with this is called determinantal Monte  Carlo, which implements local changes in $\phi$, followed by a simple Metropolis accept/reject step. This process can be rather inefficient due to the local updates. An alternative possibility is Hybrid Monte Carlo, commonly used for lattice QCD calculations, in which global updates of the field are produced using molecular dynamics as a guiding principle. Note that the field $\phi$ must be continuous in order to use this algorithm due to the use of classical differential equations when generating changes in the field. Also common in lattice QCD calculations is the use of pseudofermion fields as a means for estimating the fermion determinant. Here the determinant is rewritten in terms of a Gaussian integral over bosonic fields, $\chi$,
\beq
\det K[\phi] \propto \int \calD \chi^{\dagger}\calD \chi e^{-\chi^{\dagger}K^{-1}[\phi]\chi} \ .
\eeq
This integral is then evaluated stochastically. These are just a sample of the available algorithms. For more details on these and others in the context of non-relativistic lattice field theory, see \cite{Drut:2012md}.

\subsubsection{Example formulation}
\begin{figure}
\begin{center}
\includegraphics[width=\linewidth]{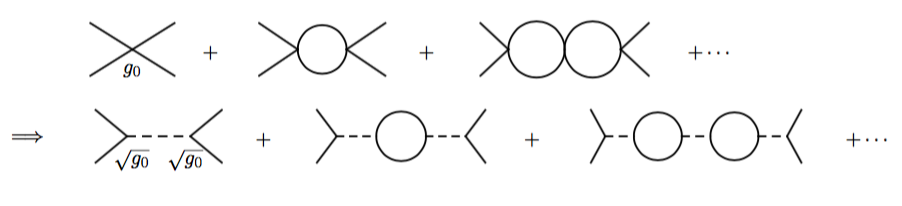}
\end{center}
\caption{\label{fig:dimer}Two-body scattering amplitude of \Fig{bubblesum}, where the contact interaction has been replaced in the second line by exchange of a dimer auxiliary field via a Hubbard-Stratonovich transformation.}
\end{figure}

Now that we have developed a general framework for lattice EFT, let's be explicit and make a few choices in order to further our understanding and make calculations simpler. The first choice I'm going to make is to use a $Z_2$ $\phi$ field, so that $\rho[\phi]$ is trivial. The next simplification I'm going to make is to allow the $\phi$ fields to live only on temporal links,
\beq
\label{eq:pointint}
\calL_{\mbox{\tiny int}} = \sum_{\bfx}\sqrt{b_{\tau}g_0}\phi_{\bfx,\tau}\psidag_{\bfx,\tau}\psi_{\bfx,\tau-1} \ .
\eeq
Note that we are free to make this choice, so long as the proper four-fermion interaction is regained in the continuum limit. This choice renders the interaction separable, as it was in our continuum effective theory. This means we may analytically sum two-body bubble chain diagrams as we did previously in order to set the coupling $g_0$ using some physical observable (see \Fig{dimer}). 

With this choice we can now write the $K$-matrix explicitly as
\beq
K[\phi,N_{\tau}] \equiv \left(\begin{array}{ccccccc}
D & -X(\phi_{N_{\tau}-1}) & 0 & 0 & . & . & . \\
0 & D & -X(\phi_{N_{\tau}-2}) & 0 & . & . & . \\
. & .& .& .&  & & \\ 
. & .&. & &. &  & \\ 
. &. & .& & & D & X(\phi_0)\\ 
X(\phi_{N_{\tau}}) & .&. & & &0 & D \\ 
\end{array} \right) \ ,
\eeq
where $X(\phi_\tau) \equiv 1-\sqrt{g_0}\phi_\tau$. Now the $\phi$-dependence exists only on the upper diagonal, as well as the lower left due to the boundary condition. This block will be eliminated through our final choice: open boundary conditions in time for the $\psi$ fields, $X(\phi_{N_{\tau}})=0$. As mentioned previously, we are free to choose the temporal boundary conditions as we please, so long as we only consider zero temperature (and zero chemical potential) observables. 

With this set of choices the matrix $K$ consists purely of diagonal elements, $D$, and upper diagonal elements, $X(\phi_\tau)$. One property of such a matrix is that the determinant, which is part of the probability distribution, is simply the product of diagonal elements, $\det K = \prod_{\tau} D$. Note that $D$ is completely independent of the field $\phi$. This means that the determinant in this formulation has no impact on the probability distribution $P[\phi]$, and therefore never needs to be explicitly computed, greatly reducing the computational burden. Thus in all of our calculations, performing the path integral over $\phi$ simply amounts to summing over $\phi = \pm 1$ at each lattice site. 

Finally, this form of $K$ also makes the calculation of propagators very simple. The propagator from time 0 to $\tau$ may be written,
\beq
K^{-1}(\tau,0) &=& D^{-1} X(\phi_{\tau-1})D^{-1}X(\phi_{\tau-2})D^{-1} \cdots X(\phi_0)D^{-1} \cr
&=& D^{-1}X(\phi_{\tau-1})K^{-1}(\tau-1,0) \ ,
\eeq
where $K^{-1}(0,0) = D^{-1}$, and all entries are $V\times V \ , (V=L^3)$ matrices which may be projected onto the desired state. This form suggests a simple iterative approach to calculating propagators: start with a source (a spatial vector projecting onto some desired quantum numbers and interpolating wavefunction), hit it with the kinetic energy operator corresponding to free propagation on the time slice, then hit it with the $\phi$ field operator on the next time link, then another free kinetic energy operator, and so on, finally projecting onto a chosen sink vector. 

As will be discussed further in Sec{systematic}, it is often preferable to calculate the kinetic energy operator in momentum space, while the auxiliary field in $X(\phi)$ must be generated in position space. Thus, Fast Fourier Transforms (FFTs) may be used between each operation to quickly translate between the bases. Example code for generating source vectors, kinetic operators, and interaction operators will be provided in later Sections.

\begin{figure}
\begin{center}
\includegraphics[width=0.4\linewidth]{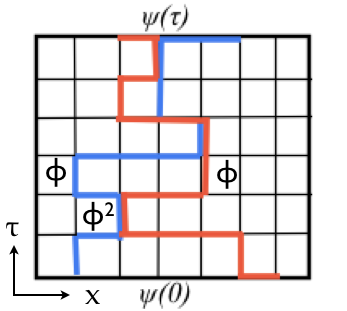}
\end{center}
\caption{\label{fig:lat}Schematic of a lattice calculation for a two-particle correlation function. The two particles (red and blue lines) propagate through the lattice between source $\psi(0)$ and sink $\psi(\tau)$, seeing particular values of the auxiliary field, $\phi$, on each time link. If two particles occupy the same temporal link, then upon summation over all possible values of $\phi$ at each link, a non-zero contribution is generated by the interaction term because $\langle \phi^2 \rangle \neq 0$.}
\end{figure}

A cartoon of this process on the lattice is shown in \Fig{lat}. The choice of $Z_2$ auxiliary fields also simplifies the understanding of how four-fermion interactions are generated. On every time link, imagine performing the sum over $\phi = \pm 1$. If there is only a single fermion propagator on a given link this gives zero contribution because the term is proportional to $\sum_{\phi=\pm 1} \sqrt{g_0} \phi = 0$. However, on time slices where two propagators overlap, we have instead $\sum_{\phi = \pm 1} g_0 \phi^2 = 2 g_0$. In sum, anywhere two fermions exist at the same spacetime point a factor of $g_0$ contributes, corresponding to an interaction.

\subsubsection{\label{sec:tuning}Tuning the two-body interaction}

There are several ways to set the two-body coupling. Here we will explore two methods, using different two-body observables. The first involves calculating the two-particle scattering amplitude, and tuning the coupling to reproduce known scattering parameters, to make a connection with our previous calculation for the effective theory. The second method uses instead the energy spectrum of a two-particle system in a box. This powerful method will be useful later when we begin to improve the theory in order to reduce systematic errors.

We have calculated the scattering amplitude previously for our effective theory using a momentum cutoff. For the first method for tuning the coupling, we will calculate it again using our lattice theory with the lattice cutoff as a regulator. First we need the single particle free propagator:
\beq
\label{eq:oneprop}
G_0(\tau,\vec{p}) &=& \langle \vec{p},\tau | \left(D^{-1}\right)^{\tau+1}|\vec{p},0\rangle = \left(1+\frac{\Delta(p)}{M}\right)^{-(\tau+1)} \ , \cr
\Delta(p) &\equiv& -\frac{1}{2}\langle \vec{p}|\nabla_L^2 | \vec{p}\rangle \cr
&=& \sum_i \sin^2 \frac{p_i}{2} \ ,
\eeq
where I've set $b_s=1$ (we will use this convention from now on until we begin to discuss systematic errors), and have used the previously defined discretized Laplacian operator. I've written the propagator in a mixed $\vec{p},\tau$ representation, as this is often useful in lattice calculations for calculating correlation functions in time when the kinetic operator, $D$, is diagonal in momentum space. 

The diagrammatic two-particle scattering amplitude is shown on the bottom line in \Fig{dimer}. Because we have chosen the interaction to be separable, the amplitude can be factorized:
\beq
\label{eq:aint}
A=g_0\left[1+\sum_n(g_0\hat{L})^n\right] = \frac{g_0}{1-g_0\hat{L}} \ ,
\eeq
where the one loop integral, $\hat{L}$, will be defined below. As before, in order to set a single coupling we need one observable, so we use the effective range expansion for the scattering phase shift to leading order,
\beq
\label{eq:aERE}
A = \frac{4\pi}{M} \frac{1}{p\cot\delta-i p} \approx -\frac{4\pi a}{M} \ .
\eeq
Relating \Eqs{aint}{aERE}, we find
\beq
\label{eq:eigeqscat}
\frac{1}{g_0} = -\frac{M}{4\pi a} + \hat{L} \ .
\eeq

We will now evaluate the loop integral using the free single particle propagators, \Eq{oneprop},
\beq
\hat{L} &=& \frac{1}{V} \sum_{\vec{p}}\sum_{\tau=0}^{\infty} \left[G_0(\tau,\vec{p})\right]^2 \cr
&=& \frac{1}{V} \sum_{\vec{p}}\sum_{\tau=0}^{\infty} \frac{1}{\left(1+\frac{\Delta(p)}{M}\right)^{2\tau+2}} \cr
&=& \frac{1}{V} \sum_{\vec{p}} \frac{1}{\left(1+\frac{\Delta(p)}{M}\right)^2}\left[1+\sum_{\tau=0}^{\infty}\frac{1}{\left[\left(1+\frac{\Delta(p)}{M}\right)^2\right]^\tau}\right] \cr
&=&\frac{1}{V} \sum_{\vec{p}}\frac{M}{2}\frac{1}{\Delta(p)\left(1+\frac{\Delta(p)}{2M}\right)} \ .
\eeq
This final sum may be calculated numerically for a given $M$ and $L$ (governing the values of momenta included in the sum), as well as for different possible definitions of the derivative operators contained in $\Delta$, giving the desired coupling, $g_0$, via \Eq{eigeqscat}.

The second method for setting the coupling utilizes the calculation of the ground state energy of two particles. We start with the two-particle correlation function,
\beq
C_2(\tau) = \frac{1}{Z}\int \calD \phi \calD \psidag \calD \psi e^{-S[\psidag, \psi, \phi]} \Psi^{\dagger}_{\mbox{\tiny src,2}} \Psi_{\mbox{\tiny snk,2}} \ ,
\eeq
where $\Psi_{\mbox{\tiny src,2(snk,2)}}$ is a source (sink) wavefunction involving one spin up and one spin down particle. Integrating out the fermion fields gives,
\beq
C_2(\tau) &=& \frac{1}{Z_{\phi}}\int \calD \phi P[\phi] \langle \Psi_{\mbox{\tiny snk,2}}|K^{-1}(\tau,0) \otimes K^{-1}(\tau,0) | \Psi_{\mbox{\tiny src,2}}\rangle \cr
&=& \frac{1}{4\tau}\sum_{\phi=\pm 1}\langle \Psi_{\mbox{\tiny snk,2}} | D^{-1} \otimes D^{-1} X(\phi_\tau)\otimes X(\phi_{\tau})D^{-1}\otimes D^{-1} X(\phi_{\tau-1})\otimes X(\phi_{\tau-1}) \cdots |\Psi_{\mbox{\tiny src,2}} \rangle \ . \cr
\eeq
I will now write out the components of the matrices explicitly:
\beq
\label{eq:c2xspace}
C_2(\tau) &=& \frac{1}{4\tau}\sum_{x_1 x_2 x_1' x_2' \cdots y_1y_2}\sum_{\phi_{x_1} \phi_{x_1'}\cdots = \pm 1} \langle \Psi_{\mbox{\tiny snk,2}}|x_1 x_2\rangle D^{-1}_{x_1 x_1'}D^{-1}_{x_2 x_2'}(\delta_{x_1 x_1'} + \sqrt{g_0}\phi_{x_1}\delta_{x_1 x_1'})(\delta_{x_2 x_2'} + \sqrt{g_0}\phi_{x_2}\delta_{x_2 x_2'}) \cr
&& \times D^{-1}_{x_1' x_1''}D^{-1}_{x_2 x_2''} \cdots \langle y_1 y_2|\Psi_{\mbox{\tiny src,2}}\rangle  \ .
\eeq
The first (last) piece in angle brackets represents the position space wavefunction created by the sink (source). All $\phi$ fields in \Eq{c2xspace} are uncorrelated, so we can perform the sum for each time slice independently. One such sum is given by,
\beq
&&\frac{1}{4}\sum_{x_1x_1'x_2x_2'}\sum_{\phi_{x_1}\phi_{x_2}} \delta_{x_1x_1'}\delta_{x_2x_2'}(1+\sqrt{g_0}\phi_{x_1}+\sqrt{g_0}\phi_{x_2}+g_0 \phi_{x_1}\phi_{x_2}) \cr
&=&\sum_{x_1x_2}(1+g_0\delta_{x_1x_2}) \ ,
\eeq
where the cross terms vanish upon performing the sum. If we make the following definitions,
\beq
\langle x_1 x_1' | \calD ^{-1} | x_2 x_2' \rangle \equiv D^{-1}_{x_1 x_1'} D^{-1}_{x_2 x_2'} \ ,  \qquad \langle x_1 x_2 | \calV | x_1' x_2' \rangle \equiv g_0 \delta_{x_1x_1'}\delta_{x_2x_2'}\delta_{x_1x_2}  \ ,
\eeq
then we can write the two-particle correlation function as,
\beq
C_2(\tau) &=& \langle \Psi_{\mbox{\tiny snk,2}}| \calD^{-1}(1+\calV)\calD^{-1}(1+\calV) \cdots \calD^{-1}(1+\calV)\calD^{-1} | \Psi_{\mbox{\tiny src}} \rangle \cr
&=& \langle \Psi_{\mbox{\tiny snk}} | \calD^{-1/2} \calT \calD^{-1/2} | \Psi_{\mbox{\tiny src,2}} \rangle \ ,
\eeq
where I have made the definition
\beq
\label{eq:transmat}
\calT \equiv \calD^{-1/2}(1+\calV) \calD^{-1/2} \ .
\eeq
Recall from statistical mechanics that correlation functions may be written as $\tau$ insertions of the transfer matrix, $e^{-H}$, acting between two states,
\beq
C(\tau) &=& \langle \Psi_{\mbox{\tiny snk,2}} | e^{-H\tau} | \Psi_{\mbox{\tiny src,2}} \rangle \cr 
&=& \langle \Psi_{\mbox{\tiny snk,2}} | \left[e^{-H}\right]^{\tau} | \Psi_{\mbox{\tiny src,2}} \rangle \ .
\eeq 
Then we may identify $\calT$ in \Eq{transmat} as the transfer matrix of the theory, $\calT = e^{-H}$. This in turn implies that the logarithm of the eigenvalues of $\calT$ give the energies of the two-particle system.

We will now evaluate the transfer matrix in momentum space:
\beq
\label{eq:transexplicit}
\langle p q| \calT | p'q'\rangle &=& \sum_{kk'll'}\langle pq|\calD^{-1/2}|kl\rangle \langle kl | 1+\calV | k'l' \rangle \langle k'l' |\calD^{-1/2}|p'q'\rangle \cr
&=& \sum_{kk'll'} \delta_{k'p'}\delta_{l'q'}\delta_{pk}\delta_{ql} \left(\delta_{kk'}\delta_{ll'}+\delta_{k+l,k'+l'} \frac{g_0}{V}\right) \cr
&\times& \left[\frac{1}{\left(1+\frac{\Delta(p)}{M}\right)\left(1+\frac{\Delta(q)}{M}\right)\left(1+\frac{\Delta(p')}{M}\right)\left(1+\frac{\Delta(q')}{M}\right)}\right]^{1/2} \cr
&=&\frac{\delta_{pp'}\delta_{qq'}+\frac{g_0}{V}\delta_{p+q,p'+q'}}{\sqrt{\xi(p)\xi(q)\xi(q')\xi(p')}} \ ,
\eeq
where I have made the definition,
\beq
\xi(p) \equiv 1+\frac{\Delta(q)}{M} \ .
\eeq
The eigenvalues of the matrix $\calT$ may be evaluated numerically to reproduce the entire two-particle spectrum. However, for the moment we only need to set a single coupling, $g_0$, so one eigenvalue will be sufficient. The largest eigenvalue of the transfer matrix, corresponding to the ground state, may be found using a simple variational analysis\footnote{Many thanks to Michael Endres for the following variational argument.}. Choosing a simple trial state wavefunction,
\beq
\langle pq| \Psi \rangle = \frac{\psi(p)}{\sqrt{V}}\delta_{p,-q} \ ,
\eeq
subject to the normalization constraint,
\beq
\frac{1}{V}\sum_p |\psi(p)|^2 = 1 \ ,
\eeq
we now need to maximize the following functional:
\beq
\langle \Psi |\calT | \Psi \rangle = \left[\frac{1}{V} \sum_p \frac{|\psi(p)|^2}{\xi^2(p)} + \frac{g_0}{V^2}\left| \sum_p \frac{\psi(p)}{\xi(p)} \right|^2 + \lambda \left(1-\frac{1}{V}\sum_p |\psi(p)|^2\right)\right] \ ,
\eeq
where $\lambda$ is a Lagrange multiplier enforcing the normalization constraint, and I have used the fact that $\xi(p)$ is symmetric in $p$ to simplify the expression. Taking a functional derivative with respect to $\psidag(q)$ on both sides gives
\beq
-\lambda \psi(q) + \frac{\psi(q)}{\xi^2(q)} + \frac{g_0}{V}\sum_p \frac{\psi(p)}{\xi(p)\xi(q)} = 0 \ ,
\eeq
where I have set the expression equal to zero in order to locate the extrema. Rearranging this equation, then taking a sum over $q$ on both sides gives
\beq
\sum_q\frac{\psi(q)}{\xi(q)} &=& \sum_q \frac{g_0}{V} \frac{1}{\lambda \xi^2(q)-1} \sum_p \frac{\psi(p)}{\xi(p)} \ ,
\eeq
finally resulting in
\beq
\label{eq:eigeqlambda}
1 = \frac{g_0}{V}\sum_q \frac{1}{\lambda \xi^2(q)-1} \ .
\eeq
We now have an equation involving two unknowns, $\lambda$ and $g_0$. We need a second equation in order to determine these two parameters. We may use the constraint equation to solve for $\psi(p)$, giving
\beq
\label{eq:varpsi}
\psi(p) = \calN \frac{\xi(p)}{\lambda \xi^2(p)-1} \ , \qquad \frac{1}{\calN^2} = \frac{1}{V} \sum_p \frac{\xi^2(p)}{\left[\lambda \xi^2(p)-1\right]^2} \ .
\eeq
Plugging this back in to our transfer matrix we find,
\beq
\langle \Psi | \calT | \Psi \rangle = \lambda \ .
\eeq
This tells us that $\lambda$ is equivalent to the eigenvalue we sought, $E_0 = -\ln \lambda(g_0)$. As a check, we can compare \Eqs{eigeqscat}{eigeqlambda} in the unitary limit: $a \to \infty, \lambda \to 1$, giving
\beq
\frac{1}{g_0} = \frac{M}{2V}\sum_p \frac{1}{\Delta\left(1+\frac{\Delta}{2M}\right)}
\eeq
for both Equations.

In \Sec{tuning} we will discuss a simple formalism for determining the exact two particle spectrum in a box for any given scattering phase shift. This will allow us to eliminate certain finite volume systematic errors automatically. The transfer matrix method is also powerful because it gives us access to the entire two particle, finite-volume spectrum. When we discuss improvement in \Sec{improve}, we will add more operators and couplings to the interaction in order to match not only the ground state energy we desire, but higher eigenvalues as well. This will allow us to control the interaction between particles with non-zero relative momentum. To gain access to higher eigenvalues, the transfer matrix must be solved numerically, however, this may be accomplished quickly and easily for a finite volume system. 

\section{\label{sec:observables}Calculating observables}
Perhaps the simplest observable to calculate using lattice (or any imaginary time) methods is the ground-state energy. While the two-body system may be solved exactly and used to set the couplings for two-body interactions, correlation functions for $N$-body systems can then be used to make predictions. However, the transfer matrix for $N\gtrsim 4$ cannot in general be solved exactly, because the dimension of the matrix increases with particle number. For this reason we form instead $N$-body correlation functions,
\beq
C_N(\tau)=\frac{1}{Z}\int \calD\phi\calD\psidag\calD\psi e^{-S[\psidag,\psi,\phi]}\Psi_{b_1 \cdots b_N}^{(b)}(\tau)\Psi_{a_1 \cdots a_N}^{\dagger (a)}(0) \ ,
\eeq
where 
\beq
\Psi_{a_1 \cdots a_N}^{(a)\dagger}(\tau) = \int dx_1\cdots dx_N A^{(a)}(x_1\cdots x_N)\psi_{a_1}(x_1,\tau) \cdots \psi_{a_N}(x_N,\tau)
\eeq
is a source for $N$ particles with spin/flavor indices $a_1 \cdots a_N$, and a spatial wavefunction $A^{(a)}(x_1 \cdots x_N)$. For the moment the only requirement we will make of the wavefunction is that it has non-zero overlap with the ground-state wavefunction (i.e. it must have the correct quantum numbers for the state of interest).

Recall that a correlation function consists of $\tau$ insertions of the transfer matrix between source and sink. We can then expand the correlation function in a basis of eigenstates,
\beq
C_N(\tau) &=& \frac{1}{Z} \langle \tilde{\Psi}_{a_1\cdots a_N}^{(a)} | e^{-H\tau}|\tilde{\Psi}_{b_1\cdots b_N}^{(b)} \rangle = \frac{1}{Z} \sum_{m,n} \langle \tilde{\Psi}_{a_1\cdots a_N}^{(a)} |m\rangle \langle m | e^{-H\tau} | n \rangle \langle n | \tilde{\Psi}_{b_1\cdots b_N}^{(b)} \rangle \cr
&=&\sum_m Z_m^{(a)} Z_m^{*(b)} e^{-E_n\tau} \ ,
\eeq
where $Z_m^{(a)}$ is the overlap of wavefunction $a$ with the energy eigenstate $m$, and $E_n$ is the $n$th eigenvalue of the Hamiltonian. In the limit of large Euclidean time (zero temperature), the ground state dominates,
\beq
C_N(\tau) \tautoinfty Z_0^{(a)} Z_0^{*(b)}e^{-E_0 \tau} \ ,
\eeq
with higher excited states exponentially suppressed by $\sim e^{-\Delta_{n0}\tau}$, where $\Delta_{n0} \equiv E_n - E_0$ is the energy splitting between the $n$th state and the ground state. It should be noted that for a non-relativistic theory the rest masses of the particles do not contribute to these energies, so the ground state energy of a single particle at rest is $E_0=0$, in contrast to lattice QCD formulations.

In this way, we can think of the transfer matrix as acting as a filter for the ground state, removing more excited state contamination with each application in time. A common method for determining the ground state energy from a correlation function is to construct the so-called effective mass function,
\beq
M_{\mbox{\tiny eff}}(\tau) \equiv \ln \frac{C(\tau)}{C(\tau+1)} \tautoinfty E_0 \ ,
\eeq
and look for a plateau at long times, whose value corresponds to the ground-state energy.

Once the ground state has been isolated, we can calculate matrix elements with the ground state as follows,
\beq
\langle \Psi_{a_1\cdots a_N}^{(a)} | A(\tau')|\Psi_{b_1\cdots b_N}^{(b)} \rangle &=& \sum_{lmnq} \langle \Psi_{a_1\cdots a_N}^{(a)} | l \rangle \langle l | e^{-H(\tau-\tau')} | m \rangle \langle m | A | n \rangle \langle n | e^{-H\tau'} | q \rangle \langle q | \Psi_{b_1\cdots b_N}^{(b)} \rangle \cr
&=& \sum_{ln} Z_l^{(a)} Z_n^{*(b)} e^{-E_l(\tau-\tau')}e^{-E_n\tau'} \langle m | A | n \rangle \ .
\eeq
To filter out the ground state, the matrix element insertion $A$ must be placed sufficiently far in time from both source and sink, $\{ \Delta_{l0}(\tau-\tau'), \Delta_{n0} \tau' \} \gg 1$,
\beq
\underset{\tau,\tau'\to\infty}{\longrightarrow} Z_0^{(a)} Z_0^{*(b)} e^{-E_0 \tau} \langle 0 | A | 0 \rangle \ .
\eeq
In order to isolate the matrix element and remove unknown $Z$ factors and ground state energies, ratios may be formed with correlation functions at various times, similar to the effective mass function.

Another observable one may calculate using lattice methods is the scattering phase shift between interacting particles. Because all lattice calculations are performed in a finite volume, which cannot accommodate true asymptotic scattering states, direct scattering measurements are not possible. However, a method has been devised by L\"uscher which uses finite volume energy shifts to infer the interaction, and therefore, the infinite volume scattering phase shift. The L\"uscher method will be discussed further in \Sec{Luscher}. Because the inputs into the L\"uscher formalism are simply energies, correlation functions may be used in the same way as described above to produce this data.

\subsection{\label{sec:SNR}Signal-to-noise}
Recall that we must use Monte Carlo methods to approximate the partition function using importance sampling,
\beq
C(\tau) \approx \frac{1}{\Ncfg} \sum_{i=1}^{\Ncfg} C(\phi_i,\tau) \tautoinfty Z_0 e^{-E_0\tau}\ ,
\eeq
where $C(\phi_i,\tau)$ is the operator for some correlation function of interest evaluated on a single configuration $\phi_i$, and the set of all fields, $\phi$, are generated according to the appropriate probability distribution. In the long Euclidean time limit we expect that this quantity will give us an accurate value for the ground state energy. As stated previously, if the ensemble is large enough for the central limit theorem to hold, then the error on the mean (noise) will be governed by the sample standard deviation,
\beq
\sigma_C^2(\tau) = \frac{1}{\Ncfg}\left[ \sum_{i=1}^{\Ncfg}|C(\phi_i,\tau)|^2 - \left|\sum_{i=1}^{\Ncfg}C(\phi_i,\tau)\right|^2\right] \ .
\eeq

As an example of how to estimate the size of the fluctuations relative to the signal, let's consider a single particle correlation function, consisting of a single propagator,
\beq
\frac{1}{Z_{\phi}}\int \calD \phi P(\phi) \langle \Psi_a | K^{-1}(\phi,\tau) | \Psi_b \rangle \approx \frac{1}{\Ncfg}\sum_{i=1}^{\Ncfg} K_{ab}^{-1}(\phi_i,\tau) \ ,
\eeq 
where the indices $\{ ab\}$ indicate projection onto the states specified by the source/sink. In the large Euclidean time limit, this object will approach a constant, $Z_0$, because the ground state energy for a single particle is $E_0=0$. For the non-relativistic theory as we have set it up, the matrix $K$ is real so long as $g_0>0$ (attractive interaction). The standard deviation is then given by
\beq
\label{eq:sig1part}
\sigma_{C_1}^2(\tau) = \frac{1}{\Ncfg}\left[ \sum_{i=1}^{\Ncfg} \left(K_{ab}^{-1}(\phi_i,\tau)\right)^2 - \left( \sum_{i=1}^{\Ncfg} K_{ab}^{-1}(\phi_i,\tau)\right)^2 \right] \ .
\eeq
The second term on the right hand side of the above equation is simply the square of the single particle correlation function, and will therefore also go to a constant, $Z_0^2$, for large Euclidean time. To gain an idea of how large the first term of $\sigma_{C_1}^2$ is, let's take a look at a correlation function for one spin up and one spin down particle,
\beq
C_2(\tau)=\frac{1}{Z}\int \calD\phi\calD\psidag\calD\psi e^{-S[\psidag,\psi,\phi]}\psi_{\uparrow}^{(b)}(\tau)\psi_{\downarrow}^{(b)}(\tau)\psi_{\uparrow}^{\dagger (a)}(0)\psi_{\downarrow}^{\dagger (a)}(0) \ ,
\eeq
where I have chosen the same single particle source (sink), $\psi^{(a)}$ ($\psi^{(b)}$), for both particles (this is only allowed for bosons or for fermions with different spin/flavor labels). After integrating out the $\psi$ fields we have
\beq
C_2(\tau) = \frac{1}{Z_{\phi}}\int \calD\phi P(\phi) K_{ab}^{-1}(\phi,\tau) K_{ab}^{-1}(\phi,\tau) \ ,
\eeq
which is approximately given by
\beq
C_2(\tau) \approx \frac{1}{\Ncfg}\sum_{i=1}^{\Ncfg}\left[ K_{ab}^{-1}(\phi_i,\tau)\right]^2 \ .
\eeq
This is precisely what we have for the first term on the right hand side of \Eq{sig1part}. Therefore, this term should be considered a two-particle correlation function, whose long Euclidean time behavior is known. Note that we must interpret this quantity as a two-particle correlation function whose particles are either bosons or fermions with different spin/flavor labels due to the lack of anti-symmetrization. 

We may now write the long-time dependence of the variance of the single particle correlator as
\beq
\sigma_{C_1}^2(\tau) \approx C_2(\tau) - \left( C_1(\tau) \right)^2 \tautoinfty Z_2e^{-E_0^{(2)} \tau} - Z_1^2 \ ,
\eeq
where $E_0^{(2)}$ is the ground state energy of the two-particle system. For a two-body system with an attractive interaction in a finite volume, $E_0^{(2)}< 0$, and we may write
\beq
\sigma_{C_1}^2(\tau) \tautoinfty Z_2e^{E_B^{(2)} \tau} - Z_1^2 \ ,
\eeq
where I've defined $E_B^{(2)} \equiv -E_0^{(2)}$. This tells us that $\sigma_{C_1}^2$, and therefore the noise, grows exponentially with time. We can write the signal-to-noise ratio $\calR_{C_1}(\tau)$ as
\beq
\calR_{C_1}(\tau) \equiv \frac{C_1(\tau)}{\frac{1}{\sqrt{\Ncfg}} \sigma_{C_1}(\tau)} \tautoinfty \sqrt{\Ncfg} \frac{Z_1}{\sqrt{Z_2} e^{E_B^{(2)}\tau/2}} = \sqrt{\Ncfg}\frac{Z_1}{\sqrt{Z_2}}e^{-E_B^{(2)} \tau/2} \ ,
\eeq
where I've dropped the constant term in $\sigma_{C_1}^2$, because it is suppressed in time relative to the exponentially growing term. This expression indicates that the signal-to-noise ratio itself grows exponentially with time, and therefore an exponentially large $\Ncfg$ will be necessary to extract a signal at large Euclidean time. Unfortunately, large Euclidean time is necessary in order to isolate the ground state. 

This exponential signal-to-noise problem is currently the limiting factor in system size for the use of any lattice method for nuclear physics. Here, we will discuss it in some detail because in many cases understanding the physical basis behind the problem can lead to methods for alleviation. One method we can use is to employ knowledge of the wavefunction of the signal and/or the wavefunction of the undesired noise in order to maximize the ratio of $Z$-factors, $Z_1/\sqrt{Z_2}$. For example, choosing a plane wave source for our single particle correlator gives perfect overlap with the desired signal, but will give poor overlap with the bound state expected in the noise. This leads to what has been referred to as a ``golden window" in time where the ground-state dominates before the noise begins to turn on \cite{Beane:2010em}. In general, choosing a perfect source for the signal is not possible, however, a proposal for simultaneously maximizing the overlap with the desired state as well as reducing the overlap with the noise using a variational principle has been proposed in \cite{Detmold:2014hla,Detmold:2014rfa}. We will discuss other methods for choosing good interpolating fields in \Sec{interp}, in order to allow us to extract a signal at earlier times where the signal-to-noise problem is less severe.

Another situation where understanding of the noise may allow us to reduce the noise is when the auxiliary fields and couplings used to generate the interactions can often be introduced in different ways, for instance, via the density channel vs. the BCS channel as mentioned previously. While different formulations can give the same effective interaction, they may lead to different sizes of the fluctuations. Understanding what types of interactions generate the most noise is therefore crucial. This will become particularly relevant when we discuss adding interactions beyond leading order to our EFT in \Sec{NLO}, where different combinations of interactions can be tuned to give the same physical observables.

Let's now discuss what happens to $\sigma_{C_1}^2$ if we have a repulsive interaction ($g_0<0$). Because nuclear potentials have repulsive cores, such a scenario occurs for interactions at large energy. Since the auxiliary-field-mediated interaction is given by $\sqrt{g_0}\phi\psidag\psi$, this implies that the interaction is complex. Our noise is now given by
\beq
\sigma_{C_1}^2(\tau) = \frac{1}{\Ncfg} \sum_{i=1}^{\Ncfg} K_{ab}^{-1}(\phi_i,\tau) \left[K_{ab}^{-1}(\phi_i,\tau)\right]^{\dagger} - |C_1(\tau)|^2 \ .
\eeq
Recall that the single particle propagator can be written
\beq
K^{-1}(\phi_i,\tau) = D^{-1} X(\phi_{i,\tau})D^{-1} X(\phi_{i,\tau-1}) \cdots \qquad X(\phi_{i,\tau}) = 1+\sqrt{g_0}\phi_{i,\tau} \ .
\eeq
The complex conjugate of the propagator then corresponds to taking $\phi \to -\phi$,
\beq
\left[ K^{-1}(\phi_i,\tau)\right]^{\dagger} = D^{-1} X(-\phi_{i,\tau})D^{-1} X(-\phi_{i,\tau-1})  \ .
\eeq
Again, $\phi$ fields on different time slices are independent, so we may perform each sum over $\phi = \pm 1$ separately. Each sum that we will encounter in the two-particle correlator consists of the product of $X(\phi_{\tau})X(-\phi_{\tau})$,
\beq
\sum_{\phi}(1+\sqrt{g_0}\phi)(1-\sqrt{g_0}\phi) = 1-g_0^2 = 1+|g_0|^2 \ ,
\eeq
which is exactly the same as we had for the attractive interaction. This implies that even though the interaction in the theory we're using to calculate the correlation function is repulsive, the noise is controlled by the energy of two particles with an attractive interaction, which we have already investigated. In this particular case for a single particle propagator, the signal-to-noise ratio is the same regardless of the sign of the interaction\footnote{This argument is somewhat simplified by our particular lattice setup in which we have no fermion determinant as part of the probability measure. For cases where there is a fermion determinant, there will be a mismatch between the interaction that the particles created by the operators see (attractive) and the interaction specified by the determinant used in the probability measure (repulsive). This is known as a partially quenched theory, and is unphysical. However, one may calculate a spectrum using an effective theory in which valence (operator) and sea (determinant) particles are treated differently. Often it is sufficient to ignore the effects from partial quenching because any differences contribute only to loop diagrams and may be suppressed.}. 

In general, however, signal-to-noise problems for systems with repulsive interactions are exponentially worse than those for attractive interactions. This is because generically the signal-to-noise ratio falls off as,
\beq
\calR \sim e^{-\left(E_{\mathcal{S}} - E_{\mathcal{N}}/2\right)\tau} \ ,
\eeq
where $E_{\mathcal{S}(\mathcal{N})}$ is the ground-state energy associated with the signal (noise). Because the signal corresponds to a repulsive system while the noise corresponds to an attractive system, the energy difference in the exponential will be greater than for a signal corresponding to an attractive system. 

\subsubsection{\label{sec:sign}Sign Problems}
A related but generally more insidious problem can occur in formulations having fermion determinants in the probability measure, known as a sign problem. A sign problem occurs when the determinant is complex, for example, in our case of a repulsive interaction. While we were able to eliminate the fermion determinant in one particular formulation, there are situations when having a fermion determinant in the probability measure may be beneficial, for example, when using forms of favorable reweighting, as will be discussed later on, or may be necessary, such as for non-zero chemical potential or finite temperature, when the boundary conditions in time may not be altered. For these reasons, we will now briefly discuss sign problems. 

The basic issue behind a sign problem is that a probability measure, by definition, must be real and positive. Therefore, a complex determinant cannot be used for importance sampling. Methods to get around the sign problem often result in exponentially large fluctuations of the observable when calculated on a finite sample, similar to the signal-to-noise problem (the two usually result from the same physical mechanism). One particular method is called reweighting, in which a reshuffling occurs between what is considered the ``observable" and what is considered the ``probability measure". For example, when calculating an observable,
\beq
\langle \calO \rangle = \frac{1}{Z_{\phi}}\int \calD\phi P(\phi)\calO(\phi) \ ,
\eeq
when $P(\phi)$ is complex, we can multiply and divide by the magnitude of $P(\phi)$ in both numerator and denominator,
\beq
\langle \calO \rangle = \frac{\int \calD\phi |P(\phi)| \frac{P(\phi)\calO(\phi)}{|P(\phi)|}}{\int \calD\phi |P(\phi)| \frac{P(\phi)}{|P(\phi)|}} \ ,
\eeq
as well as multiply and divide by $\tilde{Z}_{\phi} \equiv \int \calD \phi |P(\phi)|$,
\beq
\label{eq:reweight}
\langle \calO \rangle = \frac{\int \calD\phi |P(\phi)| \frac{P(\phi)\calO(\phi)}{|P(\phi)|}}{\tilde{Z}_{\phi}}\left/\frac{\int \calD\phi |P(\phi)| \frac{P(\phi)}{|P(\phi)|}} {\tilde{Z}_{\phi}} = \langle \calO' \rangle_{|P|}\left/\langle \calO''\rangle_{|P|}\right.\right. \ ,
\eeq
where
\beq
\calO' \equiv \frac{P(\phi)\calO(\phi)}{|P(\phi)|} \ , \qquad \calO'' \equiv \frac{P(\phi)}{|P(\phi)|} \ ,
\eeq
and $\langle \cdots \rangle_{|P|}$ implies that the path integrals in the expectation values use the measure $|P(\phi)|$. The advantage is that now the probability measure used for sampling is real and positive, at the cost of having to calculate two observables, $\calO', \calO''$. The real disadvantage, however, is that the second observable, $\calO''$ corresponds to the complex phase of the original measure, $P(\phi)$, which is highly oscillatory from field configuration to field configuration. 

We can measure the size of the fluctuations of the phase of $P(\phi)= \left[\det K(\phi)\right]^2$, corresponding to a two-spin (or flavor) theory with a repulsive interaction,
\beq
\langle \calO'' \rangle_{|P|} = \frac{\int \calD \phi \det K(\phi)\det K^*(\phi)}{\int \calD \phi \left[\det K(\phi)\right]^2} \ .
\eeq
The denominator of the above ratio corresponds to the partition function of the original theory which has two spins of particles interacting via a repulsive interaction. The numerator also corresponds to the partition function of a two-spin theory. However, recall that $K^*(\phi)$ corresponds to a propagator with the opposite sign on the interaction term. Because fermions of the same spin don't interact (Pauli principle), the only interaction in this theory is that between two particles of opposite spin, which we established previously will be an attractive interaction due to the sign flip on $K^*(\phi)$. Thus, the numerator corresponds to the partition function of a two-spin theory with an attractive interaction. 

A partition function is simply the logarithm of the free energy, $Z=e^{-\beta F}$. For a system in a finite volume at zero temperature this becomes $Z=e^{-V \calE_0}$, where $\calE_0$ is the energy density of the ground state of the theory. This implies that 
\beq
\label{eq:expectationsign}
\langle \calO'' \rangle_{|P|} \underset{\tau\to\infty}{\sim} e^{-V(\calE_0^{(\mbox{\tiny rep})}-\calE_0^{(\mbox{\tiny att})})} \ ,
\eeq
where $\calE_0^{(\mbox{\tiny rep})}$ ($\calE_0^{(\mbox{\tiny att})}$) is the energy density of the ground state of the repulsive (attractive) theory. Generically, $\calE_0^{(\mbox{\tiny att})} \leq \calE_0^{(\mbox{\tiny rep})} $, for theories which are identical up to the sign of their interaction. This may be shown using the Cauchy-Schwarz theorem,
\beq
\langle | \det K(\phi)| \rangle \leq | \langle \det K(\phi) \rangle | \ .
\eeq
Therefore, $\langle \calO'' \rangle_{|P|}$ will be exponentially small for large Euclidean times so long as $\calE_0^{(\mbox{\tiny rep})} \neq \calE_0^{(\mbox{\tiny att})}$. The variance, on the other hand, is
\beq
\langle |\calO'' |^2\rangle_{|P|} - |\langle \calO'' \rangle_{|P|}|^2 = \langle 1 \rangle - |\langle \calO'' \rangle_{|P|}|^2 \underset{\tau\to\infty}{\sim} 1-e^{-2V(\calE_0^{(\mbox{\tiny rep})}-\calE_0^{(\mbox{\tiny att})})} \sim 1 \ .
\eeq

So again, we have an exponentially small signal-to-noise ratio at large Euclidean time for the observable $\calO''$. This argument is very similar to our signal-to-noise argument for correlation functions. In general, if a theory has a sign problem there will be a corresponding signal-to-noise problem for correlation functions. The reverse is not always true, however, because reweighting is only necessary when the integration measure is complex, so even if there is a signal-to-noise problem in calculating correlation functions (as there is for an attractive interaction), a sign problem may not arise. Sign problems are in general far more problematic due to the exponential scaling with the volume, and because correlation functions give us the additional freedom of choosing interpolating fields in order to try to minimize the noise. In some cases, however, it may be possible to use knowledge learned from signal-to-noise problems in order to solve or reduce sign problems, and vice-versa \cite{Grabowska:2012ik,Nicholson:2012xt,EKLN5}.

\subsubsection{Noise in Many-Body Systems}

Let us now discuss signal-to-noise ratios for $N$-body correlation functions. First, we'll look at the two-particle case. We have already defined the correlation function for two particles with different spin/flavor labels,
\beq
C_2(\tau) = \langle\left[K_{ab}^{-1}(\phi_i,\tau)\right]^2\rangle \ .
\eeq
The variance is given by
\beq
\sigma_{C_2}^2(\tau) = \langle \left[K_{ab}^{-1}(\phi_i,\tau)\right]^4\rangle - \left(C_2(\tau)\right)^2 \ .
\eeq
It is simple to see that the first term in this expression corresponds to a four-particle correlation function, where each particle has a different flavor/spin index (because there is no anti-symmetrization of the fermion fields). Thus, we can write,
\beq
\sigma_{C_2}^2(\tau) = C_4(\tau) - \left(C_2(\tau)\right)^2 \ ,
\eeq
where $C_4(\tau)$ corresponds to a correlator with four particles having different flavors. This is much like a correlator for an alpha particle in the spin/flavor $SU(4)$ limit, thus, it will be dominated at large times by the binding energy, $E_B^{(4)}$, of a state with a large amount of binding energy per particle. Our signal-to-noise ratio is then,
\beq
\calR_{C_2}(\tau) \underset{\tau\to\infty}{\sim} \frac{e^{E_B^{(2)} \tau}}{e^{E_B^{(4)}\tau/2}} \ ,
\eeq
where, $E_B^{(4)}/2 > E_B^{(2)}$. Therefore, the signal-to-noise ratio is again falling off exponentially in time; this problem clearly becomes worse as the coupling becomes stronger. Finally, we can consider a many-body correlator composed of a Slater determinant over $N$ single-particle states in a two spin/flavor theory,
\beq
\label{2Ncorr}
C_{2N}(\tau) = \langle \left[ \det K^{-1}(\phi_i,\tau)\right]^2 \rangle \ .
\eeq
The ground state of this correlator will be either a BEC or BCS state, as discussed earlier in \Sec{scatamp}. The noise, on the other hand, will be dominated by a system of alpha-like clusters, since the number of flavors in the noise is always double that of the signal, which can bind to form nuclei. The ground-state energy of this bound state will clearly be much lower than that of a dilute BEC/BCS state, and our signal-to-noise ratio will be exponentially small in the large time limit. 

In general this pattern continues for fermion correlators with any number of particles, spins, and flavors. This is because doubling the number of flavors reduces the amount of Pauli repulsion in the resulting expression for the variance. Even for bosonic systems signal-to-noise can be a problem, simply as a result of the Cauchy-Schwarz triangle inequality, which tells you that, at best, your signal-to-noise ratio can be $1$, corresponding to a non-interacting system. Turning on interactions then generally leads to exponential decay of the signal-to-noise ratio. Signal-to-noise problems also generally scale exponentially with the system size, leading to limitations on system size based on computational resources. Thus, understanding and combatting signal-to-noise problems is paramount to further development in the field.

\subsection{\label{sec:overlap}Statistical Overlap}

For the lattice formulations we have thus far explored one generates configurations according to the probability distribution associated with the vacuum. One then introduces sources to create particles, which are considered part of the ``observable". However, the configurations which are the most important for creating the vacuum may not necessarily be the most important for the observable one wishes to calculate. 

\begin{figure}
\begin{center}
\includegraphics[width=0.3\linewidth]{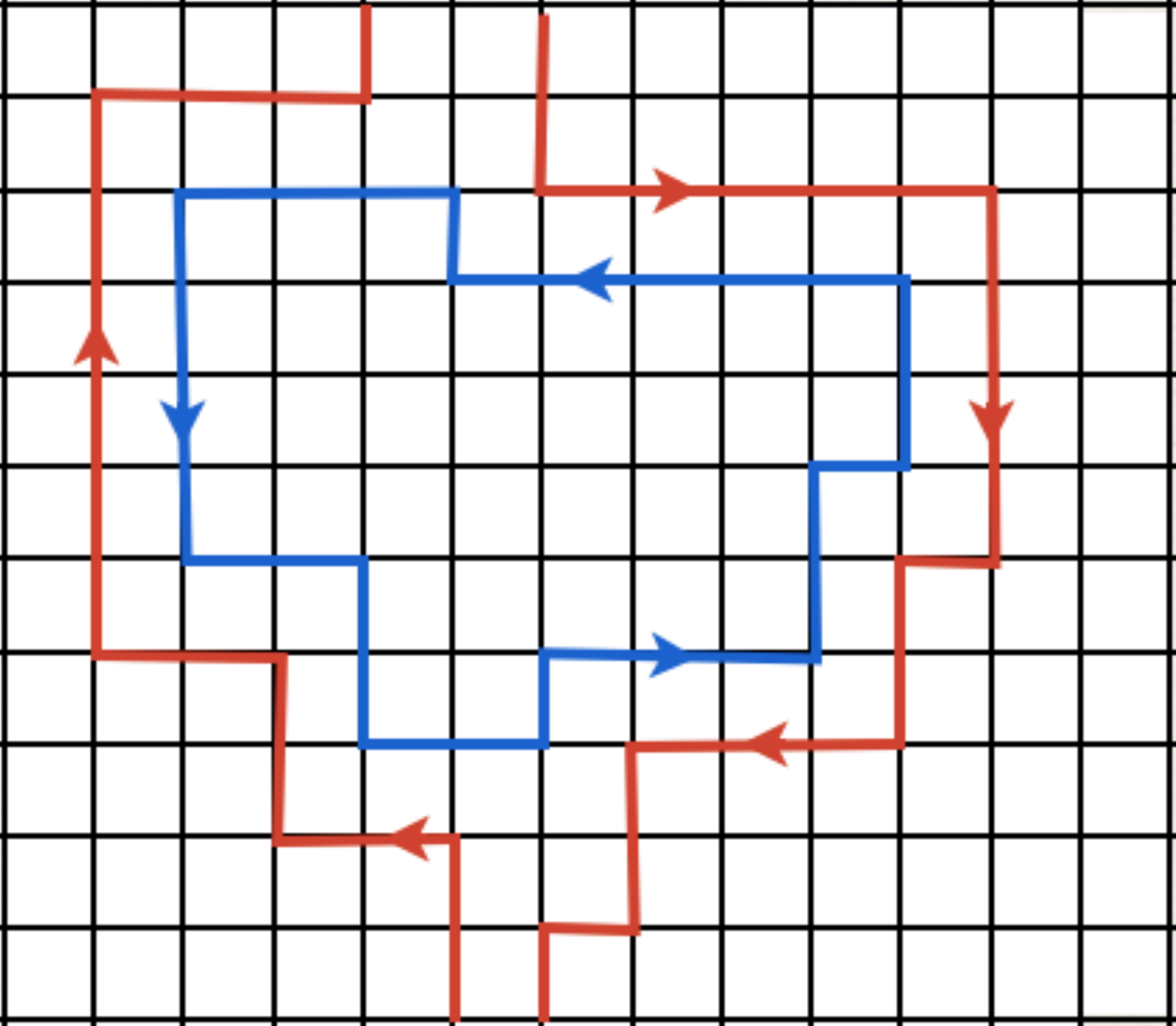}
\end{center}
\caption{\label{fig:QCDbubble}A schematic of an example configuration in LQCD which may lead to a statistical overlap problem. Red propagators correspond to valence quarks (quarks created by the sources/sinks in the operator), while blue corresponds to sea quarks (vacuum bubbles generated via Monte Carlo). Due to confinement, large bubbles (determined by the area enclosed by the blue propagator) are suppressed in the QCD vacuum and thus will likely be thrown out during importance sampling. In the presence of quark sources, however, these configurations are very important in the calculation of the observable (due to the small area enclosed between the red and blue propagators).}
\end{figure}

We can look to lattice QCD for a pedagogical example. In QCD, the fermion determinant encodes vacuum bubbles created by quark/anti-quark pairs. According to the tenets of confinement, bubbles with large spacetime area require a large energy to produce, and are therefore highly suppressed in the partition function. When doing importance sampling, small vacuum bubbles will dominate. On the other hand, if we now calculate an observable which introduces particle sources, a configuration involving a large vacuum bubble may become very important to the calculation. This is because the total relevant spacetime area of the given configuration, taking into account the particles created by the sources, can in fact be small (see \Fig{QCDbubble}). However, by sampling according to the vacuum probability, this configuration will be missed, skewing the calculation in an unknown manner. The farther the observable takes us from the vacuum, the worse this problem becomes, making this a particularly troublesome issue for many-body calculations.

Such problems are referred to as statistical overlap problems. Another situation where these overlap problems can often occur is when doing reweighting to evade a sign problem, as discussed in \Sec{sign}. For example, if the distribution being sampled corresponds to a theory with an attractive interaction, but the desired observable has a repulsive interaction, the Monte Carlo sampling will be unlikely to pick up the most relevant configurations, affecting the numerator of \Eq{reweight}.

\begin{figure}
\begin{center}
\includegraphics[width=0.5\linewidth]{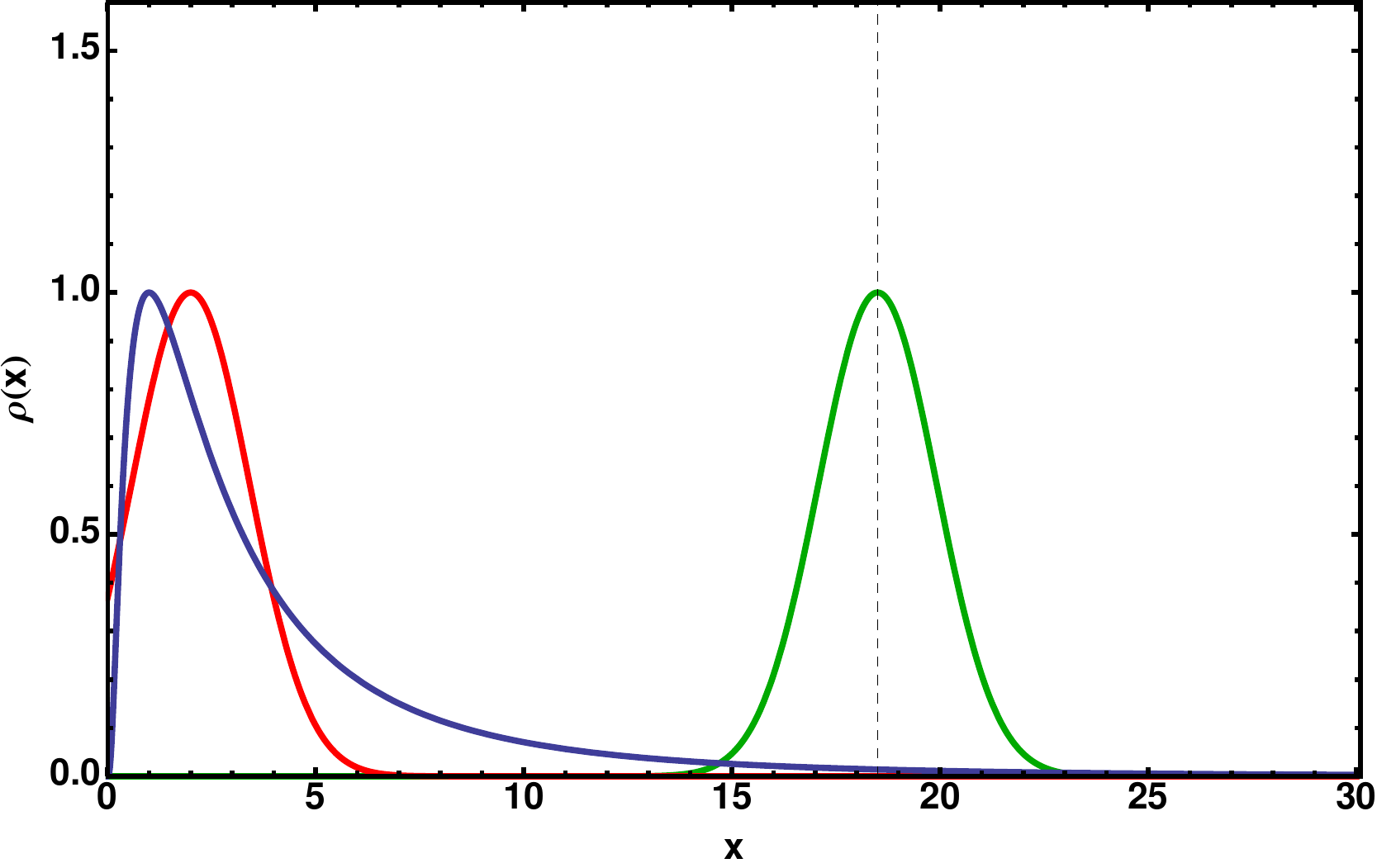}
\end{center}
\caption{\label{fig:overlap}Schematic drawing of a long-tailed probability distribution (blue) which leads to an overlap problem. Monte Carlo sampling leads to a sample distribution which is centered around the peak of the underlying distribution (red), far from the mean. The ideal probability distribution one would like to sample is narrow and centered around the mean (green).}
\end{figure}

We can understand the problem further by studying probability distributions of observables. While the distribution of the sampled field, $\phi$ in our case, may be peaked around the mean value of $\phi$, the distribution of the observable as calculated over the sample may not be peaked near the true mean of the observable. Such a distribution necessarily has a long tail. Plotting histograms of the values of the observable as calculated over the sample, $\{ C(\phi_1), C(\phi_2), \cdots C(\phi_{\Ncfg}) \}$, can allow us to gain an idea of the shape of the distribution for that observable. An example of a distribution with a statistical overlap problem is plotted in \Fig{overlap}. In this case, the peak of the distribution is far from the true mean. Values in the tail of the distribution have small weight, and are likely to be thrown out during importance sampling, skewing the sample mean without a corresponding increase in the error bar. The error bar is instead largely set by the width of the distribution near the peak. One way to determine whether there is an overlap problem is to recalculate the observable on a different sample size; if the mean value fluctuates significantly outside the original error bar this indicates an overlap problem.

The central limit theorem tells us that regardless of the initial distribution we pull from, the distribution of the mean should approach a Gaussian for a large enough sample size, so in principle we should be able to combat an overlap problem by brute force. However, what constitutes a ``large enough" sample size is dictated by the shape of the original distribution. The Berry-Esseen theorem \cite{BerryEsseen1,BerryEsseen2} can be used to determine that the number of configurations necessary to assume the central limit theorem applies is governed by
\beq
\sqrt{\Ncfg} \sim \frac{\langle \mathcal{X}^3\rangle}{\langle \mathcal{X}^2\rangle^{3/2}} \ ,
\eeq
where $\langle \mathcal{X}^n \rangle$ is the $n$th moment of the distribution of an observable, $\mathcal{X}$. Thus, a large skewness, or long tail, increases the number of configurations necessary before the central limit theorem applies, and therefore, to trust an error bar determined by the standard deviation of the distribution of the mean.

One could imagine repeating an argument similar to that made for estimating the variance of our correlation functions in order to estimate the third moment. For example, if our observable is the two-particle correlation function, $C_2(\tau)$, then the third moment will be
\beq
\langle \mathcal{X}^3 \rangle \sim \langle \left[K_{ab}(\phi_i,\tau)\right]^6 \rangle \ ,
\eeq
corresponding to a correlation function containing six particles of different flavors. Again, increasing the number of flavors generally increases the binding energy per particle of the system, leading to a third moment which is exponentially large compared to the appropriately scaled second moment. This implies that an exponentially large number of configurations will be necessary before the central limit theorem applies to the distribution of the mean of correlation functions calculated using this formulation. 

While we mentioned that using reweighting to avoid a sign problem is one situation where overlap problems often occur, it is also possible to use reverse reweighting in order to lessen an overlap problem. Here instead we would like to reweight in order to make the distribution of $\phi$ have \underline{more} overlap with the configurations that are important for the observable. An example that is commonly used is to include the desired correlation function itself, calculated at some fixed time, to be part of the probability measure. This may be accomplished using ratios of correlators at different times,
\beq
\frac{C_N(\tau'+\tau)}{C_N(\tau')} = \frac{\int \calD\phi \tilde{P}(\phi)\tilde{\calO}(\phi,\tau)}{\int\calD\phi \tilde{P}(\phi)} \ ,
\eeq
where
\beq
\tilde{P}(\phi) \equiv P(\phi)C_N(\tau',\phi) \ , \qquad \tilde{\calO}(\phi,\tau) \equiv \frac{C_N(\tau'+\tau,\phi)}{C_N(\tau',\phi)} \ .
\eeq
Now the probability distribution incorporates an $N$-body correlator at one time, $\tau'$, and will therefore do a much better job of generating configurations relevant for the $N$-body correlator at different times. A drawback of this method is that it is much more computationally expensive to require the calculation of propagators for the generation of eaach configuration. Furthermore, the configurations that are generated will be operator-dependent, so that calculating the correlator $C_{N+1}$ will require the generation of a whole new set of field configurations.

Another method for overcoming a statistical overlap problem is to try to get a more faithful estimate of the mean from the long-tailed distribution itself. To try to better understand the distribution, let's use our signal-to-noise argument to estimate higher moments of the distribution. We can easily estimate the $N$th moment of the correlation function for a single particle,
\beq
\calM_N \sim C_{N} \underset{\tau\to\infty}{\sim} e^{-E_0^{(N)}\tau} \ ,
\eeq
where $E_0^{(N)}$ is the ground-state energy of $N$ particles with different flavors. Let's consider the theory to be weakly coupled (small scattering length, $a/L \ll 1$). In this case the two-body interaction dominates and we can use perturbation theory to estimate the energy of two particles in a box: $E_0^{(2)} \approx \frac{4\pi a}{ML^3}$. A weakly coupled system of $N$ particles interacting via the two-body interaction is given by simply counting the number of possible pairs of interacting particles, $E_0^{(N)} \approx N(N-1) \frac{4\pi a}{ML^3}$, leading to the following expression for the moments \cite{DeGrand:2012ik}:
\beq
\label{eq:lnmoments}
\calM_N \sim e^{-N(N-1) \frac{4\pi a}{ML^3}} \ .
\eeq
Distributions with the particular $N$ dependence seen in \Eq{lnmoments} are called log-normal distributions, so named because the distribution of the logarithm of a log-normally distributed quantity is normal. While we derived this expression for theories near weak coupling, there is also evidence that the log-normal distribution occurs for correlators near unitarity as well \cite{Nicholson:2012zp,Nicholson:2015zxa}. 

The central limit theorem implies that normal distributions occur generically for large sums of random numbers; the same argument leads to the conclusion that log-normal distributions occur for large products of random numbers. Let's think about how correlation functions are calculated on the lattice: particles are created, then propagate through random fields from one time slice to the next until reaching a sink. Each application of the random field is multiplied by the previous one,
\beq
K^{-1}(\tau) = D^{-1}X(\tau)D^{-1}X(\tau-1) \cdots \ ,
\eeq
and then products of these propagators may be used to form correlation functions for multiple particles. Thus, one might expect that in the $\tau\to\infty$ limit (or for large numbers of particles), the distributions of these correlation functions might flow toward the log-normal distribution. More precisely though, each block $X(\tau)$ is actually a matrix of random numbers, and products of random matrices are far less well understand than products of random numbers. Nonetheless, products of random link variables are used to form most observables in nearly all lattice calculations, and approximately log-normal distributions appear to be ubiquitous as well, including in lattice QCD calculations.

\begin{figure}
\begin{center}
\includegraphics[width=0.5\linewidth]{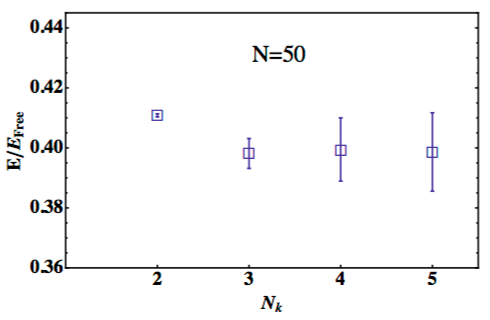}
\end{center}
\caption{\label{fig:cumulant} Results for the energy of 50 two-component fermions at unitarity using the cumulant expansion (\Eq{cumulantexp}) cut off at $\calO(N_k)$. Figure from \cite{EKLN4}.}
\end{figure}

If it is $\ln C$ that is nearly Gaussian rather than $C$, then it may be better to sample $\ln C$ as our observable instead. Without asserting any assumptions about the actual form of the distribution, we can expand around the log-normal distribution using what is known as a cumulant expansion,
\beq
\label{eq:cumulantexp}
\ln \langle \calO \rangle = \sum_{n=1}^{\infty} \frac{1}{n!} \kappa_n(\ln \calO) \ ,
\eeq
where $\kappa_n$ is the $n$th cumulant, or connected moment. The cumulants may be calculated using the following recursion relation:
\beq
\kappa_n(\mathcal{X}) = \langle \mathcal{X}^n \rangle -\sum_{m=1}^{n-1}\left(\begin{array}{c}
n-1 \\
m-1 
\end{array} \right)\kappa_m(\mathcal{X}) \langle \mathcal{X}^{n-m}\rangle \ .
\eeq
Note that the expansion in \Eq{cumulantexp} is an exact equality for an observable obeying any distribution. We may now expand the correlation function as
\beq
\ln \langle C \rangle \tautoinfty -E_0 \tau = \langle \ln C\rangle + \frac{1}{2}\left(\langle ( \ln C)^2 \rangle -\langle \ln C \rangle^2 \right) + \frac{1}{6} \kappa_3(\ln C) + \cdots \ .
\eeq
Again, this expansion is true for a correlation function obeying any distribution. However, if the distribution of $\ln C$ is exactly log-normal, then $\kappa_{n\geq 3}(\ln C) = 0$. If the distribution is approximately log-normal, then the third and higher cumulants are small corrections, further suppressed in the cumulant expansion by $1/n!$. This suggests that we may cut off the expansion after including a finite number of cumulants without significantly affecting the result (see \Fig{cumulant}). We may also include the next higher order cumulant in order to estimate any systematic error associated with our cutoff.

The benefit of using the cumulant expansion to estimate the mean rather than using the standard method is that for a finite sample size, high-order cumulants of $\ln C$ are poorly measured, which is the culprit behind the overlap problem. However, for approximately log-normal distributions these high-order cumulants should be small in the infinite statistics limit. Thus, by not including them in the expansion we do a better job at estimating the true mean on a finite sample size. In other words, by sampling $\ln C$ rather than $C$, we have shifted the overlap problem into high, irrelevant moments which we may neglect.

The cumulant expansion avoids some of the drawbacks of reweighting, such as greatly increased computational effort in importance sampling. However, the farther the distribution is from log-normal, the higher one must go in the cumulant expansion, which can be particularly difficult to do with noisy data. Thus, for some observables it may be difficult to show convergence of the series on a small sample. Which method is best given the competition between the computational effort used in generating samples via the reweighting method versus the large number of samples which may be required to show convergence of the cumulant expansion is unclear and probably observable dependent. 

\subsection{\label{sec:interp}Interpolating Fields}

The previous section highlights the importance of gaining access to the ground state as early in time as possible, since the number of configurations required grows exponentially with time. Returning to our expression for the expansion of a correlation function in terms of energy eigenstates,
\beq
C(\tau) &=& Z_0 e^{-E_0 \tau}+Z_1 e^{-E_1 \tau} + \cdots \cr
&=& Z_0 e^{-E_0\tau}\left[1+\frac{Z_1}{Z_0}e^{-(E_1-E_0)\tau} + \cdots \right] \ ,
\eeq 
we see that the condition that must be met in order to successfully suppress the leading contribution from excited state contamination is
\beq
\label{eq:taucond}
\tau \gg \frac{\ln \left(\frac{Z_1}{Z_0 E_0}\right)}{E_1-E_0} \ ,
\eeq
where $E_0,Z_0$ ($E_1,Z_1$) are the ground (first excited) state energy and wavefunction overlap factor, respectively. Assuming we have properly eliminated excited states corresponding to unwanted quantum numbers through the choice of our source/sink, we have no further control over the energy difference $E_1 - E_0$ in the denominator, because this is set by the theory. Unfortunately, this makes the calculation of many-body observables extremely difficult as this energy splitting can become arbitrarily small due to collective excitations. Therefore, our only recourse is to choose excellent interpolating fields in order to reduce the numerator of \Eq{taucond}.

The simplest possible choice for a many-body interpolating field is composed of non-interacting single particle states. A Slater determinant over the included states takes care of fermion antisymmetrization. For example, a correlation function for $N_{\uparrow}$ ($N_{\downarrow}$) spin up (spin down) particles can be written,
\beq
\label{eq:slaterdet}
C_{N_{\uparrow},N_{\downarrow}}(\tau) = \langle \det S^{\downarrow}(\tau) \det S^{\uparrow}(\tau) \rangle \ ,
\eeq
where
\beq
S_{ij}^{\sigma} (\phi,\tau) \equiv \langle \alpha_i^{\sigma} | K^{-1}(\phi,\tau) | \alpha_j^{\sigma} \rangle \ ,
\eeq
and $\langle \alpha_j^{\sigma}|$ corresponds to single particle state $i$ with spin $\sigma$. As an example, we may use a plane wave basis for the single particle states,
\beq
| \alpha_j^{\uparrow}\rangle = |\vec{p}_j \rangle \ , \qquad | \alpha_j^{\downarrow}\rangle = |-\vec{p}_j \rangle \ ,
\eeq
where I've chosen equal and opposite momenta for the different spin labels in order to enforce zero total momentum (this condition may be relaxed to attain boosted systems). 

Though the interpolating field chosen in \Eq{slaterdet} has non-zero overlap with the ground state of interest, if the overlap is small it may take an inordinately long time to remove excited state contributions. Consider a system involving only two-particle correlations, as in our two-spin fermion system, and make the simplification that the ground state consists of non-interacting two-body pairs having wavefunction $\Psi_{\mbox{\tiny 2-body}}$, and overlap with a product of two non-interacting single particle states given by
\beq
\langle \Psi_{\mbox{\tiny 2-body}} | \left( |\vec{p}\rangle \otimes | - \vec{p} \rangle \right) = \epsilon < 1.
\eeq
Then the corresponding overlap of the Slater determinant in \Eq{slaterdet} with the ground state wavefunction scales as
\beq
\left( \langle \Psi_{\mbox{\tiny 2-body}} | \otimes \cdots \otimes \langle \Psi_{\mbox{\tiny 2-body}} | \right) \left( |\vec{p}_1 \rangle \otimes | -\vec{p}_1 \rangle \otimes \cdots \otimes |\vec{p}_N \rangle \otimes | -\vec{p}_N \rangle \right) \sim \epsilon^N \ .
\eeq
Thus the overlap of single-particle states with an interacting $2N$-body state is exponentially small with $N$. This condition worsens for systems with $3$- and higher-body correlations.

\begin{figure}
\begin{center}
\includegraphics[width=\linewidth]{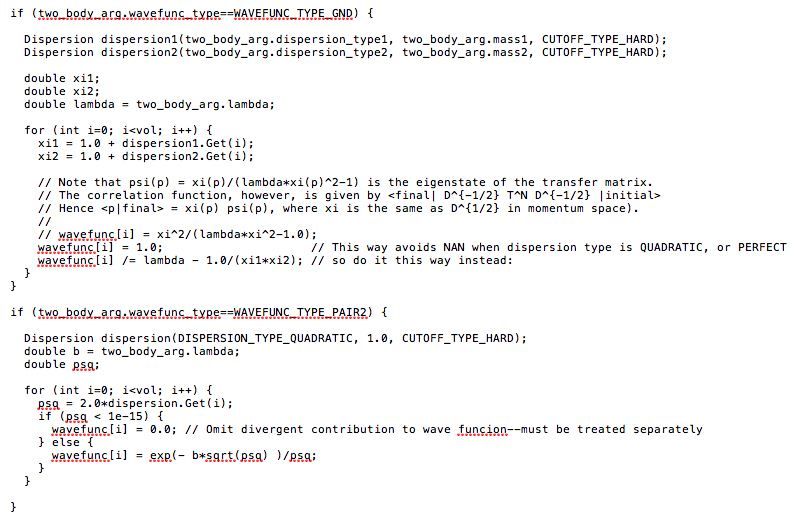}
\end{center}
\caption{\label{fig:wfcode}Portion of c++ code for implementing two types of two-body source vector: \Eq{varpsi} (GND) and \Eq{pairing} (PAIR2). Note that these vectors are computed in momentum space. The first operator applied to a source is the kinetic operator, $D^{-1}$, which is also computed in momentum space.}
\end{figure}

In order to do a better job we can incorporate two-body correlations into the sinks as follows: first, we construct a two particle propagator,
\beq
S_{ij}^{\uparrow}{\downarrow}(\phi,\tau) &=& \langle \Psi_2|K^{-1}(\phi,\tau) \otimes K^{-1}(\phi,\tau)\left( | \alpha_i^{\uparrow}\rangle \otimes | \alpha_j^{\downarrow} \rangle \right)\cr
&=& \sum_{\vec{p}} \Psi(\vec{p}) \langle \vec{p}| K^{-1}(\phi,\tau) | \alpha_i^{\uparrow} \rangle \langle -\vec{p}| K^{-1}(\phi,\tau) | \alpha_j^{\downarrow} \rangle \ ,
\eeq
where $\Psi_2(\vec{p})$ is some two-body wavefunction (this process could equally well be performed in position space). As an example, to incorporate BCS pairing, we may use a wavefunction of the form:
\beq
\label{eq:pairing}
\Psi_2(\vec{p}) \sim \frac{e^{-b|\vec{p}|}}{|\vec{p}|^2} \ ,
\eeq 
where $b$ is some parameter which may be tuned to maximize the overlap of the wavefunction. We may also use the wavefunction derived in \Eq{varpsi} for a lattice version of such a wavefunction. An example code fragment for implementing such wavefunctions is given in \Fig{wfcode}.

To ensure Pauli exclusion, it is sufficient to antisymmetrize only the sources, $|\alpha_i \rangle$, leading to the following many-body correlation function,
\beq
C_{N_{\uparrow},N_{\downarrow}}(\tau) = \langle \det S^{\uparrow \downarrow}(\tau) \rangle \ ,
\eeq 
where the determinant runs over the two sink indices. For correlation functions having an odd number of particles, one may replace a row $i$ of $S^{\uparrow\downarrow}$ with the corresponding row of the single particle object, $S^{\uparrow}$. The benefit of folding the wavefunction in at the sinks only is an $\calO(V^2)$ savings in computational cost: to fold a two-body wavefunction in at both source and sink requires the calculation of propagators from all possible spatial points on the lattice to all possible spatial points in order to perform the resulting double sum. 

Higher-body correlations may also be important and can be incorporated using similar methods. However, these will lead to further $\calO(V)$ increases in computation time. Finally, the entire system should be projected onto the desired parity, lattice cubic irreducible representation (which we will now briefly discuss), etc. in order to eliminate any contamination from excited states having different quantum numbers. 

\subsubsection{Angular momentum in a box}

The projection onto the cubic irreps is the lattice equivalent of a partial wave decomposition in infinite volume (and the continuum limit). The cubic group is finite, and therefore has a finite number of irreps, reflecting the reduced rotational symmetry of the box. The eigenstates of the systems calculated on the lattice will have good quantum numbers corresponding to the cubic irreps. When mapping these states onto angular momenta associated with infinite volume, there will necessarily be copies of the same irrep corresponding to the same angular momentum due to the reduced symmetry. This means that the box mixes angular momenta, as displayed in \Tab{cubicirreps}. For example, an energy level calculated in a finite volume that has been projected onto the positive parity $A_1$ irrep will have overlap with $j=0,4,\cdots$. For low energies it may be possible to argue that contributions from high partial waves are kinematically suppressed, since the scattering amplitude scales with $p^{2l+1}$, but in general the different partial wave contributions must be disentangled using multiple data points from different cubic irreps. 

\begin{table}[h!]
\label{tab:cubicirreps}
\begin{center}
\begin{tabular}{cc}
 j \hspace{1mm} & cubic irreps \\
 \hline
 0  \hspace{1mm}& $A_1$ \\
 1 \hspace{1mm} & $T_1$ \\
 2 \hspace{1mm} & $E+T_2$ \\
 3 \hspace{1mm} & $A_2 + T_1 + T_2$ \\
 4 \hspace{1mm} & $A_1 + E + T_1 + T_2$ \\
 \end{tabular}
 \end{center}
 \caption{Decomposition of the cubic group onto total angular momentum, $j$.}
 \end{table}
 
 A pedagogical method for projecting two-particle states onto the desired cubic irrep involves first projecting the system onto a particular spin state: for example, a two nucleon system may be projected onto either a spin singlet (symmetric) or spin triplet (anti-symmetric) state. The wavefunctions may then be given an ``orbital angular momentum" label by performing a partial projection using spherical harmonics confined to only the allowed rotations in the box. For example, we could fix the position of one of the particles at the origin $(0,0,0)$, then displace the second particle to a position $(x_0,y_0,z_0)$. This configuration will be labeled by the wavefunction $\psi_{s,m_s}\left[(x_0,y_0,z_0)\right]$, where $s,m_s$ are the total and $z$-component of the spin. We can then perform the partial projection, 
 \beq
 \tilde{\psi}_{l,m_l;s,m_s} = \sum_i Y_{l,m_l}\left[R_i(x_0,y_0,z_0)\right]\psi_{s,m_s}\left[R_i(x_0,y_0,z_0)\right] \ ,
 \eeq
 where the $R_i$ are cubic rotation matrices. Essentially, the set $R_i(x,y,z)$ correspond to all possible lattice vectors of the same magnitude. For example, if our original vector was $(1,0,0)$, then we would sum over the set of displacements $\{ (\pm 1,0,0),(0,\pm1,0),(0,0,\pm1)\}$. I want to emphasize that the $l,m_l$ are only wavefunction labels and do not correspond to good quantum numbers due to the reduced rotational symmetry.
 
 Now that the wavefunctions have spin and orbital momentum labels, these may be combined into total angular momentum labels $j,m_j$ using the usual Clebsch-Gordan coefficients. Finally, these wavefunctions are projected onto cubic irreps using so-called subduction matrices \cite{Dudek:2010wm}. As an example, a wavefunction labeled with $j=2$ (having five possible $m_j$ labels) will have overlap with two cubic irreps, $T_2,E$. The subduction matrices are:
 \beq
 T_2: 
\begin{array}{c}
 \overbrace{\rule{3.2cm}{0pt}}^{m_j=-2,-1,0 ,1,2} \\
 \left(\begin{array}{ccccc}
 0 & 1 & 0 & 0 & 0 \\
 1/\sqrt{2} & 0 & 0 & 0 & -1/\sqrt{2} \\
 0 & 0 & 0 & 1 & 0 \\
 \end{array} \right)  \end{array} \ , \qquad E: \left( \begin{array}{ccccc}
 0 & 0 & 1 & 0 & 0 \\
 1/\sqrt{2} & 0 & 0 & 0 & 1/\sqrt{2} \\
 \end{array}\right) \ .
 \eeq
 Note that the $T_2$ irrep has three degenerate states, while the $E$ irrep has two, matching the total of five degenerate states for $j=2$ in infinite volume. 
 
 Using this method for projection onto the cubic irreps has several benefits, including ease of bookkeeping and extension to higher-body systems using pairwise combinations onto a given $j,m_j$, followed by subduction of the total resulting wavefunction. Furthermore, in cases where more than one partial wave has overlap onto the chosen cubic irrep, wavefunctions with different partial wave labels may have different overlap onto the ground- and excited states of the system. Therefore, they can be used as a handle for determining the best source for the state of interest. We will discuss methods for using multiple sources for disentangling low-lying states and allowing for measurements at earlier times in the next subsection.
 
 \subsection{\label{sec:analysis}Analysis methods}

Having done our best to come up with interpolating wavefunctions, we can attempt to extract the ground state energy (and possibly excited state energies) earlier in time by performing multiple exponential fits to take into account any remaining excited state contamination. Using the known functional form for the correlator,
\beq
\label{eq:funcformC}
y(\tau) = \sum_n^{\Lambda}Z_n e^{-E_n \tau} \ ,
\eeq
where $\Lambda$ is a cutoff in the number of exponentials included in the fit, we may perform a correlated $\chi^2$ minimization,
\beq
\chi_{\Lambda}^2 = \sum_{\tau,\tau'}\left[ C(\tau)-y(\tau)\right] \left(\calC^{-1}\right)_{\tau\tau'}\left[ C(\tau')-y(\tau')\right]  \ ,
\eeq
where $\calC$ is the covariance matrix taking into account the correlation between different time steps. Because the correlation function at a given time is built directly upon the correlation function for the previous time step, there is large correlation between times that must be taken into account. 

We can go further by noting that correlation functions formed using different sources, but having the same quantum numbers, will lead to the same spectrum in \Eq{funcformC}, but with different overlap factors, $Z_n$. Thus, the $\chi^2$ minimization can be expanded to include different sources $s$, with only a modest increase in the number of parameters to be fit. Different sources may be produced, for example, by varying some parameter in the wavefunction, such as $b$ in \Eq{pairing}, through a different basis of non-interacting single particle states, such as plane waves vs. harmonic oscillator states, or through different constructions of the same cubic irrep, as discussed in the previous subsection. The resulting $\chi^2$ minimization is
\beq
y_s(\tau) = \sum_n^{\Lambda}Z_n^{(s)} e^{-E_n \tau} \ , \qquad \chi_{\Lambda}^2 = \sum_{\tau,\tau',s,s'}\left[ C_s(\tau)-y_s(\tau)\right] \left(\calC^{-1}\right)_{\tau\tau'}^{ss'}\left[ C_{s'}(\tau')-y_{s'}(\tau')\right]  \ ,
\eeq
where the covariance matrix now takes into account the correlation between different sources calculated on the same ensembles. 

In general, multiple parameter fits require high precision from the data in order to extract several parameters. The use of priors through Bayesian analysis techniques may be beneficial in some circumstances when performing multi-exponential fits to noisy data.

A more elegant approach using a set of correlation functions created using different operators is based on a variational principle \cite{MICHAEL1983433,Luscher:1990ck}. A basic variational argument proceeds as follows \cite{Blossier:2009kd}: starting with some set of operators $\calO_i$ which produce states $|\phi_i\rangle = \calO_i |0\rangle$ from the vacuum, we can evolve the state to some time $\tau_0$, $|\tilde{\phi}_i\rangle = e^{-\tau_0H/2}|\phi_i\rangle$ in order to eliminate the highest excited states, but leaving a finite set of states contributing to the correlation function. We would like to find some wavefunction $|\psi \rangle = \sum_{i=1}^N \alpha_i | \tilde{\phi}_i \rangle$ which is a linear combination of our set of operators parameterized by $\{\alpha_i\}$, that maximizes the following quantity for $\tau>\tau_0$:
\beq
\lambda_0(\tau,\tau_0) = \underset{\{\alpha_i\}}{\mbox{Max}}\frac{\langle \psi|e^{-(\tau-\tau_0)H}|\psi \rangle}{\langle\psi | \psi \rangle} \ ,
\eeq
so that 
\beq
\lambda_0(\tau,\tau_0) \approx e^{-E_0(\tau-\tau_0)} \ .
\eeq

A powerful method for finding the appropriate linear combination of states satisfying the variational principle uses a generalized eigenvalue problem (GEVP). For this method we form a matrix of correlation functions using all combinations of sources and sinks formed from a set of operators,
\beq
C_{ij}(\tau) = \langle \calO_i(\tau) \calO^*_j(0)\rangle = \sum_n e^{-E_n \tau}Z_i^{(n)}Z_j^{(n)} \ .
\eeq
The GEVP may be stated as:
\beq
\label{eq:GEVP}
C(\tau)v_n(\tau,\tau_0) = \lambda_n(\tau,\tau_0)C(\tau_0)v_n(\tau,\tau_0) \ ,
\eeq
where $v_n$ ($\lambda_n$) are a set of eigenvectors (eigenvalues) to be determined as follows: assume we choose $\tau_0$ to be far out enough in time such that only $N$ states contribute to the correlation function,
\beq
C_{ij}(\tau) = \sum_n^N e^{-E_n \tau}Z_i^{(n)}Z_j^{(n)} \ .
\eeq
Let's introduce a set of dual vectors $u_i^{(n)}$ such that
\beq
\sum_i u_i^{(n)} Z_i^{(m)} = \delta_{mn} \ .
\eeq
Applying $u_i$ to $C_{ij}$ gives
\beq
\sum_jC_{ij}(\tau)u_j^{(m)} = \sum_j \sum_n e^{-E_n \tau} Z_i^{(n)}Z_j^{(n)}u_j^{(m)} = e^{-E_m \tau}Z_i^{(m)} \ .
\eeq
Going back to our original GEVP, \Eq{GEVP},
\beq
C(\tau) u^{(m)} = \lambda_m(\tau,\tau_0) C(\tau_0)u^{(m)} \ ,
\eeq
we can now identify,
\beq
\lambda_m(\tau,\tau_0) = e^{-E_m(\tau-\tau_0)} \ .
\eeq
Thus, the energies may be found from the eigenvalues of the matrix, $C^{-1}(\tau_0)C(\tau)$. Solving this GEVP gives us access to not only the ground state, but some of the lowest excited states as well. 

Any remaining contributions from states corresponding to $E_n, n>N$ can be shown to be exponentially suppressed as $e^{-(E_{N+1}-E_n)\tau_0}$, where $E_{N+1}$ is the first state neglected in the analysis. We should define a new effective mass function to study the time dependence of each of the extracted states,
\beq
E_n^{(\mbox{eff})}(\tau,\tau_0) \equiv \ln \frac{\lambda_n(\tau,\tau_0)}{\lambda_n(\tau+1,\tau_0)} \ ,
\eeq
and look for a plateau,
\beq
\underset{\tau\to\infty}{\mbox{lim}}E_n^{(\mbox{eff})}(\tau,\tau_0) = E_n \ ,
\eeq
to indicate convergence to the desired state. The reference time $\tau_0$ may be chosen to optimize this convergence, and should generally be close to the beginning of the plateau of the standard effective mass. 

The GEVP method works very well in many situations and has been used extensively for LQCD spectroscopy. The main determining factor on the applicability of the method is whether one is able to construct a basis of operators which encapsulates the full low-lying spectrum sufficiently well. One major drawback is that the GEVP assumes a symmetric correlator matrix, meaning that the same set of operators must be used at both source and sink. As discussed in \Sec{interp}, this may be difficult to do numerically due to increases in computational time which scale with the volume when projecting onto a given wavefunction (unless the wavefunction is simply a delta function; however, this operator generally has extremely poor overlap with any physical states of interest). This is particularly a problem for noisy systems where large amounts of statistics are necessary.

There are a few alternatives to the GEVP which do not require a symmetric correlator matrix, such as the generalized pencil of functions (GPof) method \cite{Aubin:2011zz,GPOF1,GPOF2}, and the matrix Prony method \cite{Beane:2009kya,Fleming:2009wb}. We will now briefly discuss the latter, following the discussion of \cite{Beane:2009kya}. 

The Prony method uses the idea of a generalized effective mass,
\beq
M_{\tau_0}^{(\mbox{eff})}(\tau) = \frac{1}{\tau_0}\ln \frac{C(\tau)}{C(\tau+\tau_0)} \tautoinfty E_0 \ ,
\eeq
for some, in principle arbitrary, offset $\tau_0$. Because the correlator $C(\tau)$ is a sum of exponentials, it follows certain recursion relations. As an example, for times where only a single exponential contributes we have,
\beq
C(\tau+\tau_0)+\alpha C(\tau) &=& 0 \ .
\eeq
Plugging in our single exponential for the correlator we can solve for $\alpha$, then plug it back in to our original expression,
\beq
e^{-E_0\tau_0} + \alpha &=& 0 \cr
\longrightarrow C(\tau-\tau_0)-e^{E_0\tau_0}C(\tau) &=& 0 \ .
\eeq
Solving for the ground state energy gives us the same expression as the generalized effective mass at large times,
\beq
E_0 = \frac{1}{\tau_0}\ln \frac{C(\tau)}{C(\tau+\tau_0)} \ .
\eeq
This recursion relation may be generalized for times with contributions from multiple states using the correlation function at different time separations,
\beq
C(\tau+\tau_0k) + \alpha_k C(\tau+\tau_0(k-1))+ \cdots + \alpha_1 C(\tau) = 0 \ .
\eeq

We can now generalize this method for a set of correlation functions produced using different operators. Let $C_i(\tau)$ be an $N$-component vector of correlation functions corresponding to different sources and/or sinks. The correlators then obey the following matrix recursion relation,
\beq
\label{eq:MProny}
M C(\tau+\tau_0)-V C(\tau) = 0 \ ,
\eeq
for some matrices, $M,V$, to be determined. Assume the correlator has contributions from $\Lambda$ states,
\beq
C(\tau) = \sum_n^{\Lambda} \alpha_n u_n \lambda_n^{-\tau} \ ,
\eeq
where $\lambda_n = e^{E_n}$, and $u_n$ is a normalized vector, then we have the following modified GEVP,
\beq
Mu=\lambda^{\tau_0}Vu \ .
\eeq
A solution for $M$ and $V$ may be found by applying $\sum_{t=\tau}^{\tau+t_W}C(t)^T$ to both sides of \Eq{MProny},
\beq
M\sum_{t=\tau}^{\tau+t_W}C(t+\tau_0)C(t)^T - V\sum_{t=\tau}^{\tau+t_W}C(t)C(t)^T = 0 \ ,
\eeq
leading to the solution,
\beq
M = \left[\sum_{t=\tau}^{\tau+t_W}C(t+\tau_0)C(t)^T\right]^{-1} \ , \qquad V=\left[\sum_{t=\tau}^{\tau+t_W}C(t)C(t)^T\right] \ .
\eeq
The parameter $t_W$ is essentially free and may be tuned for optimization, but must obey $t_W \geq \Lambda-1$ in order to ensure that the matrices are full rank. The $\lambda_n$ may then be found from the eigenvalues of $V^{-1}M$. 

Here we have only used a single recursion relation, which is useful for finding the ground state at earlier times than traditional methods. However, this method is generally less effective for calculating excited states than the symmetric GEVP described previously. It may be possible to construct higher order recursion relations for the matrix Prony method in order to get more reliable access to excited states.

\section{\label{sec:systematic}Systematic errors and improvement}
\subsection{Improving the kinetic energy operator}
The first systematic effect we will examine comes from the discretization of the kinetic operator, first discussed in \Sec{LEFT}. In this section I will show the lattice spacing dependence explicitly so that we may see how discretization errors scale. The kinetic term depends on the definition of the Laplacian operator, which we originally defined to be, 
\beq
\nabla_{L}^2 f_j = \sum_{k=1,2,3} \frac{1}{b_s^2} \left[ f_{j+\hat{k}}+f_{j-\hat{k}}-2f_j \right] \ ,
\eeq
leading to the following kinetic term in momentum space,
\beq
\label{eq:Deltasin}
\Delta(p) = \frac{1}{b_s^2}\sum_i \sin^2\frac{b_s p_i}{2} \approx -\frac{p^2}{2}  + \frac{p^4}{24}b_s^2 + \cdots \ .
\eeq
The transfer matrix for the non-interacting system is given by
\beq
\calT = e^{-b_{\tau}H} = 1+b_{\tau} \frac{\Delta({p)}}{M}  \ ,
\eeq
leading to the energy, 
\beq
E=\frac{p^2}{2M} + \calO\left(\frac{p^4}{M} b_s^2 \right) \ .
\eeq
Therefore, discretization errors in this observable appear at $\calO\left(b_s^2\right)$ using this particular discretization. To be more precise, the errors scale with the dimensionless combination $(pb_s)^2$, reflecting the fact that the errors grow as higher momentum scales are probed. As we will discuss in \Sec{NLO}, small lattice spacings can lead to computational difficulties beyond the obvious scaling with the number of lattice sites, and taking the continuum limit may prove to be quite difficult. Therefore, it would be beneficial to have an improved operator whose discretization errors come in at a higher order in $pb_s$. One way to determine such an operator is to examine the relation between the finite difference and the continuum derivative in more detail using a Taylor expansion of the finite difference operator acting on a generic function, $f(x)$,
\beq
f(x+b_s)-f(x) = b_s f'(x) + \frac{b_s^2}{2}f''(x) + \frac{b_s^3}{6} f'''(x) + \frac{b_s^4}{24}f''''(x) + \cdots  \ .
\eeq
Using this expansion, the expression we used previously for the discretized Laplacian can be written,
\beq
\nabla_L^2 f(x) = \frac{1}{b_s^2} \left( f(x+b_s)+f(x-b_s)-2f(x) \right)= f''(x) + \frac{b_s^2}{12} f''''(x)+ \cdots \ .
\eeq
We see that the leading error comes in at $\calO(b_s^2)$, as expected. One method for eliminating the leading error is to add terms involving multiple hops,
\beq
\label{eq:improvkinetic}
\tilde{\nabla}_L^2 f(x) = \frac{1}{b_s^2} \left( f(x+b_s)+f(x-b_s)-2f(x) +c_1 f(x+2b_s) + c_2 f(x-2b_s) \right) \ ,
\eeq
where $c_1,c_2$ must be fixed in such a way as to eliminate the leading error. From symmetry, we must have $c_1=c_2$. We can then Taylor expand these new terms in our action, and determine the resulting energy as a function of $c_1$,
\beq
E(c_1) = \frac{p^2}{2M} + h(c_1) \frac{p^4}{M} b_s^2 + \cdots \ .
\eeq
By solving $h(c_1)=0$ for $c_1$, discretization errors will only enter at $\calO(b_s^4)$, implying a faster approach to the continuum as $b_s$ is decreased. Perhaps more importantly, in cases where decreasing the lattice spacing is difficult or impossible, the resulting systematic errors at finite lattice spacing will be significantly reduced.

This is our first, very simple, example of improvement. A more general method for improving the action in order to reduce discretization effects utilizes an EFT-like approach \cite{Symanzik1,Symanzik2,Symanzik3,Symanzik4,EKLN4}: we add higher dimension operators consistent with the symmetries of the theory and having unknown coefficients. The coefficients are then fixed by matching onto known physical quantities. The dimension of the operator added determines the order at which discretization errors have been eliminated.

In principle, one would need an infinite number of operators in order to eliminate all discretization errors. We are, of course, limited in the number of displacements we can add, as in \Eq{improvkinetic}, by the number of lattice sites. Therefore, the best possible kinetic operator, utilizing all possible spatial hops allowed by the lattice, may still only exactly reproduce the non-interacting spectrum up to the momentum cutoff set by the edge of the first Brillouin zone. Because the kinetic operator $\Delta$ is diagonal in momentum space, we may determine this ``perfect" operator directly by setting the transfer matrix,
\beq
\calT= 1+\frac{b_{\tau}\Delta(p)}{M} = e^{-\frac{b_{\tau}p^2}{2M}} \ ,
\eeq
up to a cutoff, leading to the operator,
\beq
\label{eq:perfect}
\Delta_{\mbox{\tiny perf}}(p) = M\left(e^{\frac{b_{\tau}p^2}{2M}}-1\right) \ , \qquad p< \frac{\pi}{b_s} \ .
\eeq

\begin{figure}
\begin{center}
\includegraphics[width=\linewidth]{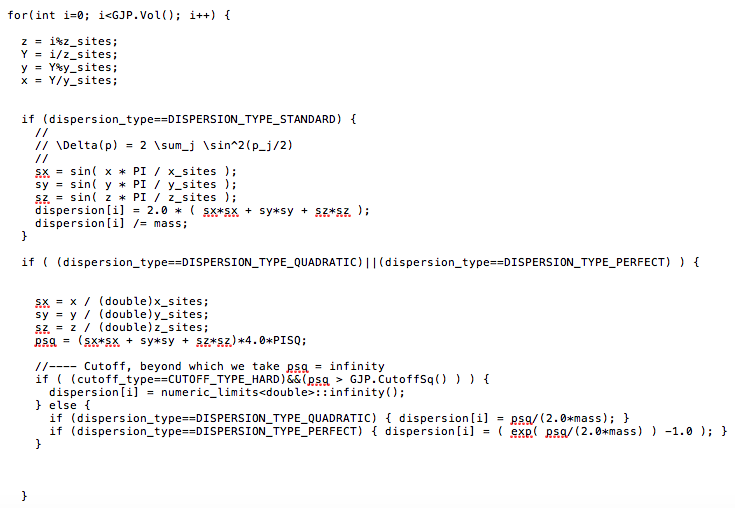}
\end{center}
\caption{\label{fig:dispersion}Example c++ code fragment for computing various lattice Laplacian operators: \Eq{Deltasin} (STANDARD), \Eq{perfect} (PERFECT), as well as a simple quadratic in momentum (QUADRATIC). Note that these are computed in momentum space, and they may be used to calculate the kinetic operator $D^{-1}$, then directly applied to the momentum space vectors computed in \Fig{wfcode}.}
\end{figure}

While this operator is simple in momentum space, it is highly non-local in position space, as expected, and would be unwieldy to use in a typical lattice calculation. However, another benefit of having a non-relativistic formulation with a separable interaction is that the form of the propagator, 
\beq
K^{-1}(\tau) &=& D^{-1}X(\tau)D^{-1}X(\tau-1)\cdots D^{-1} \cr
&=& D^{-1}X(\tau)K^{-1}(\tau-1) \ ,
\eeq
suggests that the kinetic ($D^{-1}$) and interaction ($X$) operators may each be applied separately in whatever basis is most convenient. So, we may choose to start with a source in momentum space (which is often preferable), then apply an exact kinetic operator, $D^{-1}$, also in momentum space, perform a FFT to position space, hit the resulting vector with the $X$ operator, which is most easily specified in position space, FFT again back to momentum space to perform a kinetic operation, and so on until finally the sink is applied. Example code for calculating various forms of inverse kinetic operator in momentum space is shown in \Fig{dispersion}.

The benefit to using the FFT repeatedly rather than simply converting the kinetic operator into position space is that modern FFT libraries are highly optimized and cheap to use. For comparison, if we used the ``perfect" kinetic operator in position space it would be a dense $V\times V$ matrix. The operation of applying such an object to a $V$-dimensional vector,
\beq
D^{-1}(x) | \psi(x) \rangle \ ,
\eeq
scales like $V^2$. On the other hand, using the FFT to convert the $V$-dimensional vector to momentum space, then applying a diagonal matrix to it,
\beq
D^{-1}(p) \left( \mbox{FFT} | \psi(x) \rangle = | \tilde{\psi}(p) \rangle \right) \ ,
\eeq
scales like $V \log V$. This is a method referred to as ``Fourier acceleration" (see e.g. \cite{Batrouni,Daviesetal1,Daviesetal2,Katzetal1}). 

For formulations lacking separability of the kinetic and interaction operations, this method cannot generally be applied. In such cases, the kinetic operator should be kept relatively sparse in position space. Such a condition disfavors the use of \Eq{perfect} for a more modestly improved operator, composed of only a few spatial displacements, using the method outlined in the beginning of this Section.

\subsection{\label{sec:improve}Improving the interaction}
To discuss systematic errors and improvement of the interaction, we will focus on systems tuned to unitarity. Because unitarity corresponds to a conformal fixed-point, the systems we will study only depend on a single scale, the density, $n$. The finite lattice spacing necessarily breaks this conformal symmetry, and we can consider dependence on any new scales to stem from systematic errors. Systems having multiple intrinsic scales contain more complicated dependences of systematic errors, and will be discussed later on. 

Recall that the scattering phase shift for two particles at unitarity is,
\beq
p\cot\delta = 0 \ , 
\eeq
implying that the inverse scattering length, effective range, and all other shape parameters vanish. In \Sec{tuning}, we discussed how to tune the two-particle coupling in order to reproduce infinite scattering length. The lattice, however, naturally induces an effective range for the interactions, which have been generated via auxiliary fields extending across a lattice link, of size $b_s$. In order to improve the interaction and eliminate the unwanted effective range contribution stemming from discretization, we may add a higher-order interaction operator,
\beq
\sum_{\bfx}\sqrt{g_2} \phi \psidag_{\bfx} \nabla_L^2 \psi_{\bfx} \ ,
\eeq
recalculate the scattering amplitude, $A$, as a function of $g_0, g_2$, and tune $g_2$ to eliminate the $r_0$ term in the effective range expansion. In principle, one may further generalize the interaction operator,
\beq
\calL_{\mbox{\tiny int}} = \sum_n \sqrt{g_{2n}}  \phi \psidag \nabla_L^{2n} \psi \ ,
\eeq
where we will now suppress spacetime indices, and use the $g_{2n}$ to tune away successive terms in the effective range expansion. In practice this may be difficult because the interaction is generally no longer separable, so that loops can't be summed analytically. An easier method may be to use the transfer matrix, as we did in \Sec{LEFT}, to determine the two particle energy spectrum in a box, then tune the couplings in order to reproduce the desired energies. The target energies may be determined for systems obeying any known physical scattering phase shift using an approach known as the L\"uscher method, which we will now briefly review.

\subsubsection{\label{sec:Luscher}L\"uscher's method}

L\"uscher's method (\cite{Luscher:1986pf,Luscher:1990ux}) was originally developed as a tool for extracting physical scattering phase shifts from finite volume, Euclidean space observables produced by lattice QCD. The concept of asymptotic ``in" and ``out" scattering states does not exist in a finite volume, making direct scattering ``experiments" impossible on the lattice. Furthermore, the issue of analytic continuation from Euclidean to Minkowski time is a tricky one, particularly when utilizing stochastic techniques. Thus, L\"uscher proposed utilizing a different observable, finite volume energy shifts, and inferring the infinite volume scattering phase shift that would lead to the observed finite volume spectrum. In this section, we will largely follow the discussion in \cite{Beane:2003da}.

First let's recap how to calculate the infinite volume $s$-wave scattering phase shift in our effective theory assuming the following generic tree-level interaction: $\calL_2 = \sum_n g_{2n}p^{2n}$. The scattering amplitude is given by,
\beq
\label{eq:Aluscher}
A_{\infty} = \frac{\sum_{n}g_{2n} p^{2n}}{1-\sum_n g_{2n}p^{2n}I_0^{\infty}} = \frac{4\pi}{M}\frac{1}{p\cot\delta-ip} \ ,
\eeq
where I will now include the super/subscript ``$\infty$" to indicate infinite volume quantities, and $I_0^{\infty}$ is defined as,
\beq
I_0^{\infty}=\int \frac{d^3q}{(2\pi)^3} \frac{1}{E-q^2/M} \ .
\eeq
Note that I have assumed that the interaction is separable in deriving \Eq{Aluscher}. This would not be possible using a momentum cutoff as a regulator, so we will use dimensional regularization for this integral. By investigating the inverse scattering amplitude,
\beq
A^{-1}_{\infty} = \frac{1}{\sum_n g_{2n}p^{2n}} - I_0^{\infty} = \frac{M}{4\pi}(p\cot\delta - i p) \ ,
\eeq 
we can identify
\beq
\label{eq:C2n}
\sum_ng_{2n}p^{2n} = \left[I_0^{\infty} + \frac{M}{4\pi}(p\cot\delta - ip) \right]^{-1} \ .
\eeq
the quantity on the right can be expanded using the effective range expansion; the couplings are then determined by the scattering parameters, as we have seen previously.

Now that we have a relation between the couplings and the physical scattering parameters, let's now use this same effective theory to determine its finite volume spectrum. In a finite volume, there is no continuum of scattering states, but rather a discrete spectrum corresponding to poles in the finite volume analogue of the scattering amplitude, $A_{\mbox{\tiny FV}}$,
\beq
\label{eq:luschereig}
\mbox{Re}\left[A_{\mbox{\tiny FV}}^{-1}\right]=0 \ .
\eeq
Because the imposition of a finite volume can affect only the IR behavior of the theory, the interactions, and therefore the couplings, $g_{2n}$, remain unchanged. Any differences come from loops, where intermediate particles may go on shell and explore the finite boundary. Therefore, our finite volume analogue of the scattering amplitude may be written,
where
\beq
A_{\mbox{\tiny FV}}^{-1} = \frac{1}{\sum_ng_{2n}p^{2n}} - I_0^{\mbox{\tiny FV}} \ ,
\eeq
where the loop integral has been replaced by a finite volume sum over the allowed quantized momenta in a box,
\beq
I_0^{\mbox{\tiny FV}} = \frac{1}{L^3} \sum_{\vec{n}}^{\Lambda} \frac{1}{E-\left(\frac{2\pi n}{L}\right)^2/M} \ .
\eeq

\begin{figure}
\begin{center}
\includegraphics[width=0.5\linewidth]{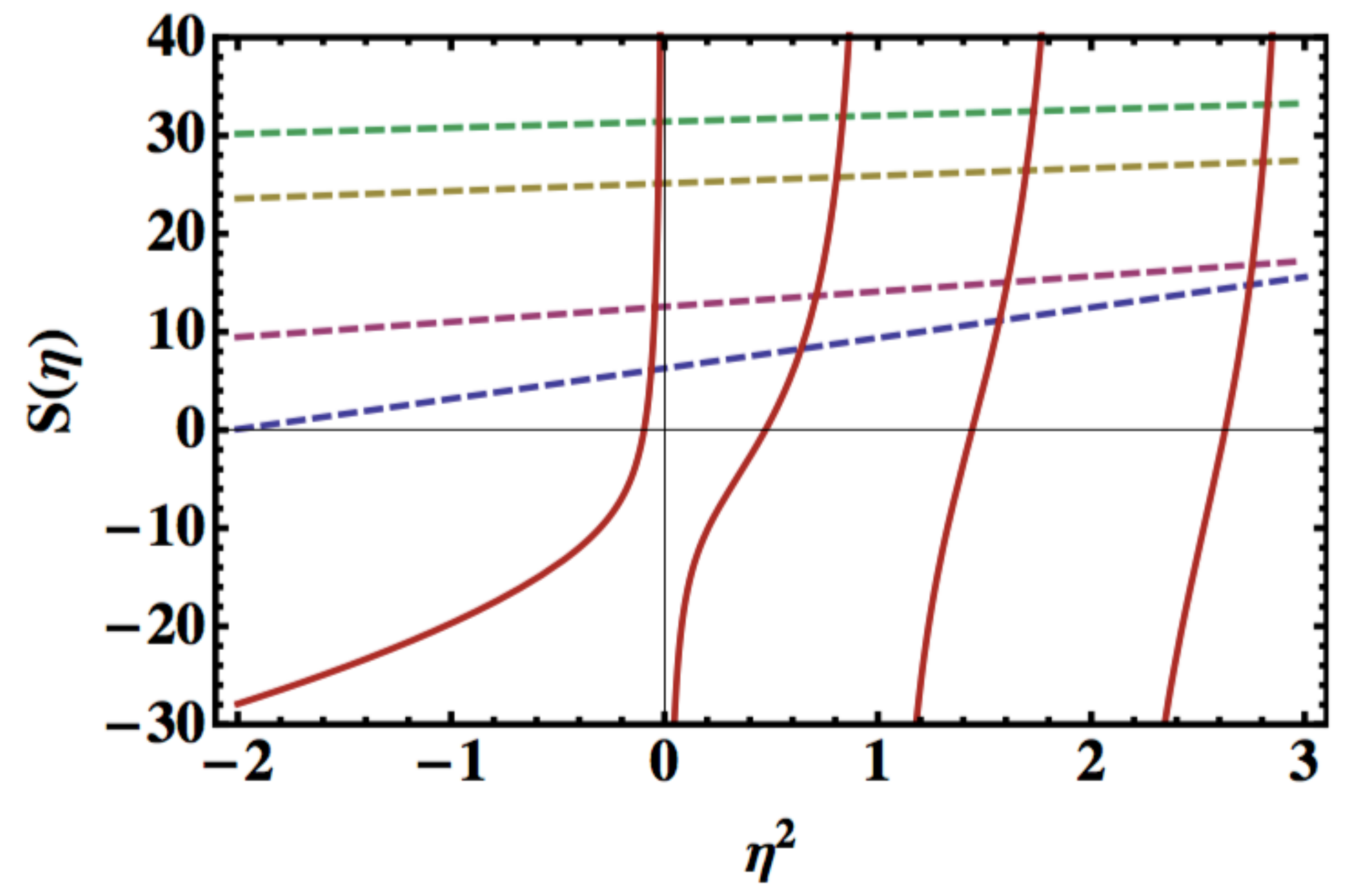}
\end{center}
\caption{\label{fig:luscher}$S(\eta)$ (solid red) and $\pi L p \cot \delta$ (dashed) as a function of $\eta \equiv \left(\frac{p L}{2\pi}\right)^2$. The $\pi L p \cot \delta$ correspond to $r_0/a=-0.1$, for the following volumes: $L/|a|=2$ (blue), $L/|a|=4$ (pink), $L/|a|=8$ (yellow), $L/|a|=10$ (green).  The energy eigenstates for the corresponding volumes are given by the intercepts of $S(\eta)$ with the dashed lines. Figure from \cite{Drut:2012md}.}
\end{figure}

Again, because the couplings are unchanged by the finite volume we are free to use \Eq{C2n} to replace them with the physical infinite volume phase shift, resulting in,
\beq
A_{\mbox{\tiny FV}}^{-1} = \frac{M}{4\pi}(p\cot\delta - ip) + I_0^{\infty} -I_0^{\mbox{\tiny FV}} \ .
\eeq
This leads to the eigenvalue equation,
\beq
\mbox{Re}\left[A_{\mbox{\tiny FV}}^{-1}\right]= \frac{M}{4\pi}p\cot\delta + \mbox{Re}\left[I_0^{\infty} - I_0^{\mbox{\tiny FV}}\right] = 0 \ .
\eeq
I have specified taking the real part of the inverse amplitude merely for calculational simplicity; this quantity is, in fact, already purely real because there are no integrals, and therefore, no $i\epsilon$ prescription. Furthermore, the difference between the infinite volume integral and the finite volume sum must be finite because the two encode the same UV behavior. Finally, we have the result,
\beq
\label{eq:pcotdeltaeig}
p\cot\delta = \frac{4\pi}{M}\left[-\frac{M}{4\pi^2L} \sum_{\vec{n}}^{\Lambda}\frac{1}{\left(\frac{pL}{2\pi}\right)^2 - n^2} - \frac{M\Lambda}{\pi L}\right] = \frac{1}{\pi L}S(\eta) \ ,
\eeq
where $\eta \equiv \left(\frac{pL}{2\pi}\right)^2$, and
\beq
S(\eta) \equiv \sum_{\vec{n}}^{\Lambda} \frac{1}{n^2 - \eta} -4\pi \Lambda \ ,
\eeq
is related to the Riemann zeta function. The cutoff on the sum, $\Lambda$, may be interpreted as an upper limit on the allowed momenta due to the finite lattice spacing, however, in practice it is taken to $\infty$ so that discretization and finite volume effects may be separately accounted for (note that we haven't used our lattice propagators in this derivation, which would be necessary for a proper treatment of discretization effects). Values of momenta which solve this eigenvalue equation for a given phase shift and volume correspond to the predicted finite volume spectrum. This is illustrated in \Fig{luscher}, where the function $S(\eta)$ has been plotted, along with several representative phase shifts, corresponding to positive and negative scattering lengths. The locations of the intersections give the energy eigenvalues for that volume. The poles of the $S$ function give the locations of the energies of a non-interacting system in a box, while the zeroes give the energies for systems at unitarity. 

Many extensions of L\"uscher's method exist for more complicated systems, such as multi-channel processes \cite{Briceno:2012yi,Hansen:2012tf,Li:2014wga,Briceno:2014oea,Briceno:2015tza,Briceno:2015csa,Briceno:2015axa,Briceno:2014uqa}, higher partial waves \cite{Luu:2011ep,Konig:2011nz,Koenig:2011ti}, moving frames \cite{Rummukainen:1995vs,Kim:2005gf}, moving bound states \cite{Bour:2011ef,Davoudi:2011md}, asymmetric boxes \cite{Li:2003jn,Feng:2004ua}, and three-body systems \cite{Hansen:2016fzj,Hansen:2015zga,Hansen:2014eka,Briceno:2012rv}, as well as perturbative expansions for many-boson systems \cite{Beane:2007qr,Detmold:2008gh,Smigielski:2008pa}. Formulations for general systems involving two nucleons may be found in \cite{Briceno:2013lba,Briceno:2013bda}. These formulations have been successfully applied in Lattice QCD for the determination of scattering phase shifts of nucleon-nucleon \cite{Beane:2006mx,Beane:2011iw,Beane:2012vq,Detmold:2015daa,Beane:2015yha,Chang:2015qxa,Orginos:2015aya,Yamazaki:2012hi,Yamazaki:2015asa,Berkowitz:2015eaa,Murano:2013xxa}, meson-meson \cite{Wilson:2014cna,Wilson:2015dqa,Dudek:2012gj,Dudek:2012xn,Dudek:2014qha,Beane:2011sc,Aoki:2007rd,Aoki:2011yj,Pelissier:2012pi,Feng:2010es,Torres:2014vna,Bolton:2015psa,Briceno:2015dca,Lang:2012sv,Prelovsek:2013ela,Lang:2014yfa,Lang:2015hza,Lang:2011mn,Briceno:2016kkp}, meson-baryon \cite{Verduci:2014csa,Lang:2012db,Torok:2009dg,Detmold:2015qwf}, and hyperon-nucleon \cite{Beane:2012ey,Beane:2009py,Beane:2006gf} systems.

\subsubsection{Applying L\"uscher's method to tune the two-body couplings}
The prescription for a lattice QCD calculation of nucleon-nucleon phase shifts is to start with quark interpolating fields to create a two nucleon correlation function, measure a set of finite volume energies, then use the eigenvalue equation, \Eq{pcotdeltaeig}, to infer the infinite volume two nucleon phase shift that produces those energies. For our lattice EFT, however, two nucleon phase shifts are used as input into the coefficients in the Lagrangian. Thus, we can use the L\"uscher method in reverse to calculate what we expect the two nucleon energies in a box to be given a known phase shift, then tune the couplings to reproduce those same energies in our lattice calculations. Having tuned the two-body sector, we can then make predictions about 3- and higher-body systems.

\begin{figure}
\begin{center}
\includegraphics[width=\linewidth]{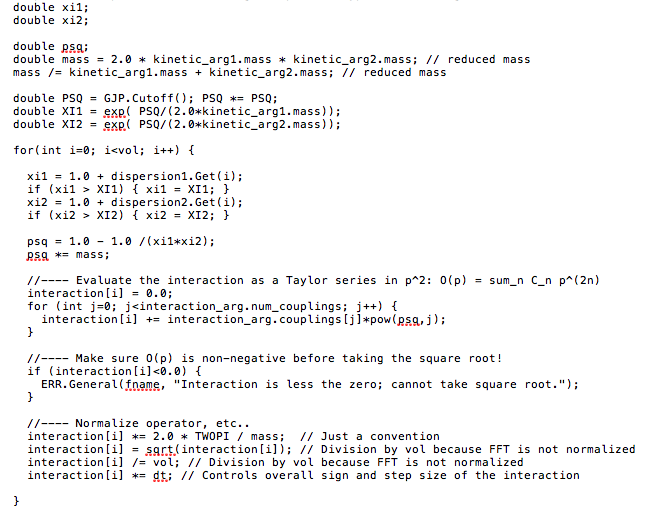}
\end{center}
\caption{\label{fig:interaction}C++ code fragment for calculating the interaction given in \Eq{tuningcoef}, using the operators \Eq{Ofunc}, given some set of input coefficients interaction\_arg.couplings$[\Lambda_n]$. Note that this operator is calculated in momentum space. It may be applied directly to the momentum space vector resulting from the first operation of the kinetic operator, $D^{-1}$. A FFT must then be performed before applying the random auxiliary field, $\phi_x$. A final FFT must then be performed to return to momentum space before applying the next operation of $D^{-1}$ in order to propagate the system forward in time.}
\end{figure}

Our prescription for tuning the coefficients will be to construct the two-body transfer matrix with some set of operators,
\beq
\label{eq:tuningcoef}
\mathcal{G}(\vec{p}) = \sum_n^{\Lambda_n} g_{2n}\calO_{2n}(\vec{p}) \ ,
\eeq
which satisfy the low energy expansion $\calO_{2n}(\vec{p}) = \vec{p}^{2n} \left[ 1 + \calO(\vec{p}^2) \right]$ at low momenta, and should be chosen to depend only on the relative momentum of the two particle system in order to ensure Galilean invariance. This is important so that once the interaction is tuned boosted pairs of particles will see the same interaction. A convenient choice for the operators is given by,
\beq
\label{eq:Ofunc}
\calO_{2n}(\vec{p}) =  M^n \left(1-e^{-\hat{\vec{p}}^2 /M} \right)^n\ ,
\eeq
where $\hat{\vec{p}}$ is taken to be a periodic function of $\vec{p}$ and satisfies the relation $\hat{\vec{p}}^2 = \vec{p}^2 \theta(\Lambda-|\vec{p}|) + \Lambda^2  \theta(|\vec{p}| - \Lambda)$ for $\vec{p}$ in the first Brillouin zone. Sample code for calculating this interaction operator is shown in \Fig{interaction}.

\begin{figure}
\begin{center}
\includegraphics[width=0.5\linewidth]{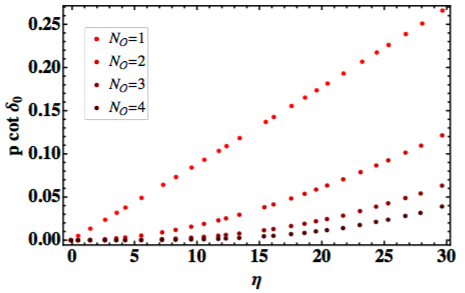}
\end{center}
\caption{\label{fig:tuning}Effective scattering phase shifts $p \cot \delta$ vs. $\eta$ produced by a set of contact interactions of the form in \Eq{tuningcoef}, with $N_{\calO}$ coefficients tuned to unitarity. Figure from \cite{EKLN1}.}
\end{figure}

The transfer matrix may then be diagonalized numerically to determine the energy eigenvalues. The $g_{2n}$ should then be tuned until the energies match the first $\Lambda_n$ eigenvalues given by the L\"uscher method. This process serves a dual purpose: tuning multiple couplings helps reduce lattice spacing effects like the effective range, as we discussed previously, and also takes into account finite volume effects by correctly translating the exact infinite volume phase shifts into a finite volume. The process of tuning for the case of unitarity is illustrated in \Fig{tuning}. Here, $N_{\calO}$ coefficients have been tuned to correctly reproduce the first $N_{\calO}$ L\"uscher eigenvalues. The entire two-body spectrum is then calculated using these coefficients, and the resulting energies are plugged back into \Eq{pcotdeltaeig} to determine the effective phase shift seen by pairs of particles with different momenta. To be truly at unitarity, we should have $p\cot\delta = 0$ for all momenta. Clearly, tuning more coefficients brings us closer to unitarity for larger and larger momenta. This is particularly important for calculations involving many-body systems, where the average momentum grows with the density, $\langle p \rangle \sim n^{1/3}$.

A quantitative prediction can be made for the error remaining in higher, untuned two-body energy levels \cite{EKLN1}. Assuming $N_{\calO}$ terms in the effective range expansion have been tuned to zero,
\beq
p\cot\delta \sim r_{N_{\calO}-1}p^{2N_{\calO}} =\left(\frac{2\pi}{L}\right)^{2N_{\calO}} r_{N_{\calO}-1} \eta^{N_{\calO}} \ ,
\eeq
we can then use L\"uscher's relation for the first untuned eigenvalue $\eta_k$,
\beq
\left(\frac{2\pi}{L}\right)^{2N_{\calO}} r_{N_{\calO}-1} \eta_k^{N_{\calO}} = \frac{1}{\pi L}S(\eta_k) \ .
\eeq
Let's suppose $\eta_k^*$ is the eigenvalue one would expect in the true unitary limit. We can then Taylor expand the function $S(\eta_k)$ around $\eta_k^*$,
\beq
S(\eta_k) \approx c_k (\eta_k-\eta_k^*)\ ,
\eeq
where $c_k$ is the slope near $\eta_k^*$. The error is then estimated as,
\beq
\frac{\eta_k}{\eta_k^*}-1 \approx \frac{\pi L}{\eta_k^* c_k} \left(\frac{2\pi}{L}\right)^{2N_{\calO}}r_{N_{\calO}-1} \left(\eta_k^{*}\right)^{N_{\calO}} \sim \calO\left(L^{1-2N_{\calO}}\right) \sim \calO\left((b_s n^{1/3})^{2N_{\calO}-1}\right) \ ,
\eeq
where on the right I have rewritten the scaling with the volume as a scaling with the density to remind you that though the errors scale with the volume, these are not actually finite volume errors we are investigating, but discretization effects scaling with the dimensionless quantity $b_s n^{1/3} \sim b_s /L_{\mbox \tiny phys} = 1/L$ for systems at unitarity. The L\"uscher method takes into account finite volume effects automatically. 

\subsection{Scaling of discretization errors for many-body systems}
Having tuned our two-body interaction, we can now also predict the scaling of errors that we should expect to find in an $N$-body calculation. Let us suppose that the first untuned operator contains at most $2N_{\calO}$ derivatives,
\beq
\label{eq:errop}
\calO_{2N_{\calO}} \sim \left(\psi\psi\right)^{\dagger} \psi \nabla^{2N_{\calO}}\psi \ .
\eeq
The leading error results when any pair of particles interacts via this operator, and should scale with the dimension of this operator. 

To determine the operator dimension, first let me briefly recap how scaling dimensions are determined in a non-relativistic theory (see \cite{Kaplan:2005es} for more details). We expect the action, $S$, to be a dimensionless quantity, so we will consider the action for a non-interacting theory to determine how the fields and derivatives must scale,
\beq
S=\int d\tau d^3x \psidag \left( \partial_{\tau}-\frac{\nabla^2}{2M} \right) \psi \ .
\eeq
First, note that the mass, $M$, carries zero scaling dimension in a non-relativistic theory because it is considered to be much larger than any scale of interest. Then, from the expression in parentheses, we see that time and space must scale differently, $[\partial_{\tau}] = 2 [\nabla]$. Using the convention $[\nabla]=1$, we can then determine that the dimension of the fermion field must be $[\psi]=3/2$. 

Now let us return to the operator, \Eq{errop}, and determine its scaling dimension relative to the energy,
\beq
\left[ \left(\psi\psi\right)^{\dagger} \psi \nabla^{2N_{\calO}}\psi \right] - \left[\psidag \partial_{\tau} \psi \right] =( 6+2N_{\calO} ) - ( 5 ) = 1+2N_{\calO} \ .
\eeq
This indicates that the error from such an operator will scale as $\sim \calO(b_s p)^{1+2N_{\calO}}$, or $\sim \calO\left((b_s n^{1/3})^{1+2N_{\calO}}\right)$ for unitary fermions. This is similar scaling that we saw for higher two-body states, however, here the dependence on the number of particles is also important.

One may in principle tune as many operators as possible in order to perfect the interaction for higher energies. In practice, however, as more and more operators are tuned, the coefficients in front of higher dimensional operators which are still untuned can become very large. This can cause interactions seen by pairs of particles far in the tail of the momentum distribution to generate large errors. Thus, similar to the case of the kinetic operator, there is a limit to how ``perfect" the interaction can be made. 

On the other hand, these $s$-wave two-body interactions are not the only possible errors that are induced by the lattice, so we should not expect to see much improvement by tuning more operators corresponding to errors which are higher order than the leading operator which is not accounted for. For example, an unfortunate consequence of our tuning program is the introduction of interactions in the $p$-wave channel, as well as in higher partial waves. While a simple interaction which is point-like in space has no $p$-wave contribution, the introduction of spatial derivatives in our tuning operators gives rise to these new $p$-wave interactions. The leading $p$-wave operator has the form,
\beq
\calO_{p\mbox{\tiny -wave}} \sim \psidag \vec{\nabla} \psi \cdot \psidag \vec{\nabla} \psi \ ,
\eeq
and induces errors at $\calO\left(b_s n^{1/3})^{3}\right)$. In order to cancel this operator we could in principle add a $\phi$ field which carries momentum and carry out a similar program for tuning the coefficients as we used for the $s$-wave interaction. This destroys the separability of our interaction, however, and may be difficult to implement, in addition to introducing a new source of noise.

In general, we can determine all possible sources of discretization error as well as their scaling using a method referred to as the Symanzik effective action \cite{Symanzik1,Symanzik2,Symanzik3,Symanzik4,EKLN4}. The basic procedure begins through considering any possible operators (that have not been explicitly tuned) which are allowed by the symmetry of the theory. Because these operators may only be induced through discretization and must disappear in the continuum limit, they should be multiplied by the lattice spacing raised to the appropriate scaling dimension of the operator. We can then determine at what order in $b_s$, relative to the energy, we can expect systematic errors to arise.

Let's take a look another interesting operator which arises due to discretization, corresponding to a three-body interaction. While there can be no point-like 3-body interaction in the continuum limit for 2-component fermions due to the Pauli exclusion principle, three particles separated by a lattice spacing may interact via $\phi$-field exchange because they don't all lie on the same spacetime point. Thus, we should include in our Symanzik effective action an operator,
\beq
\calO_{\mbox{\tiny 3-body}} \sim \left(\psi \psi \psi \right)^{\dagger} \psi \psi \psi \ .
\eeq
Na\"ively, the dimension of this operator is 9, and therefore should contribute errors of $\calO\left((b_sn^{1/3})^4\right)$. So far, all of the operators we've discussed obey this simple scaling, corresponding to na\"ive dimensional analysis. However, our theory is strongly interacting, which can in general lead to large anomalous dimensions of certain operators. 

\begin{figure}
\begin{center}
\includegraphics[width=\linewidth]{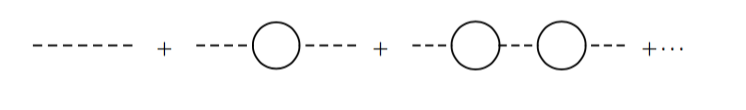}
\end{center}
\caption{\label{fig:phiprop}Propagator for the bosonic field $\phi$, dressed by fermionic loops.}
\end{figure}

As an example, let's consider the scaling dimension of a very basic operator, the field $\phi$. The canonical (non-interacting) dimension for a generic bosonic field in a non-relativistic theory can be deduced by looking at the kinetic term in the action,
\beq
S_{\mbox{\tiny kin}} = \int d\tau d^3x \nabla^2 \phi^2 \ ,
\eeq
leading to a scaling dimension, $[\phi] = 3/2$. However, once interactions with the $\psi$ fields are included, the $\phi$ propagator is renormalized through loop diagrams (see \Fig{phiprop}). For a non-perturbative interaction, we must sum all possible loop diagrams. However, there is a simpler way to determine the scaling dimension of the strongly interacting $\phi$ field. The key is to recognize that near unitarity the $\phi$ field represents a bound state of two $\psi$ fields at threshold. We can therefore write $\phi$ as a local operator,
\beq
\phi(x) = \underset{x\to y}{\lim} |x-y| \psidag(x)\psi(y) \ ,
\eeq
where $|x-y|$ must be included to ensure that matrix elements of the operator are finite (the wavefunction for two particles at unitarity must scale as $|x-y|^{-1}$ at short distances \cite{NishidaSonConformal}). Using our previous analysis for the scaling dimension of the $\psi$ field, we find,
\beq
[\phi]_{\mbox{\tiny int}} = 2 \ ,
\eeq
which implies a very strong wavefunction renormalization.

In general it can be very difficult to calculate anomalous dimensions directly in a non-perturbative fashion. However, for non-relativistic conformal field theories (CFT), there exists an operator-state correspondence (similar to an ADS/CFT correspondence), which relates the scaling dimension of an operator in the CFT (e.g. for unitary fermions) to the energy of the corresponding state in a harmonic potential \cite{NishidaSonConformal}. For example, we have already determined the dimension of the field $\psi$ to be 3/2, and the energy of a single fermion in a harmonic potential with oscillator frequency $\omega$ is $3/2 \omega$. The energy of two unitary fermions in a harmonic potential is $2\omega$, corresponding to the dimension of the $\phi$ field, $[\phi]=2$. 

Returning now to our 3-body operator, we can use numerical results for the energy of three fermions in a total $l=0$ state in a harmonic potential \cite{2007PhRvL..99w3201B,2011CRPhy..12...86B} to determine that,
\beq
\left[\psi\psi\psi\right] = 4.67 \ .
\eeq
The error-inducing operator in the Symanzik effective action both creates and destroys this 3-body state, resulting in
\beq
\left[\left(\psi\psi\psi\right)^{\dagger}\psi\psi\psi\right] = 9.34 \ .
\eeq
The relative error in the energy will then be $\calO\left(L^{-(9.34-5)}\right) = \calO\left(L^{-4.34}\right)$.

It turns out that the ground state of three fermions in a harmonic potential is actually not the $s$-wave state, but a $p$-wave state with energy $\sim 4.27\omega$. Thus, we should expect an additional systematic error corresponding to a 3-body $p$-wave operator that contributes at $\calO\left(L^{-3.55}\right)$ \cite{2006PhRvL..97o0401W}. Finally, at approximately the same order as the 3-body $s$-wave there is a 2-body $d$-wave operator (four derivatives) with zero anomalous dimension, and therefore contributing at $\calO\left(L^{-5}\right)$.

While certainly only the leading error ($\calO\left(L^{-3}\right)$) will dominate very close to the continuum limit, at a finite lattice spacing we have just demonstrated that there are several sources of error scaling with very similar powers of the lattice spacing. If we wish to eliminate discretization errors through extrapolation to the continuum limit, we must include all possible non-negligible contributions in our extrapolation function. For example, we could employ the following function:
\beq
E(L) = E_0\left[1+a L^{-3} + b L^{-3.55} + c L^{-4.34} + d L^{-5} + \cdots \right] \ ,
\eeq
and fit the coefficients $\{a,b,c,d\}$ using data at several volumes, in order to extract the continuum energy, $E_0$ \cite{EKLN4}.

\subsection{Additional sources of systematic error}
It should be pretty clear by now that understanding and controlling systematic errors can be quite complicated, even for conformal systems! For more complex systems with contributions from multiple scales, such as nucleii, things become even messier. As a simple example of a system with more than one scale we can consider trapping our unitary fermions in a harmonic potential, which will allow us to discuss finite volume errors that are not accounted for by the L\"uscher method. This is clearly relevant for cold atom experiments, which utilize traps, but may also be useful for calculating the energies needed to use the operator-state correspondence discussed in the previous subsection.

The new characteristic length scale contributed by the introduction of the harmonic trap is given by the size of the trap, $L_0$. We now have two different dimensionless quantities which determine the scaling of systematic errors due to discretization, $b_s/L_0$, and finite volume, $L_0/L_{\mbox{\tiny phys}}$, individually. To determine the size of discretization errors we may use the Symanzik effective action method as previously described, with the average momentum scale replaced by $n^{1/3} \to N^{1/3}/L_0$. Finite volume errors may be estimated by examining the long distance behavior of the wavefunction of the system of interest, where distortions due to the finite boundary can occur. For a system in a harmonic trap with local interactions, wavefunctions behave as Gaussians at large distance, so we might consider using a function $E(L_{\mbox{\tiny phys}}) = E_0\left(1+a e^{-\left(L_0/L_{\mbox{\tiny phys}}\right)^2}\right)$ to extrapolate to the infinite volume limit. 

For the case of nuclei, which are bound states whose wavefunctions fall off exponentially at long distance, we might expect systematic errors to scale as $e^{-R/L_{\mbox{\tiny phys}}}$, where $R$ is the characteristic size of the bound state. In general, one may also need to consider effects from interactions between images produced due to the periodic boundary conditions. For example, if the interaction between images is mediated at long distances by the exchange of a light particle, such as a pion, then we might expect systematic errors to fall off exponentially with $\sim \left(m_{\pi} L_{\mbox{\tiny phys}}\right)$. Note that this type of finite volume effect is not accounted for by the L\"uscher formalism; this is because in order to derive \Eq{pcotdeltaeig} we had to assume that all interactions were point-like.

Finally, we should briefly discuss systematic errors associated with temporal discretization. These tend to be far less worrisome for zero temperature results for several reasons. The first is due to the relation $b_{\tau} = \frac{b_s^2}{M}$ for non-relativistic theories, indicating that temporal discretization errors are of lower order than spatial discretization errors. Furthermore, our tuning method for improving the kinetic and interaction operators also translates into an improved temporal derivative operator. The lattice temporal derivative is given by the finite difference,
\beq
\partial_{\tau}\psi \sim \psi_{\tau+1} - \psi_{\tau} \sim \left(\calT -1\right) \psi_{\tau} \ ,
\eeq
where on the right hand side I have used the knowledge that the transfer matrix $\calT$ is our time-translation operator. By perfecting the transfer matrix with our tuning method, we are in turn perfecting the single time hop operation, thereby reducing temporal discretization errors.

We also have the freedom to use the anisotropy parameter $M$ to tune the temporal lattice spacing to be intrinsically smaller than the spatial lattice spacing. However, it should be noted that because the temperature is controlled by the physical Euclidean time length, $1/\left(b_{\tau}N_{\tau}\right)$, increasing the anisotropy parameter $M$ will necessitate an increase in the number of temporal lattice points to reach the zero temperature limit. On the other hand, having a finer temporal lattice spacing may also help to better resolve plateaus occurring within a short ``golden window" before the noise begins to set in, due to the increase in the number of points available for fitting. For this reason, anisotropic lattices are sometimes used in lattice QCD for noisy systems. However, points corresponding to a finer temporal lattice spacing are also more correlated, so it is currently unclear whether anisotropic lattices are actually beneficial for resolving noisy signals.

\section{\label{sec:NLO}Beyond leading order EFT}

The first step away from unitarity and toward real nuclear physics that we can easily take is to introduce a four-component nucleon field, $N$, containing two flavors of spin up and spin down fermions. The nucleons have two allowed $s$-wave scattering channels, $^1S_0$ and $^3S_1$, which should be tuned independently (breaking the approximate $SU(4)$ symmetry between the nucleons) to give the physical nucleon-nucleon scattering lengths. One possible way to achieve this is to introduce two four-fermion interactions corresponding to,
\beq
\label{eq:STint}
\calL_{\mbox{\tiny int}} = -\frac{1}{2} g_S \left(N^{\dagger}N\right)^2 - \frac{1}{2} g_T \left(N^{\dagger}\vec{\sigma}N\right)^2 \ ,
\eeq
where $\sigma_i$ is a Pauli matrix acting on the spin indices, and $g_S,g_T$ are couplings for the spin singlet and spin triplet channel, respectively. The lattice version of this interaction requires the introduction of two independent auxiliary fields, $\phi_{S},\phi_{T}$. One possibility is,
\beq
\label{eq:noSU4}
\calL_{\mbox{\tiny int}}^{(L)} = \sqrt{g_S} \phi_S N^{\dagger}N+\sqrt{g_T} \phi_T \vec{\sigma} \cdot N^{\dagger} \vec{\sigma} N \ .
\eeq

There are, in fact, many ways to implement the same interactions, and the different implementations will affect the signal-to-noise ratios of observables. For example, one could imagine having one of the $\phi$ fields couple to both channels equally (the $SU(4)$ limit), tuned to give the scattering length of the more attractive channel, $^3S_1$, then adding a second auxiliary field coupling only to the $^1S_0$ channel and tuning this coupling to be repulsive, making this channel more weakly attractive as desired. As we learned in \Sec{SNR}, repulsive interactions cause severe sign and noise problems, so this would clearly be a poor choice of implementation.

Let's look at the signal-to-noise ratio for a two-particle correlator in the $^1S_0$ channel using the interaction shown above, \Eq{noSU4}, where neither interaction is repulsive, but their relative strengths are different. The signal goes like,
\beq
\langle K^{\uparrow -1}_n(\tau) K^{\downarrow -1}_n(\tau) \rangle \sim e^{-E_0^{(^1S_0)}\tau} \ .
\eeq
while the noise is given by,
\beq
\label{eq:noSU4noise}
\sigma^2 \sim \langle K^{\uparrow -1}_n(\tau) K^{\downarrow -1}_n(\tau) \rangle K^{\uparrow -1}_{n'}(\tau) K^{\downarrow -1}_{n'}(\tau) \rangle \sim e^{E_B^{(4)}\tau} \ ,
\eeq
where $n'$ denotes a particle of different flavor from $n$, and $E_B^{(4)}$ is the binding energy of a four particle, four flavor state. This causes a signal-to-noise problem which is similar to our original two-body correlator, however, in this case the problem is exacerbated by the fact that particles in \Eq{noSU4noise} having different flavor index interact through the most attractive channel, $^3S_1$. This results in a greater disparity between the energies governing the signal and the noise, leading to more severe exponential decay of the signal-to-noise ratio. Unequal interactions can also lead to problems with reweighting methods designed to alleviate an overlap problem if the desired reweighting factor is no longer real or positive.

One method, devised by the Bonn-Raleigh group (for a review, see e.g. \cite{Lee:2008fa}), for avoiding the extra noise caused by unequal interactions in the two $s$-wave channels, is to use an $SU(4)$ symmetric transfer matrix, $\calT_{SU(4)}$, to evolve the system for several time steps before applying the full asymmetric transfer matrix. This process may be thought of as utilizing several applications of $\calT_{SU(4)}$ in order to produce a better interpolating wavefunction from some initial guess wavefunctions, $\Psi_{i,f}$, which is then used as a source for the correlation function,
\beq
C(\tau) = \langle \Psi_f | \calT_{SU(4)}^{\tau'} \calT^{\tau} \calT_{SU(4)}^{\tau'}|\Psi_i \rangle = \langle \tilde{\Psi}_f | \calT^{\tau} | \tilde{\Psi}_i\rangle \ ,
\eeq
where $| \tilde{\Psi}_i\rangle  \equiv \calT_{SU(4)}^{\tau'}|\Psi_i \rangle$. Using this method reduces the number of times the noisier $\calT$ must be used because the system begins in a state that is already closer to the true ground state.

Another method used by the same group to reduce noise is to perform a Fierz transformation on the four-fermion interactions in order to define interactions with more symmetric couplings \cite{Borasoy:2006qn}. Using the identity,
\beq
\left(N^{\dagger}N\right)^2 = -\frac{1}{2} \left(N^{\dagger}\vec{\sigma}N\right)^2 - \frac{1}{2} \left(N^{\dagger}\vec{\tau}N\right)^2 \ ,
\eeq
we can rewrite the four-fermion interactions, \Eq{STint}, to give the following,
\beq
\tilde{\calL}_{\mbox{\tiny int}} = -\frac{1}{2} g_0 \left(N^{\dagger}N\right)^2 - \frac{1}{2} g_I \left(N^{\dagger}\vec{\tau}N\right)^2 \ ,
\eeq
where $\tau_i$ is a Pauli matrix acting on the flavor components of $N$, and the couplings $g_{0,I}$ are related to the original couplings by,
\beq
g_0 = g_S-2g_T \ , \qquad g_I = - g_T \ .
\eeq

\subsection{Tuning the effective range}
The method outlined in \Sec{tuning} was devised as a way to allow us to tune our couplings to reproduce any physical scattering phase shift using the L\"uscher finite volume method. We were able to successfully tune the system to unitarity, where the effective range and all higher shape parameters vanish. For nucleon scattering, the effective ranges in the $s$-wave channels are given roughly by the Compton wavelength of the pion, so the next logical step in our quest toward nuclear physics should be to try to tune our coefficients to give the physical effective ranges. Unfortunately, a problem arises for producing a non-zero effective range non-perturbatively using point-like interactions in combination with a lattice regulator. 

The choice of regulator is relevant when attempting to perform non-perturbative calculations because EFTs in general are non-renormalizable. However, they should be renormalizable order by order in perturbation theory, because at each order we introduce a new operator having the correct dimensions and symmetries to act as a counterterm, absorbing infinities from loops containing lower order interactions. Lattice methods incorporate the Lagrangian of the theory non-perturbatively, effectively summing the entire subset of diagrams for each interaction. In principle, such a formulation may also require the introduction of an infinite number of counterterms to absorb the divergences from all loop diagrams. 

In certain cases, however, this situation can be avoided. An example is our non-perturbative tuning of the scattering length. Recall that all bubble diagrams involving only the coupling $g_0$ were separable; this allowed us to write the non-perturbative scattering amplitude as a geometric sum, and we were able to absorb all loop divergences into the single coupling, $g_0$. The condition of separability for loop diagrams containing interactions which carry momenta is dependent on the choice of regulator. Our choice of a lattice regulator, which is similar to a momentum cutoff, leads to a bound, known as the Wigner bound, on the allowed effective ranges one can access non-perturbatively \cite{PhysRev.98.145,Phillips:1996ae,Cohen:1996my}.

Because the general tuning method introduced in \Sec{tuning} involves the numerical calculation of the transfer matrix, understanding the Wigner bound in this context is difficult. To better illustrate the issue, let's attempt to tune the effective range instead using the first method for tuning, outlined in \Sec{couplings}. This method involves calculating the scattering amplitude and tuning the couplings to match the desired scattering parameters directly from the effective range expansion. 

We will again calculate a sum of bubble diagrams, however, we must now include an interaction of the form $\calL_{\mbox{\tiny int}} \sim g_2 \psidag \nabla^2 \psi$, which we would like to use to tune the effective range. We will largely follow the discussion of \cite{Phillips:1997xu}. A generic integral from one of these diagrams will have the form,
\beq
I_{2n} = \frac{1}{2\pi^2}\int dq \frac{q^{2+2n}}{E-q^2/M} \ ,
\eeq
where $n=0,1,2$, depending on which of the two interactions  we have at the two vertices. Since we are interested in the renormalizability of the scattering amplitude, we will separate out the divergent pieces of such an integral by expanding around $q\to\infty$,
\beq
\label{eq:intrecursion}
I_{2n} = \frac{1}{2\pi^2}\int dq \left[M q^{2n}-EM\int dq \frac{q^{2n-2}}{E-q^2/M} \right] \ ,
\eeq
and investigate the integrals using different regularization schemes. The above relation may be iterated for a given $n$ until the remaining integral is finite. The lowest order integral that we will need is given by,
\beq
I_0 = -\frac{1}{2\pi^2}\int dq \frac{q^2}{E-q^2/M} \ .
\eeq
We evaluated this integral previously using a cutoff, $\pi\Lambda/2$, to find,
\beq
I_0 = \frac{M}{4\pi}\left[\Lambda+iME\right] \qquad \mbox{(cutoff)} \ .
\eeq
Using dimensional regularization (dim reg), on the other hand, eliminates power-law divergences, so the result becomes,
\beq
I_0 = \frac{M}{4\pi}iME  \qquad \mbox{(dim reg)}\ .
\eeq
The other two integrals we will need have two and four additional powers of the momentum. Using our relation, \Eq{intrecursion}, we can write,
\beq
I_2 = ME I_0 - \lambda_2 \ ,
\eeq
where
\beq
\lambda_2 = \frac{M}{2\pi^2} \int dq q^2 = \left\{ \begin{array}{cc}
-\frac{M\pi}{48} \Lambda^3 & \mbox{cutoff} \\
0 & \mbox{dim reg} \\
\end{array} \right. \ ,
\eeq
and
\beq
I_4 = ME I_2 - \lambda_4 \ ,
\eeq
where
\beq
\lambda_4 = \frac{M}{2\pi^2} \int dq q^4 = \left\{ \begin{array}{cc}
-\frac{M\pi^3}{320} \Lambda^5 & \mbox{cutoff} \\
0 & \mbox{dim reg} \\
\end{array} \right. \ .
\eeq
From these results we see that dim reg leads to a separable interaction because each of the integrals can be written in terms of $I_0$ times some overall factor. On the other hand, the cutoff introduces new terms which cannot be factorized. 

In order to evaluate the scattering amplitude more generally for a non-separable interaction we must solve a matrix equation. We will set this up by noting that the interaction can be written,
\beq
V(p,p') = \sum_{i,j=0}^1 p'^{2i} v_{ij} p^{2j} \ ,
\eeq
where 
\beq
v=\left( \begin{array}{cc}
g_0 & g_2 \\
g_2 & 0 \\
\end{array} \right) \ .
\eeq
The amplitude is then,
\beq
A = - \sum_{i,j=0}^1 \left(ME\right)^{i+j} a_{ij} \ ,
\eeq
where
\beq
a = v + v \calI a \ , \qquad \calI = \left( \begin{array}{cc}
I_0 & I_2 \\
I_2 & I_4 \\
\end{array} \right) \ .
\eeq
We can now solve for $a$,
\beq
a = \left[1-v\calI \right]^{-1} v = \frac{1}{\lambda} \left( \begin{array}{cc}
g_0+g_2^2 I_4 & g_2(1-g_2I_2) \\
g_2(1-g_2I_2) & g_2^2 I_0 \\
\end{array} \right) \ ,
\eeq
where
\beq
\lambda \equiv 1-g_0 I_0 -2g_2 I_2 + g_2^2 (I_2^2-I_0I_4) \ .
\eeq
Finally, we have
\beq
\frac{1}{A} &=& - \frac{(g_2\lambda_2 -1)^2}{g_0+g_2[ME(2-g_2 \lambda_2) +g_2 \lambda_4 ]}+I_0 \cr
&=& \frac{M}{4\pi}\left(-1/a + 1/2 r_0 ME - i\sqrt{ME} \right) \ ,
\eeq
where I have used the effective range expansion for the inverse scattering amplitude on the right hand side. 

This expression may be used to determine the couplings $g_{0,2}$ in terms of the effective range parameters, $a,r_0$, by expanding the left hand side in powers of $ME$, and comparing the resulting coefficients to the corresponding parameters in the effective range expansion. The leading order is,
\beq
\left. \frac{1}{A}\right|_{E=0} = - \frac{(g_2\lambda_2-1)^2}{g_0+g_2^2\lambda_4}+\left. I_0 \right|_{E=0} = -\frac{M}{4\pi a}  \ ,
\eeq
while the next order gives,
\beq
\left[ \frac{\partial}{\partial (ME)}\frac{1}{A}\right]_{E=0}\frac{g_2\left( \left. I_0 \right|_{E=0} + \frac{M}{4\pi a}\right)^2)(2-g_2\lambda_2)}{(g_2\lambda_2-1)^2} = \frac{M}{8\pi} r_0 \ .
\eeq
Using these two expressions and the above relations for  $\lambda_n$ and $I_0$, we can derive the following dependence of the effective range on the couplings for a theory regularized using dim reg,
\beq
r_0 = \frac{Mg_2}{\pi a^2} \ .
\eeq
Because the effective range is proportional to the coupling $g_2$, it can be tuned arbitrarily. Thus, as expected from the separability of the interaction, there are no issues with renormalizability when using dim reg. 

Let us now see what happens for the case of a cutoff. The relation becomes,
\beq
\label{eq:r0cutoff}
r_0 &=& \frac{8\pi}{M} \left(\frac{M}{4\pi a}+\left. I_0 \right|_{E=0}\right)^2 \left[\frac{1}{(g_2\lambda_2-1)^2\lambda_2}-\frac{1}{\lambda_2}\right] \cr
&=& \frac{M}{2\pi} \left(1/a+\Lambda\right)^2 \left[ -\frac{1}{\left(g_2 \frac{M\pi}{48}\Lambda^3-1\right)^2\frac{M\pi}{48}\Lambda^3} + \frac{48}{M\pi \Lambda^3} \right] \ .
\eeq
We should now attempt to remove the cutoff by taking, $\Lambda \to \infty$,
\beq
r_0 \underset{\Lambda\to\infty}{\longrightarrow} -\frac{\frac{M}{2\pi}\Lambda^2}{(g_2\frac{M\pi}{48}\Lambda^3-1)^2\frac{M\pi}{48}\Lambda^3} \ ,
\eeq
where I have kept the first term in square brackets in \Eq{r0cutoff} because there $g_2$ may be renormalized to absorb factors of $\Lambda$. Because $g_2$ must be real to ensure a Hermitian Hamiltonian, this expression shows that if we attempt to remove the cutoff of the theory, we are only allowed to tune $r_0 \leq 0$. 

More generally, Wigner showed that for any potential which obeys $V(r,r') \to 0$ for $r,r'>R$ sufficiently quickly for some characteristic radius $R$, then
\beq
r_0 \leq 2\left(R-\frac{R^2}{a}+\frac{R^3}{3a^2}\right) \ .
\eeq
For a potential generated using delta function interactions and a momentum cutoff, $R\sim 1/\Lambda$, and we arrive at our expression $r_0\leq 0$.

In our lattice formulation the interactions are generated by an auxiliary field extending across a single time link, so that $R \sim b_s$. Therefore, if we try to tune $r_0$ non-perturbatively via the inclusion of such interactions in the Lagrangian, we are limited to $r_0 \lesssim b_s$. This was not a problem when we considered unitarity, since at this point $r_0 =0$. For nuclear physics, this bound restricts us to tuning the effective range to be smaller than the lattice spacing, implying that there is no continuum limit to the theory. On the other hand, the theory we are attempting to simulate is only an effective theory of nucleons, valid up to a physical cutoff. Thus, so long as we do not attempt to probe physics beyond scales of order $\sim 1/r_0$ there will be no inconsistencies. This is clearly a limitation, however, and also restricts our ability to vary the lattice spacing when studying discretization effects.

One possibility for avoiding this restriction is to include the effective range contribution to observables perturbatively, keeping the renormalizability of the effective theory intact. Perturbative corrections may be added by expanding the transfer matrix,
\beq
\calT \approx e^{-H_0 b_{\tau}} - b_{\tau} \delta H e^{-H_0 b_{\tau}} \ ,
\eeq
where $H = H_0 + \delta H$ is the full Hamiltonian and $\delta H$ is the piece we wish to treat perturbatively. Multiple insertions of $\delta H$ may be included to reach higher orders in the effective theory.

\subsection{Including pions}
If we wish to probe energies of order the pion mass we must include pions explicitly into the effective theory. Unfortunately, pions are notoriously difficult to include in a consistent power counting scheme. Here, we will only briefly outline some of the issues related to power counting for pion contributions. 

The KSW expansion proposed that pion exchange be treated as a series of perturbative corrections to the leading order pionless EFT \cite{Kaplan:1996xu,Kaplan:1998tg,Kaplan:1998we}. In this case, a tree level one pion exchange (1PE) diagram may be given by \cite{Fleming:1999ee},
\beq
\label{eq:tree}
\begin{array}{cc}
\includegraphics[width=0.1\linewidth]{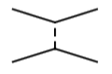} & \sim \frac{g_A^2}{2f_{\pi}^2} f\left(\frac{p}{m_{\pi}}\right) \\
\end{array} \ ,
\eeq
where $g_A$ is the axial coupling, $f_{\pi}$ is the pion decay constant, and $f(p/m_{\pi})$ is a dimensionless function. By comparison, at one loop there is a box diagram,
\beq
\begin{array}{cc}
\includegraphics[width=0.1\linewidth]{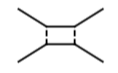} & \sim \left(\frac{g_A^2}{2f_{\pi}^2}\right)^2 \frac{M m_{\pi}}{4\pi} \tilde{f}\left(\frac{p}{m_{\pi}}\right)\\
\end{array} \ .
\eeq
Note that the factor of the nucleon mass, a large energy scale for the effective theory, comes from diagrams in which intermediate nucleons can go on-shell. This implies that an expansion parameter for the set of ladder diagrams is approximately,
\beq
\frac{g_A^2Mm_{\pi}}{8\pi f_{\pi}^2} \sim 0.5 \ ,
\eeq
and that the expansion may converge very slowly. In practice, the convergence for this formulation might be acceptable in the $^1S_0$ scattering channel, but is poor in the spin triplet channel. This is likely due to the singular tensor force contribution to the two-nucleon potential in this channel, which we will discuss in a moment \cite{Fleming:1999ee}.

Weinberg's formulation for nuclear EFT involves summing a subset of diagrams non-perturbatively, then using the resulting nucleon-nucleon potential to solve the Schrodinger equation. In doing so we can take into account higher orders in a perturbative expansion that  breaks down or converges slowly. For the pions we can iterate all possible tree level pion exchange diagrams to give the following 1PE potential \cite{Epelbaum:2010nr},
\beq
V_{\mbox{\tiny 1PE}}(\vec{r} = \left(\frac{g_A}{2f_{\pi}}\right)^2 \vec{\tau}_1 \cdot \vec{\tau}_2 \left[m_{\pi}^2\frac{e^{-m_{\pi}r}}{12\pi r}\left(S_{12}(\hat{r}) \left(1+\frac{3}{m_{\pi}r} + \frac{3}{(m_{\pi}r)^2}\right)+\vec{\sigma}_1 \cdot \vec{\sigma}_2 \right) -\frac{1}{3}\vec{\sigma}_1 \cdot \vec{\sigma}_2 \delta^3(r)\right] \ , \cr
\eeq
where $S_{12} = 3\vec{\sigma}_1 \cdot \hat{r} \vec{\sigma}_2 \cdot \hat{r} - \vec{\sigma}_1 \cdot \vec{\sigma}_2$ . 

The most divergent part of this potential, scaling like $\sim 1/r^2$, comes from the tensor force in the spin triplet channel. Attractive potentials which scale as $r^{-n}$ for $n \geq 2$ are referred to as singular potentials. Particles sitting in a singular potential eventually fall toward the center with infinite velocity, which is clearly unphysical. Thus, singular potentials can only be defined with an explicit cutoff that cannot be removed. Particles generally sit near this cutoff, rendering the system sensitive to the short-range details of the choice of boundary condition. Therefore, systems involving singular potentials are generally model dependent and we can no longer have a true effective theory because the cutoff cannot be removed. 

The reason such a singular potential arises is similar to that which led to the Wigner bound in the previous section. Again, we are attempting to sum a subset of diagrams in an effective theory non-perturbatively, which cannot in general be assumed to be a renormalizable process. In practice, nuclear theorists using so-called chiral potentials are generally able to demonstrate that the cutoff dependence is small so long as the cutoff is only varied within a particular range, typically $\Lambda \sim 300-1000$ MeV. Therefore, if we wish to include pions non-perturbatively in our lattice theory we should keep this in mind as it implies a restriction on the allowed lattice spacings, just as we found for the non-perturbative inclusion of effective range contributions. 

Pion fields may be added directly to our lattice Lagrangian in a straightforward way. The incorporation of dynamical pions, however, will likely complicate importance sampling by introducing noise and/or sign problems, and adds complexity to the Monte Carlo algorithms. Fortunately fully dynamical pions are unnecessary; all we actually seek is the addition of a term in the Lagrangian which generates the tree level diagrams between a single pion and two nucleons. The lattice formulation then non-perturbatively accounts for all possible loop diagrams involving this pion-nucleon interaction. Diagrams involving vacuum pion loops, pion self-energies, etc. are higher order in our chiral expansion and can be included perturbatively if necessary. 

One possible implementation utilized by the Bonn-Raleigh group is to use static pion auxiliary fields, $\pi_{\vec{x},\tau}^{(I)}$, with isospin $I$, and the following action \cite{Lee:2008fa,Borasoy:2006qn}:
\beq
S_{\pi\pi} = \left( \frac{m_{\pi}^2}{2} + 3\right) \sum_{\vec{x},\tau,I} \pi_{\vec{x},\tau}^{(I)} \pi_{\vec{x},\tau}^{(I)} - \sum_{\vec{x},\tau,I,k} \pi_{\vec{x},\tau}^{(I)} \pi_{\vec{x}+\hat{k},\tau}^{(I)} \ .
\eeq
Because the pions are derivatively coupled to the nucleons, the interaction term should behave like,
\beq
S_{\pi NN} \sim \frac{g_A}{2f_{\pi}}\sum_{I,k} \left[\pi^{(I)}_{\vec{x}+\hat{k}}-\pi^{(I)}_{\vec{x}-\hat{k}}\right] \psidag_{\vec{x}} \psi_{\vec{x}} \ ,
\eeq
(see \cite{Lee:2008fa} for more details on the particular interaction chosen). The pions have been chosen to only couple to the nucleons through spatial displacements. This simplifies the analysis by eliminating the renormalization of the nucleon mass through nucleon self-energy diagrams such as:

\includegraphics[width=\linewidth]{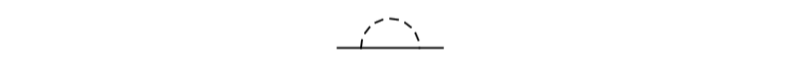}

Then we can simply utilize the physical value, $M \sim 938$ MeV, for the nucleon mass. These pions therefore act instantaneously, much the same way as they do in a pion potential picture. 

\subsection{3- and higher-body interactions}

Na\"ive dimensional analysis dictates that the leading three-body interaction should be suppressed relative to the two-body interaction by $\calO(L^3)$. We should be more cautious by this point, since we have seen dimensional analysis fail in previous cases for strongly interacting systems. For that reason, we will now inspect the three-body system more carefully. 

To begin, we will consider a system of three particles interacting via only the simplest, leading order two-body contact interaction. We will follow the discussion of \cite{Braaten:2004rn}. Let us assume that all three particles carry different quantum numbers, as they do for the triton and $^3$He, and that all pairs of particles interact via the same two-body coupling, $g_0$. To calculate the three-particle scattering amplitude for a strongly coupled system we must iterate this interaction non-perturbatively, as we did for the two-particle system. 

\begin{figure}
\begin{center}
\includegraphics[width=\linewidth]{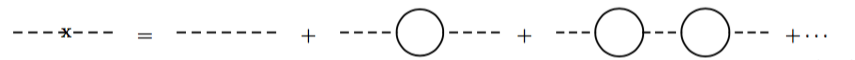}
\end{center}
\caption{\label{fig:dimerprop}Dressed propagator for the bosonic dimer field, $\phi$.}
\end{figure}

A useful trick for calculating this quantity is the addition of a bosonic dimer field, $\phi$, coupling to two fermion particles, $\psi$. This allows us to rewrite the three-particle scattering amplitude in the form of a two-particle scattering amplitude. The dimer propagator must be fully dressed by fermion loop bubbles and can be written diagrammatically as shown in \Fig{dimerprop}. This bubble sum is essentially the same as the one we have encountered several times before in these lectures. However, we must now allow external momentum, $(p_0,\vec{p})$ to flow through the diagrams, leading to the following dressed propagator for the dimer field,
\beq
D_0(p_0,\vec{p}) = \frac{1}{1-g_0 \left. I_0 \right|_{E=p_0-p^2/M}} = \frac{1/a-\Lambda}{1/a + i \sqrt{Mp_0-p^2-i\epsilon}} \ ,
\eeq 
where I've used the results from \Sec{couplings} to rewrite the coupling in terms of the scattering length, $a$, and the cutoff, $\Lambda$. We see that the dimer propagator has a pole at $p_0=\frac{p^2}{M}-\frac{1}{Ma^2}$, corresponding to a (virtual) bound state for (negative) positive scattering length with energy $E_B=\frac{1}{Ma^2}$. 

\begin{figure}
\begin{center}
\includegraphics[width=\linewidth]{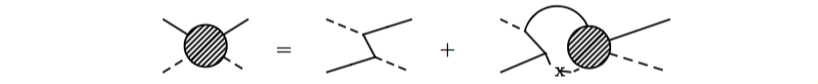}
\end{center}
\caption{\label{fig:3bodyint}Full three-particle scattering amplitude written in terms of a two-particle amplitude for a fermion scattering with a dimer field. Here we have only included two-body interactions, with no explicit three-body contact interaction.}
\end{figure}

Using this dimer field, we can write the full three-body scattering amplitude, $A_3$, as an integral equation, shown in \Fig{3bodyint}. To simplify the expression, we can set the $\psi$ fields to be on-shell, so that all off-shell properties are absorbed into the dimer propagator. The amplitude can then be written,
\beq
A_3(p,k;E,p^2/M) &=& -\frac{ g_0}{E-p^2/M-k^2/M -(p+k)^2/M+i\epsilon} \cr
&+& \frac{8\pi i}{g_0}\int\frac{d^4q}{(2\pi)^4} \left(\frac{g_0}{E-p^2/M - q_0-(p+q)^2/M+i\epsilon}\right) \cr
&\times &\left(\frac{1}{q_0-q^2/M+i\epsilon}\right)\left(\frac{A_3(q,k;E,q_0)}{1/a+i\sqrt{M(E-q_0)+q^2-i\epsilon}}\right) \ ,
\eeq
known as the Skorniakov-Ter-Martirosian (STM) integral equation. Integrating over $q_0$ and projecting the system onto the $s$-wave channel gives (see \cite{Braaten:2004rn} for more details),
\beq
\tilde{A}_3(p,k;E) &=& \frac{1}{apk}\ln \left(\frac{p^2+pk+k^2-ME-i\epsilon}{p^2-pk+k^2-ME-i\epsilon}\right) \cr
&+& \frac{1}{4\pi^2}\int^{\Lambda}dq \frac{q}{p}\ln \left( \frac{p^2+pq+q^2-ME-i\epsilon}{p^2-pq+q^2-ME-i\epsilon}\right)  \frac{\tilde{A}_3(q,k;E)}{-1/a + \sqrt{3q^2-ME-i\epsilon}} \ .
\eeq
For large scattering length (strong interaction) we have,
\beq
\tilde{A}_3(p,k;E) \underset{a\to\infty}{\longrightarrow} \frac{1}{4\pi^2}\int^{\Lambda}dq \frac{q}{p} \ln \left(\frac{p^2+pq+q^2-ME-i\epsilon}{p^2-pq+q^2-ME-i\epsilon}\right)\frac{\tilde{A}_3(q,k;E)}{\sqrt{3q^2-ME-i\epsilon}}
\eeq

This integral contains divergences, which may be renormalized by adding an explicit three-body coupling, $H$. To absorb the divergences, the coupling must have the following dependence on the momentum cutoff, $\Lambda$ \cite{Bedaque:1998km,Bedaque:1998kg,Beane:2000wh}:
\beq
H(\Lambda) = \frac{\cos \left[s_0 \ln(\Lambda/\Lambda_{*})+\tan^{-1}s_0\right]}{\cos \left[s_0 \ln(\Lambda/\Lambda_{*})-\tan^{-1}s_0\right]} \ ,
\eeq
where $s_0 \sim 1.006$ is a constant, and $\Lambda_{*}$ is some reference scale which may be set by a three-body observable, such as the triton binding energy, or the neutron-deuteron scattering length. 

\begin{figure}
\begin{center}
\includegraphics[width=0.5\linewidth]{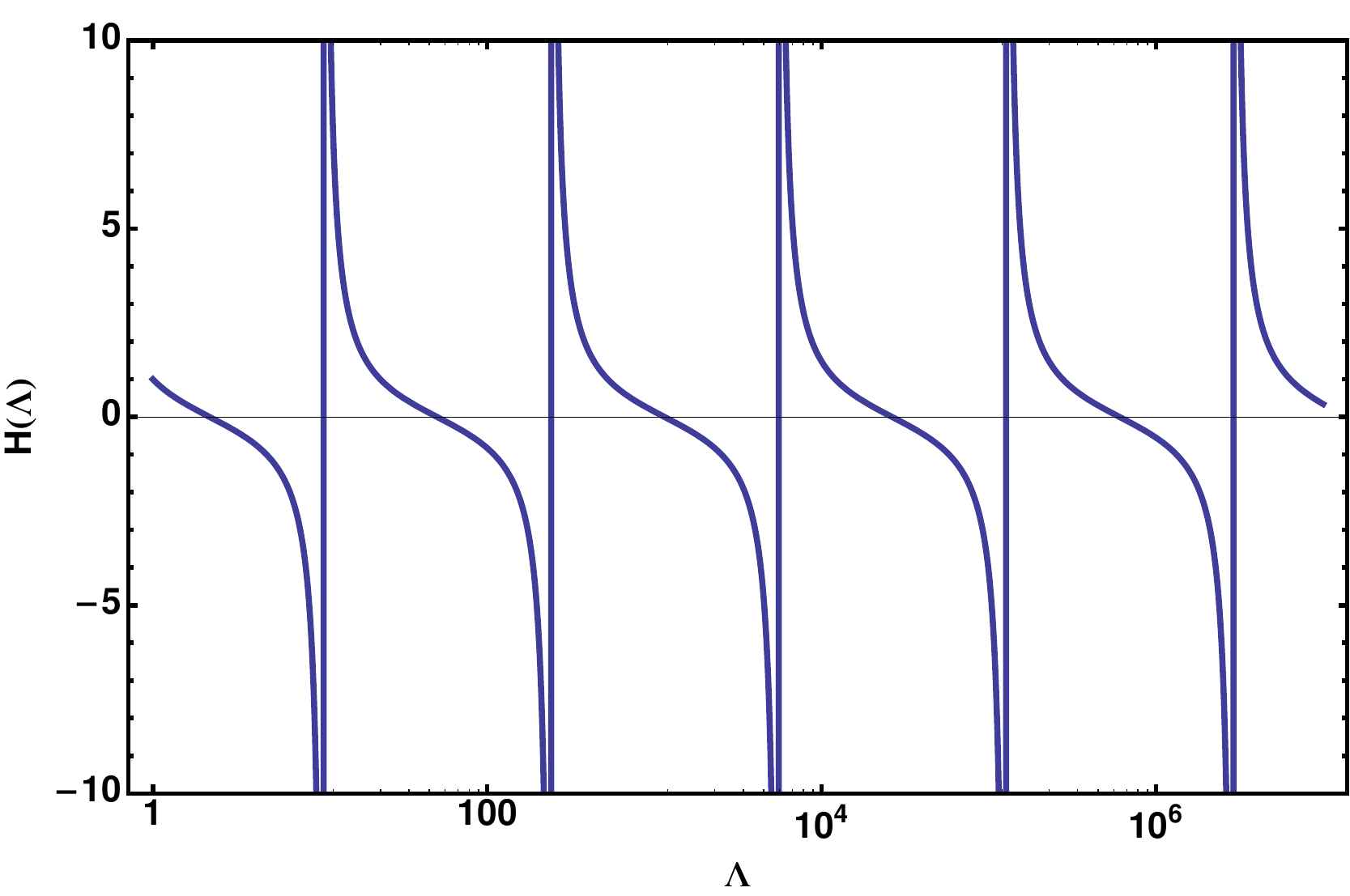}
\end{center}
\caption{\label{fig:HLambda}Running of the three-body contact interaction $H(\Lambda)$ at unitarity vs. the momentum cutoff, $\Lambda$, showing log-periodicity.}
\end{figure}

There are two remarkable things to note here: the first is that this result for the scattering amplitude is only a leading order result, yet we had to introduce a three-body coupling in order to renormalize the theory. This illustrates another case where na\"ive dimensional analysis does not work, because the three-body coupling contributes at the same order as the two-body coupling. The second is the running of the coupling $H(\Lambda)$, plotted on a logarithmic scale in \Fig{HLambda}. We see that the coupling, and therefore also observables depending on the coupling, displays a log-periodic discrete scaling symmetry, related to the so-called Efimov effect. This property arises for systems obeying a potential at the threshold of singularity, $\sim 1/r^2$, as can be shown to occur for our three-body system using hyperspherical coordinates \cite{V1970563,Efimov:1971zz}.

Because the three-body interaction has been demonstrated to be relevant at leading order, we should in general include it non-perturbatively to our lattice theory by adding an interaction term to the Lagrangian such as,
\beq
C_3 \phi_3 \psidag_{\tau} \psi_{\tau+1} \ ,
\eeq
where $C_3$ is tuned to reproduce some three-body observable, and $\phi_3 \in Z_3$ (cube roots of 1). However, $\phi_3$ is necessarily a complex field, will induce severe noise and/or sign problems. The interaction may alternatively be introduced via multiple $Z_2$ interactions, but the noise problem remains.

\Fig{HLambda} is important for our discussion because it shows how the three-body coupling runs as we change the lattice spacing. The larger the coupling, the worse the noise/sign problem will be. The solution chosen by the Bonn-Raleigh group is to tune the ratio $b_{\tau}/b_s$ until a chosen three-body observable is sufficiently well-described by tuning only the two-body interactions. This implies that the three-body interaction is small at this point, and can then be regarded as a higher-order correction and included perturbatively. A drawback to this approach is that we can no longer use the anisotropy parameter as a knob for probing temporal discretization errors. Because the spatial lattice spacing may also already be restricted by the condition of renormalizability of any pion or effective range contributions to the Lagrangian, we have forfeited most of our ability to demonstrate that discretization errors are under control.

Another possibility for reducing the contribution from the three-body interaction might be to change the short-distance behavior of the two-body sector in another way. For example, tuning different numbers of two-body interaction coefficients (\Sec{tuning}) or changing the discretization of the kinetic operator will shift the reference scale $\Lambda^*$, giving us a different value for $H(\Lambda)$ at a fixed lattice spacing.

Finally, given that the three-body sector required a reshuffling of the orders in perturbation theory at strong coupling, should we expect the same for higher $N$-body interactions? Fortunately it has been fairly well established that four- and higher body operators are not necessary to renormalize the theory at leading order and are therefore irrelevant. This means that we may treat four- and higher-body interactions as perturbative corrections.

This is observed via the so-called Tjon line (see, e.g. \cite{Hammer:2010kp}). Recall that while the two-body system at unitarity has no intrinsic scale, in order to describe the three-body system we had to introduce a single scale, $\Lambda_{*}$, to be set by some three-body observable. Once this scale is set, all other three-body observables may then be predicted. If four- and higher-body operators appear only at higher orders, then this three-body scale remains the only relevant scale in the problem, and observables must be proportional to $\Lambda_{*}$ \footnote{This single scale is also critical for the appearance of the log-normal distribution in correlators near unitarity, where the moments are given by
\beq
\calM_N \sim e^{-E_{\mbox{\tiny N-body}}\tau} \sim e^{-f(N) \Lambda_{*} \tau} \ .
\eeq
Numerical evidence was shown in \cite{Nicholson:2012zp} that $f(N)$ has the expected form for the log-normal distribution.}. This implies that varying the three-body parameter $\Lambda_{*}$, in a plot of the binding energy for the four-body system versus the binding energy of the three-body system, will result in a straight line. Any non-linear dependence on higher-order $N$-body operators contributes only within the error band predicted at this order in perturbation theory.

\subsection{\label{conclusions}Final considerations}
Perhaps the most worrisome issue we have discussed is the inability to take the continuum limit due to interactions that are included non-pertubatively and which generate new non-zero scales beyond the scattering length. The lattice spacing must also be kept reasonably large for another reason mentioned previously, related to numerical stability: if the lattice spacing becomes too small, the system will begin to probe the repulsive core of the two-body potential, leading to sign and/or signal-to-noise problems. 

Though we may not have the ability to vary the lattice spacing by significant amounts, we must still prove that our results do not depend strongly on the short-distance details of the action. This can be demonstrated instead by changing the discretization of derivatives in the action, using more or less improvement of the interaction, etc., and showing that the results do not change significantly \cite{Lee:2008fa}. 

Showing convergence of the EFT for the lattice results is also a major concern, particularly since we have no single power-counting scheme that is known to converge in all channels even in the continuum theory. One possible indication of issues with convergence in the current Bonn-Raleigh method is the need for a significant repulsive four-body interaction in order to stabilize four- and higher-body systems, which seem prone to forming four-body clusters on a single lattice site. This is akin to the particles falling to the bottom of a singular potential, and may be related to the particular tuning of the three-body interaction. However, once this interaction has been set the convergence of the results appears to be relatively stable.

Possibly the biggest open issues to be resolved are the sign/noise problems and proving convergence to the ground (or desired excited) state. Noise problems have restricted most calculations of nuclear systems to nuclei in (or near) the alpha ladder, where approximate $SU(4)$ symmetry applies. New theories and/or algorithms would be enormously helpful in this arena. The engineering of better sources or methods for extracting the desired states might be particularly beneficial for both the reduction of noise and to eliminate the need for performing long temporal extrapolations.

Despite these limitations there have been enormous successes for lattice EFT for few- and many-body states both for systems at unitarity and nuclei. As an example, at unitarity the energies of up to 50 two-component fermions have been calculated with errors comparable to state-of-the-art Green's Function Monte Carlo calculations \cite{EKLN1,EKLN2,EKLN3,EKLN4,LEKN1,NEKL1}. The Raleigh-Bonn group has calculated properties of nuclei up to $A=28$ \cite{Epelbaum:2010xt,Epelbaum:2009pd,Epelbaum:2013paa,Lahde:2013kma,Lahde:2013uqa}. Particularly exciting is their investigation of the structure of the Hoyle state, a key component of the triple alpha process necessary for Carbon production in stars \cite{Epelbaum:2011md,Epelbaum:2012qn,Epelbaum:2012iu,Epelbaum:2013wla,Lahde:2014bna}. 

\section{Reading assignments and Exercises}

\begin{enumerate}
\item Much of these lecture notes follow this review: arXiv:1208.6556. There you will also find more information about algorithms. The following is an excellent pedagogical introduction to EFT's by David B. Kaplan: arXiv:nucl-th/0510023.

\item Explore the cumulant expansion using a toy model \cite{EKLN2}: 
\beq
C(\tau,\phi) = \prod_{i=1}^{\tau}(1+g\phi_i) \ ,
\eeq
for $0\leq g \leq 1$ and $\phi \in [-1,1]$. The true mean of the correlator should be $\langle C(\tau,\phi)\rangle = 1$, corresponding to $E_0=0$. Compare the cumulant expansion cut off at various orders on a finite sample size to the mean calculated using standard methods as the sample size is varied.

\item Reading:
D. Lee: arXiv:0804.3501 \cite{Lee:2008fa}
G.P. Lepage: Analysis of algorithms for lattice field theory \cite{Lepage:1989hd}.

\item Add a term
\beq
c \psidag_{\tau} \nabla_{L}^2 \psi_{\tau-1}
\eeq
to the simple interaction, \Eq{pointint}, and derive an analytic expression for tuning the couplings, $g_0$ and $c$ in order to eliminate the effective range contribution. You may use either the scattering amplitude or the transfer matrix method.

\item Write numerical code (Mathematica will suffice) to solve the transfer matrix for two particles for a chosen set of coefficients, $g_{2n}$ (\Eq{tuningcoef}), using $L=32$, $M=5$, and tune your coefficients to match the first few expected L\"uscher eigenvalues at unitarity. Compare your results with those in Table II of Ref.~\cite{EKLN1}.
\end{enumerate}

\section*{Appendix: Compilation and running of the code}
This code requires the use of the FFTW library, which you may download and install from fftw.org. The script ``create\_lib.sh" should be run first from the head directory. Once this script is successful, you may go into the production directory, modify the script ``create\_binary.sh" to reflect your path to the FFTW library, and compile by running this script. The executable created is called ``a.out", which should be run without specifying any additional parameters in the command line. Input parameters are specified in the files included in the ``arg" folder. The parameters for each file are described in the header ``arg.h".

Output is created in the folder ``results". The file gives a list of the values (real part listed first, imaginary second) of the two-particle correlation function calculated at different values of Euclidean time, on a set of auxiliary field configurations. The organization of the output is as follows: 
\beq
\begin{array}{ccccccc}
\mathrm{Re}\left[C_{\phi_1}(\tau_1)\right] & \mathrm{Im}\left[C_{\phi_1}(\tau_1)\right] & \mathrm{Re}\left[C_{\phi_1}(\tau_2)\right] & \mathrm{Im}\left[C_{\phi_1}(\tau_2)\right] & \cdots & \mathrm{Re}\left[C_{\phi_1}(\tau_{N_{\tau}})\right] & \mathrm{Im}\left[C_{\phi_1}(\tau_{N_{\tau}})\right] \\
\mathrm{Re}\left[C_{\phi_2}(\tau_1)\right] & \mathrm{Im}\left[C_{\phi_2}(\tau_1)\right] & \mathrm{Re}\left[C_{\phi_2}(\tau_2)\right] & \mathrm{Im}\left[C_{\phi_2}(\tau_2)\right] & \cdots & \mathrm{Re}\left[C_{\phi_2}(\tau_{N_{\tau}})\right] & \mathrm{Im}\left[C_{\phi_2}(\tau_{N_{\tau}})\right] \\
&&& \vdots &&& \\
\mathrm{Re}\left[C_{\phi_{\Ncfg}}(\tau_1)\right] & \mathrm{Im}\left[C_{\phi_{\Ncfg}}(\tau_1)\right] & \mathrm{Re}\left[C_{\phi_{\Ncfg}}(\tau_2)\right] & \mathrm{Im}\left[C_{\phi_{\Ncfg}}(\tau_2)\right] & \cdots & \mathrm{Re}\left[C_{\phi_{\Ncfg}}(\tau_{N_{\tau}})\right] & \mathrm{Im}\left[C_{\phi_{\Ncfg}}(\tau_{N_{\tau}})\right] \\
\end{array} \nonumber
\eeq
where $N_{\tau}$ and $\Ncfg$ are the total number of time steps, specified in ``do.arg", and total number of configurations, specified in ``evo.arg", respectively. To calculate the correlation function at a given time, $\tau$, average over all values: $C(\tau) = \sum_i \left(\mathrm{Re}\left[C(\phi_i,\tau)\right] + i \ \mathrm{Im}\left[C(\phi_i,\tau)\right]\right)$.

\subsection{Exercises}
\begin{enumerate}
\item Set the first value in the file ``interaction.arg" to a coupling of your choice, and the remaining couplings to $0$. Use the long time behavior of the effective mass function, $\ln \frac{C(\tau)}{C(\tau+1)} \tautoinfty E_0$ (see \Sec{observables}), to determine the ground state energy for your choice of coupling, $g$. Compare this with what you expect from \Eq{eigeqlambda}, using the relation $\lambda = e^{-E_0}$, as the number of lattice points is increased. You may test the improved interaction, \Sec{tuning}, using coefficients calculated from your code developed in Prob.~4 by setting multiple couplings in the ``interaction.arg" file. Be careful to set the dispersion relation in ``kinetic.arg" to match the one used in setting up your transfer matrix for the tuning.

\item Add a harmonic potential by setting the parameters in potential.arg. The three numerical values correspond to the spring constant, $\kappa$, for the $x,y,z$-directions. Set the interaction coefficients to correspond to unitarity, then find the energies of two unitary fermions in a harmonic trap, exploring and removing finite volume and discretization effects by varying the parameters, $L,L_0=\left(\kappa M\right)^{-1/4}$, and performing extrapolations in these quantities if necessary. Compare your result to the expected value of 2$\omega$, where $\omega = \sqrt{\kappa/M}$, and the mass $M$ is set in the file ``kinetic.arg". 

\item Construct sources for three fermions in an $l=0$ and $l=1$ state and find the lowest energies corresponding to each state at unitarity. Which $l$ corresponds to the true ground state of this system?
\end{enumerate}

\begin{acknowledgements}
The author would like to thank Michael Endres, David B. Kaplan, and Jong-Wan Lee for extensive discussions, and especially M. Endres for the development of and permission to use this code. AN was supported in part by U.S. DOE grant No. DE-SC00046548. 
\end{acknowledgements}

\bibliography{chapter5}

\end{document}